\documentclass[a4paper,fleqn]{cas-sc}

\usepackage{subfigure}  % Add this in your preamble if not already included
\usepackage{nameref}

\usepackage{graphicx}
\usepackage{booktabs}

\usepackage{amsthm}
\usepackage[numbers]{natbib}

\usepackage{algorithm}
\usepackage{algorithmic}	
\usepackage[english]{babel}
\newtheorem{theorem}{Theorem}

\def\tsc#1{\csdef{#1}{\textsc{\lowercase{#1}}\xspace}}
\tsc{WGM}
\tsc{QE}
\tsc{EP}
\tsc{PMS}
\tsc{BEC}
\tsc{DE}

\begin{document}
	\let\WriteBookmarks\relax
	\def\floatpagepagefraction{1}
	\def\textpagefraction{.001}
	\shorttitle{A Novel W-PIRNNs Framework}
	\shortauthors{Barman, Ray}
	%\begin{frontmatter}
	
	\title [mode = title]{An Efficient Wavelet-based Physics Informed Residual Neural Networks for Flow Field Reconstruction with Extremely Sparse Data}                      
	%\tnotemark[1,2]
	
	%\tnotetext[1]{This document is the results of the research
%		project funded by the National Science Foundation.}
	
%	\tnotetext[2]{The second title footnote which is a longer text matter
%		to fill through the whole text width and overflow into
%		another line in the footnotes area of the first page.}
	\author[1]{Biswanath Barman}
        \ead{d23165@students.iitmandi.ac.in}
        \author[1]{Rajendra K. Ray}  % <- Same affiliation [1], plus corresponding mark
        
        \cortext[cor1]{Corresponding author}
        \ead{rajendra@iitmandi.ac.in}
        
        \credit{Conceptualization of this study, Methodology, Software (BB); Data curation, Writing - Original draft preparation (RKR)}
        
        \address[1]{School of Mathematical and Statistical Sciences, Indian Institute of Technology Mandi\\
        Mandi, Himachal Pradesh, 175005, India}

	% \author[1]{Biswanath Barman}
	% % \cormark[1]
	% % \fnmark[1]
	% % \ead{shinta@mail.ac.id}
	% %\ead[url]{www.cvr.cc, cvr@sayahna.org}
	% \credit{Conceptualization of this study, Methodology, Software}
	% \address[1]{School of Mathematical and Statistical Sciences,            Indian Institute of Technology Mandi \\
 %         Mandi, Himachal Pradesh, 175005, India}
	% \author[2]{Rajendra K. Ray}%[style=chinese]
	% % \fnmark[2]
 %        \cormark[1]
	% \ead{rajendra@iitmandi.ac.in}
	% %\ead[URL]{www.sayahna.org}
	% \credit{Data curation, Writing - Original draft preparation}
	% % \address[2]{Department of Computer }
	% 	\cortext[1]{Corresponding author}
%	\cortext[cor2]{Principal corresponding author}
%	\fntext[fn1]{This is the first author footnote. but is common to third
%		author as well.}
%	\fntext[fn2]{Another author footnote, this is a very long footnote and
%		it should be a really long footnote. But this footnote is not yet
%		sufficiently long enough to make two lines of footnote text.}
	
%	\nonumnote{This note has no numbers. In this work we demonstrate $a_b$
%		the formation Y\_1 of a new type of polariton on the interface
%		between a cuprous oxide slab and a polystyrene micro-sphere placed
%		on the slab.
%	}

\begin{abstract}
          This paper introduces wavelet-physics-informed residual neural networks (W-PIRNNs) to study complex fluid flow problems by reconstructing the flow field from highly sparse, supervised data. Our W-PIRNNs fundamentally integrate ResNet and employ the wavelet $W(t) = w_1 \sin(t) + w_2 \cos(t)$ as an activation function. Due to the vanishing and ballooning gradient problems associated with typical PINNs' deep networks, we implemented residual-based skip connections.  Our W-PIRNNs, which integrate supervised data with physical principles, demonstrate efficacy even in scenarios of sparse or partial data, enabling the reconstruction of flow fields using merely $0.05\%$ velocity data for training. The wake flow around a circular cylinder served as the test case for our proposed technique, which depends exclusively on velocity data for training. This technique facilitates the precise reconstruction of velocity, pressure, streamlines, and vorticity, requiring fewer epochs and less processing time. Significantly, our proposed W-PIRNNs effectively resolve PDEs in both forward and inverse contexts. Burger's equation served as a test case for both the forward and inverse problem configurations. Our network calculates the diffusion or viscosity coefficient ($\lambda_2$) with an absolute error of $0.065\%$ and the convection coefficient ($\lambda_1$) with an absolute error of $0.002\%$. Furthermore, the Schrödinger equation is examined in the forward setting to assess the framework's ability to handle periodic boundary conditions. To the best of our knowledge, W-PIRNNs represent the first method capable of flow reconstruction using highly sparse supervised data, as well as reconstructing streamline and vorticity, and they effectively address both forward and inverse problems with high accuracy.
	\end{abstract}


%	\begin{graphicalabstract}
	%	\includegraphics{}%{figs/grabs.pdf}
	
	\maketitle
	%============================= INTRODUCTION===============================

\section{Introduction}
	
Deep learning (DL) has opened new opportunities for modeling and forecasting the dynamics of complex multiscale systems due to its remarkable ability to approximate nonlinear functions from data. The development of various architectures, including convolutional neural networks (CNNs) \citep{lecun2002gradient} and long short-term memory (LSTM) \citep{hochreiter1997long}, has enabled multiple deep learning frameworks for modeling complex physical systems by integrating spatio-temporal dependencies into predictions. Deep learning has been applied in diverse scientific and technical domains, including astronomy \citep{waldmann2019mapping}, climate modeling \citep{ham2019deep}, solid mechanics \citep{mianroodi2021teaching}, chemistry \citep{segler2018planning}, and sustainability \citep{vinuesa2020role, larosa2023halting}. The field of fluid mechanics has recently been a prominent area of research for the development of innovative deep-learning approaches \citep{duraisamy2019turbulence, brunton2020machine, vinuesa2022enhancing}. The effective application of deep neural networks (DNNs) has been shown in data-driven turbulence closure modeling \citep{ling2016reynolds,jiang2021interpretable}, forecasting temporal dynamics of low-order turbulence models \citep{srinivasan2019predictions}, deriving turbulence theory for two-dimensional decaying isotropic turbulence \citep{jimenez2018machine}, non-intrusive sensing in turbulent flows \citep{guastoni2021convolutional}, and active flow control through deep reinforcement learning \citep{rabault2019artificial}.

The foundational work on learning in experimental fluid mechanics dates back to the early 1960s, focusing on drag reduction \citep{rechenberg1964cybernetic}. Then, in the early 1990s, diverse neural network-based approaches were developed for trajectory analysis and classification in particle tracking velocimetry (PTV) \citep{adamczyk19882} and particle image velocimetry (PIV) \citep{adrian1984scattering,grant1995investigation}. In recent decades, the development of deep learning techniques has revolutionized image processing and computer vision. These techniques have enabled their application to the processing and enhancement of PIV measurements, enabling this revolution. Enhancing spatial resolution, recovering missing data, reconstructing three-dimensional fields from sparse observations, and post-processing have been the main areas of focus for the research.

The reconstruction of high-resolution images from low-resolution counterparts, known as super-resolution reconstruction, has been a prominent research focus in computer science \citep{wang2020deep}. Convolutional neural networks (CNNs) \citep{fukami2019super,liu2020deep}, super-resolution generative adversarial networks (SRGANs) \citep{guemes2021coarse}, and unsupervised learning using cycle-consistent GANs (CycleGANs) \citep{kim2021unsupervised} have all been used in fluid mechanics and turbulence due to the effectiveness of deep learning in super-resolution tasks in computer science. Deep learning has also been utilized for the super-resolution of four-dimensional (4D) flow magnetic resonance imaging (MRI) data \citep{ferdian20204dflownet}. Supervised deep-learning super-resolution algorithms are limited by the necessity for high-resolution labels during training, requiring a paired dataset of low and high-resolution images, which is difficult to get experimentally. Recently demonstrated the ability of GANs \citep{guemes2021coarse} to produce high-resolution fields from PIV data without prior access to any high-resolution field. 

In experimental fluid mechanics, predicting and reconstructing flow data from incomplete data is a frequent problem. Gappy proper orthogonal decomposition (Gappy POD) \citep{everson1995karhunen} has demonstrated efficacy in reconstructing flow fields from incomplete data \citep{bui2004aerodynamic}; however, it cannot be directly utilized with incomplete original data as it necessitates spatial coefficients derived from reference data \citep{willcox2006unsteady}. Recently introduced an autoencoder-type CNN \citep{morimoto2021convolutional} for estimating PIV data, effectively tackling the challenge of missing regions and overcoming existing limitations. Nowadays, machine learning and deep learning have been employed for flow-field reconstruction from sparse turbomachinery data \citep{akbari2021reconstruction}. Later \citep{stulov2021neural} introduced a CNN-based architecture that uses both two-dimensional and three-dimensional convolutions to reconstruct three-dimensional data from a limited number of two-dimensional sections in a computationally efficient manner. Additionally, advances in computer vision related to the motion estimation problem \citep{cai2019particle} are driving the use of deep learning for end-to-end PIV, as demonstrated within the experimental fluid mechanics community by \citep{rabault2017performing}.

Deep-learning methodologies are powerful modeling tools for data-intensive domains such as vision, language, and audio. Nonetheless, generating interpretable information and collecting generalizable knowledge remain barriers, particularly in areas with insufficient data (Rudin, 2019; Vinuesa, 2021). Data-driven methodologies require extensive datasets for training, which may be inaccessible due to numerous technical challenges. Furthermore, these models may align well with the observational data yet fail to adhere to the fundamental rules of physics. Therefore, it is essential to incorporate governing physical laws and domain knowledge into model training, providing informative priors that enhance the model's understanding of the empirical, physical, or mathematical aspects of the system, along with the observational data. Physics-informed neural networks (PINNs), developed by Raissi, Perdikaris, and Karniadakis in 2019 \citep{raissi2019physics}, offer a framework for integrating data with physical laws represented by governing partial differential equations (PDEs) to learn. This may produce models that are robust to incomplete data, including missing or incorrect values, and can generate accurate, physically consistent predictions \citep{karniadakis2021physics}. 

PINNs are very good at addressing ill-posed and inverse problems connected with various types of PDEs. Subsequently, PINNs were employed to model vortex-induced vibrations \citep{raissi2019deep} and to tackle ill-posed inverse fluid mechanics problems \citep{raissi2020hidden}. Jin et al. (2021) \citep{jin2021nsfnets} showed that Physics-Informed Neural Networks (PINNs) are effective for direct turbulence simulation, with excellent agreement between PINNs and DNS results. Physics-Informed Neural Networks (PINNs) have been utilized to address Reynolds-averaged Navier–Stokes (RANS) equations without a predetermined model or assumptions on turbulence, employing data as the turbulence closure, as demonstrated in this article \citep{eivazi2022physics}. Moreover, PINNs have been utilized for the purpose of super-resolution and denoising of cardiovascular-flow magnetic resonance imaging (MRI) data \citep{fathi2020super}, as well as for the purpose of predicting near-wall blood flow from sparse data \citep{arzani2021uncovering}. Recently, \citep{cai2021flow} introduced a method that uses Physics-Informed Neural Networks (PINNs) to infer the complete continuous three-dimensional velocity and pressure fields from snapshots of three-dimensional temperature fields acquired via tomographic background-oriented Schlieren (Tomo-BOS) imaging. A comprehensive analysis of the current trends in integrating physics into machine-learning algorithms and various applications of Physics-Informed Neural Networks (PINNs) is presented in the study by \citep{karniadakis2021physics}, while a review of PINNs applications in fluid mechanics is offered by \citep{cai2021physics}. Eivazi et al. (2024) \citep{eivazi2024physics} demonstrate that Physics-Informed Neural Networks (PINNs) may achieve super-resolution of flow fields in both temporal and spatial dimensions utilizing a limited dataset of measurements. This is achieved without requiring high-resolution targets. Subsequently, \citep{xu2023practical} employed Physics-Informed Neural Networks (PINNs) to reconstruct velocity and pressure from sparse and absent velocity data; they achieved commendable results, although reconstruction of these parameters was not feasible when the supervised velocity data was extremely sparse. During training, the losses exhibit considerable fluctuations among epochs, indicating that the approach lacks stability in convergence. Their algorithm requires $3000$ epochs to achieve convergence in loss. Reconstructing high-resolution flow fields from sparse data is the goal of a recently proposed physics-informed convolutional network using feature fusion (FFPICN) \citep{liu2025physics}. However, these algorithms are ineffective when data sparsity is excessively high in both spatial and temporal dimensions. Additionally, they require $3000$ epochs to achieve convergence of the loss, which also exhibits fluctuations throughout the epochs. Both algorithms reconstruct only velocity and pressure, without expanding their coverage, and both are ineffective when supervised data is very sparse. Due to the scarcity of data in experimental fluid mechanics, our objective is to develop a novel algorithm that achieves exceptional performance under high data sparsity.

We are expanding this study to reconstruct additional physical phenomena, including vorticity ($\omega$). Specifically, we aim to analyze the Von Kármán street of the circular cylinder at Re=100, streamlines, pressure (p), and actual velocity, using only sparse training data comprising velocity measurements. Despite the training data sparsity reaching $0.05\%$, our proposed approach demonstrates high accuracy in reconstructing the flow field. Given the significant limitations of the available velocity data for training, our proposed W-PIRNNs rely on the model's fundamental physics, specifically its ability to incorporate key principles of fluid dynamics, such as the Navier-Stokes equations. In summary, our proposed methodology leverages data and the fundamental physical phenomena of the problems to produce a robust, accurate reconstruction of the flow field. Further, we demonstrate the efficacy of our proposed method by reconstructing the flow field past a circular cylinder at Reynolds number (Re=3900), when turbulence becomes more chaotic.

\section*{Main Contributions}

This study introduces a novel framework, \textbf{Wavelet-Physics-Informed Residual Neural Networks (W-PIRNNs)}, designed to reconstruct complex fluid flow fields from significantly sparse measurements while preserving the physical consistency of the governing equations. The main contributions of our current research work are as follows:
% The principal conclusions of our research were as follows:

 \begin{itemize}

 \item \textbf{W-PIRNNs for sparse flow field reconstruction:} We present a wavelet-based physics-informed residual network that effectively reconstructs velocity, vorticity, pressure, and streamlines, even with as little as $0.05\%$–$1\%$ velocity data. The model demonstrates exceptional performance on complex unsteady flows and significantly diminishes data requirements and training durations.

    \item \textbf{Wavelet-based activation function:} However, Uddin et al. (2023) \citep{uddin2023wavelets} first introduced core wavelet (e.g., Morlet, Haar)-based PINN applications in fluid flow problems. In our case it is not a core wavelet, but it is the single-frequency Fourier activation. An innovative activation function is implemented, defined as
    $$
    W(t) = w_1 \sin(t) + w_2 \cos(t),
    $$
    where $w_1$ and $w_2$ are learnable parameters. This wavelet-based formulation enhances the network's capacity to detect multiscale structures and periodic flow features, making it especially adept at resolving localized vortical regions and oscillatory events in fluid dynamics.

    \item \textbf{Residual architecture with skip connections:} The fundamental network architecture utilizes residual blocks with skip connections to mitigate vanishing gradient problems and enhance convergence. This approach enables the robust training of deeper networks and guarantees efficient learning within physics-informed constraints.

    \item \textbf{Adaptive learning rate strategy:} We implement adaptive learning rate scheduling to improve convergence stability. This eliminates manual tuning and expedites optimization by dynamically modifying learning rates according on gradient behavior throughout training.

    \item \textbf{Generalization to forward and inverse problems:} Beyond flow reconstruction, the proposed W-PIRNN framework is capable of handling both forward and inverse problems governed by nonlinear partial differential equations. To evaluate its broader application, we use the one-dimensional Burgers equation as a sample benchmark. In the inverse setting, the model effectively identifies the convection coefficient \(\lambda_1\) with an absolute error of \(0.002\%\), and the diffusion (viscosity) coefficient \(\lambda_2\) with an absolute error of \(0.065\%\), demonstrating excellent parameter identification accuracy with minimum supervision. Furthermore, to test the proposed approach's effectiveness in implementing periodic boundary conditions, we expand the framework to solve the Schrödinger equation as an additional validation example.
\end{itemize}

In every instance, the proposed W-PIRNNs provide precise solutions with reduced computational expense and expedited convergence relative to conventional PINN \citep{xu2023practical, raissi2019physics} and FFPICN \citep{liu2025physics}. Its versatility across tasks—spanning sparse-data-driven flow reconstruction to parameter identification in nonlinear PDEs—makes it an attractive tool for both physical modeling and data assimilation.

This article is organized as follows: In Section~\ref{sec:preliminary}, we provide an introduction to physics-informed neural networks (PINNs), residual neural networks (ResNet), the Min-Max data normalization technique, the wavelet activation function, the proposed W-PIRNNs method, and the proposed method with \textbf{Algorithm}~\ref{alg:wavelet_pinn}. Section~\ref{sec:methodology} discusses flow reconstruction problems. Section~\ref{sec:Methodological validation} analyzes our entire algorithm for solving forward and inverse problems that are governed by Burgers' equation and the Schrodinger equation. In Section~\ref{sec:conclusion}, we summarize the numerous capabilities of W-PIRNNs in reconstructing flow fields from highly sparse data, tackling both forward and inverse problems, and after that, we include the ablation study of hyperparameters and computational cost in Appendices A and B.

	%============================= RELATES WORKS===============================

\section{Methodology}
\label{sec:preliminary}
% \subsection{Overview}
\subsection{Physics-informed neural networks (PINNs)}
Various fluid flows, such as cavity flow, pipeline flow, and flow over a bluff body, can be described by the incompressible Navier-Stokes equation, expressed as follows \cite{ding2004simulation}:

% \begin{align}
% \mathcal{F}(\mathbf{u}, p) = 0 \Rightarrow 
% \begin{cases}
% \nabla \cdot \mathbf{u} = 0, & \quad \mathbf{x}, t \in \Omega_{f,t},\ \theta \in \mathbb{R}^d \\
% \frac{\partial \mathbf{u}}{\partial t} + (\mathbf{u} \cdot \nabla) \mathbf{u} + \frac{1}{\rho} \nabla p - \nu \nabla^2 \mathbf{u} = 0, & \quad \mathbf{x}, t \in \Omega_{f,t},\ \theta \in \mathbb{R}^d
% \end{cases}
% \label{eq:navier_stokes}
% \end{align}
\begin{align}
\mathcal{G}(\boldsymbol{v}, p) = 0 \Rightarrow 
\begin{cases}
\nabla \cdot \boldsymbol{v} = 0, & \quad \boldsymbol{x}, t \in \mathcal{D},\ \theta \in \mathbb{R}^n \\
\frac{\partial \boldsymbol{v}}{\partial t} + (\boldsymbol{v} \cdot \nabla) \boldsymbol{v} + \frac{1}{\rho} \nabla p - \nu \nabla^2 \boldsymbol{v} = 0, & \quad \boldsymbol{x}, t \in \mathcal{D},\ \theta \in \mathbb{R}^n
\end{cases}
\label{eq:governing_equation}
\end{align}

Where $\boldsymbol{v}$ denotes the velocity field (including streamwise and spanwise velocities); $p$ represents the pressure field; and $\rho$ and $\nu$ indicate the fluid's density and viscosity, respectively; $\mathcal{D}$ denotes the spatiotemporal domain; $\theta$ represents the parameters in the Navier-Stokes equations, encompassing boundary condition data, fluid characteristics, and domain values, among others.

The resulting solution, derived under the stated assumptions, is uniquely determined by the specified initial and boundary conditions and can be written as follows:

\begin{align}
\begin{cases}
\mathcal{I}(\boldsymbol{x}, p, \boldsymbol{v}, \theta) = 0, & \quad \boldsymbol{x} \in \mathcal{D}, t = 0,\ \theta \in \mathbb{R}^n \\
\mathcal{B}(t, \boldsymbol{x}, p, \boldsymbol{v}, \theta) = 0, & \quad \boldsymbol{x}, t \in \partial \mathcal{D} \times [0, T],\ \theta \in \mathbb{R}^n
\end{cases}
\label{eq:initial_boundary}
\end{align}

Let $\mathcal{I}$ and $\mathcal{B}$ denote the differential operators corresponding to the initial and boundary conditions, respectively. Upon determining the parameters $\theta$, the resolution of the flow dynamics (i.e., $\boldsymbol{v}(t, \boldsymbol{x})$ and $p(t, \boldsymbol{x})$) can be numerically computed by discretizing Equations (\ref{eq:governing_equation}) and (\ref{eq:initial_boundary}) using the finite difference method (FDM), finite volume method (FVM), or finite element method (FEM). This method requires the creation of large meshes and the iterative resolution of nonlinear issues, which is inherently time-intensive. Furthermore, certain parameters in fluid dynamics are challenging to ascertain. The generation of computational meshes presents a challenge when dealing with complex geometries in numerical simulations. A deep learning methodology employing Physics-Informed Neural Networks (PINNs) \citep{raissi2019physics}, implemented as a Feedforward Neural Network (FNN), has been developed to enable fast and efficient calculations, with forward/inverse uncertainty quantification and optimization modes, specifically for fluid dynamics governed by the Navier-Stokes equations. The data sample can be obtained by numerical simulations or experiments.

Since the 1980s, artificial neural networks (ANNs) have been the focus on artificial intelligence research. It builds a basic model, abstracts the human brain's neural network from an information-processing perspective, and generates several networks with different connection patterns. An artificial neural network (ANN) can be mathematically characterized as a directed graph consisting of vertices representing neurons and edges denoting connections. There are various types of neural networks, such as fully connected neural networks (FC-NN), convolutional neural networks (CNN), and recurrent neural networks (RNN), to do more complex tasks \citep{piscopo2019solving}.

This article employs a feedforward neural network, specifically a multilayer perceptron (MLP), to address the problem. A fully connected neural network (FC-NN) architecture has been defined by:

\begin{align}
{z}_i = g_i \left({W}_i{z}_{i-1} + {{b}_i}\right)
\label{eq:fc_nn}
\end{align}

Here, ${z}_i$ refers to the hidden layers located between the input and output layers; the subscript $i$ indicates the layer index; ${W}_i$ and ${b}_i$ denote the transposed weight matrix and bias vector, respectively; ${g_i}(\cdot)$ signifies the activation function (e.g., sigmoid, ReLU, tanh, etc.) that enhances nonlinear processing abilities. The fluid flow's velocity and pressure can be determined using the feedforward networks outlined in Equation~\eqref{eq:fc_nn}. The computing cost of the fully connected neural network (FC-NN) is minimal in comparison to traditional numerical simulations, as only weight matrices and bias vectors require training.

Traditionally, the deep learning approach for addressing fluid dynamics problems generates input-output connections. The solution can be determined using a black-box surrogate model, such as a fully-connected neural network (FC-NN) or a convolutional neural network (CNN), and the method is as follows:

\begin{align}
\mathrm{f}(t, x, \theta) \approx \tilde{\mathrm{f}}(t, x, \theta) \triangleq {z}_i(t, x, \theta; {W}, {b})
\label{eq:surrogate}
\end{align}

Let $\mathrm{f}(t, x, \theta)$ denote the exact solution of fluid dynamics that includes velocity and pressure fields; $\mathbf{z}_l(t, x, \theta, {W}, {b})$ indicate the neural network approximation; and $\tilde{\mathrm{f}}(t, x, \theta)$ represent the local minimizer trained by the fully connected neural network (FC-NN).

The loss function may be expressed as:

\begin{align}
\mathcal{L}_{\text{data}}({W}, {b}) = \frac{1}{\mathcal{N}} \left| \mathrm{f}^d(t, \boldsymbol{x}, \theta) - {z}_i(t, \boldsymbol{x}, \theta; {W}, {b}) \right|^2,
\quad
{W}^*, {b}^* = \underset{{W}, {b}}{\operatorname{argmin}} \ \mathcal{L}_{\text{data}}({W}, {b})
\label{eq:loss}
\end{align}

In this context, $\mathcal{L}_{\text{data}}({W}, {b})$ signifies the loss function, referred to as "data-based loss" by \cite{sun2020surrogate}; ${W}^*$ and ${b}^*$ represent the optimized weights and biases; and $\mathrm{f}^d(t, \boldsymbol{x}, \theta)$ denotes the training data.

However, the previously described classic surrogate model is a black-box model that requires a significant amount of training data to establish an input-output relationship. Furthermore, the predictions of the black-box model on unidentified data lack reliability due to its absence of physical interpretability. This research introduces the Physics-Informed Neural Network (PINNs). The loss function of Physics-Informed Neural Networks (PINNs) is defined as follows.

\begin{align}
\mathcal{L}_{\text{residual}}({W}, {b}) = 
& \underbrace{\frac{1}{\mathcal{N}_u} \sum_{i=1}^{\mathcal{N}_u} \left| {u}(t_{{u}}^i, x_{{u}}^i) - {u}^i \right|^2}_{\text{Initial or Boundary conditions}} 
+ \underbrace{\frac{1}{\mathcal{N}_f}\sum_{i=1}^{\mathcal{N}_f} \left| \frac{\partial \boldsymbol{v}}{\partial t} + (\boldsymbol{v} \cdot \nabla) \boldsymbol{v} + \frac{1}{\rho} \nabla p - v \nabla^2 \boldsymbol{v} \right|^2}_{\text{Navier-stokes Residual}} \notag \\
& + \underbrace{\frac{1}{\mathcal{N}_f} \sum_{i=1}^{\mathcal{N}_f} \left| \nabla \cdot \boldsymbol{v} \right|^2}_{\text{Mass conservation}};
\quad
{W}^*, {b}^* = \underset{{W}, {b}}{\operatorname{argmin}} \ \mathcal{L}_{\text{residual}}({W}, {b})
\label{eq:pinn_loss}
\end{align}

\begin{figure}[htbp]
    \centering
    \includegraphics[width=1.0\textwidth]{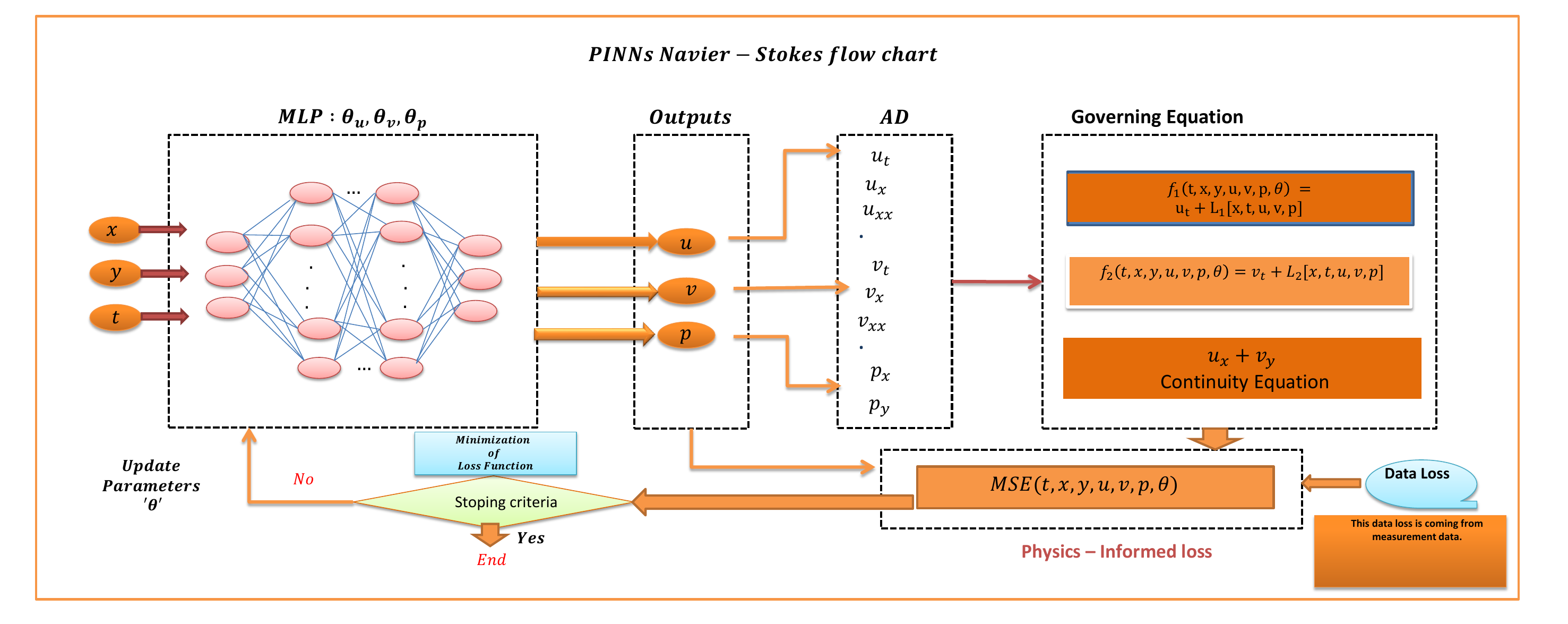}  % <-- Replace with your actual image file name
    \caption{Schematic of PINNs' architecture to solve partial differential equations (PDEs).}
    \label{fig:pinns_architecture}
\end{figure}

The loss function incorporates the mass conservation equation, the initial and boundary conditions, and the Navier-Stokes equations. The first and second derivative terms of velocity and pressure are immediately calculated using automatic differentiation (AD) \citep{baydin2018automatic}. Automatic differentiation is a technique that lies between symbolic and numerical differentiation. Numerical differentiation prioritizes the direct substitution of values into a numerical approximation first, whereas symbolic differentiation focuses on resolving algebraic expressions prior to replacing numerical values. Automatic differentiation employs symbolic differentiation on fundamental operators (such as constants, power functions, exponentials, logarithms, and trigonometric functions), subsequently replaces numerical values, retains intermediate results, and finally applies the process to the entire function. Consequently, automatic differentiation can proficiently circumvent truncation and round-off problems. This paper implements the automatic differentiation technique utilizing the PyTorch framework. The Adam algorithm is utilized to optimize the loss function, with further details provided in \citep{adam2014method} Figure~\ref{fig:pinns_architecture} illustrates a schematic illustration of the proposed technique for general PDEs.
% Burgers' equation.

% \newpage
	\subsection{Residual Neural Network (ResNet)}
Deep neural networks can be viewed as a composition of nonlinear transformations applied to the input. In the limiting case where all hidden layers learn identity mappings, a deep network effectively reduces to a shallow one. However, learning an identity mapping directly, i.e., \(\mathcal{H}(x)=x\), is often difficult for standard deep architectures. Residual learning addresses this issue by reformulating the problem to learn a residual function \(\mathcal{F}(x)=\mathcal{H}(x)-x\), such that the desired mapping is expressed as \(\mathcal{H}(x)=\mathcal{F}(x)+x\). When the residual function vanishes, the identity mapping is recovered naturally, simplifying optimization and improving gradient flow in deep networks.

\begin{figure}[htbp]
    \centering
    \includegraphics[width=0.6\textwidth]{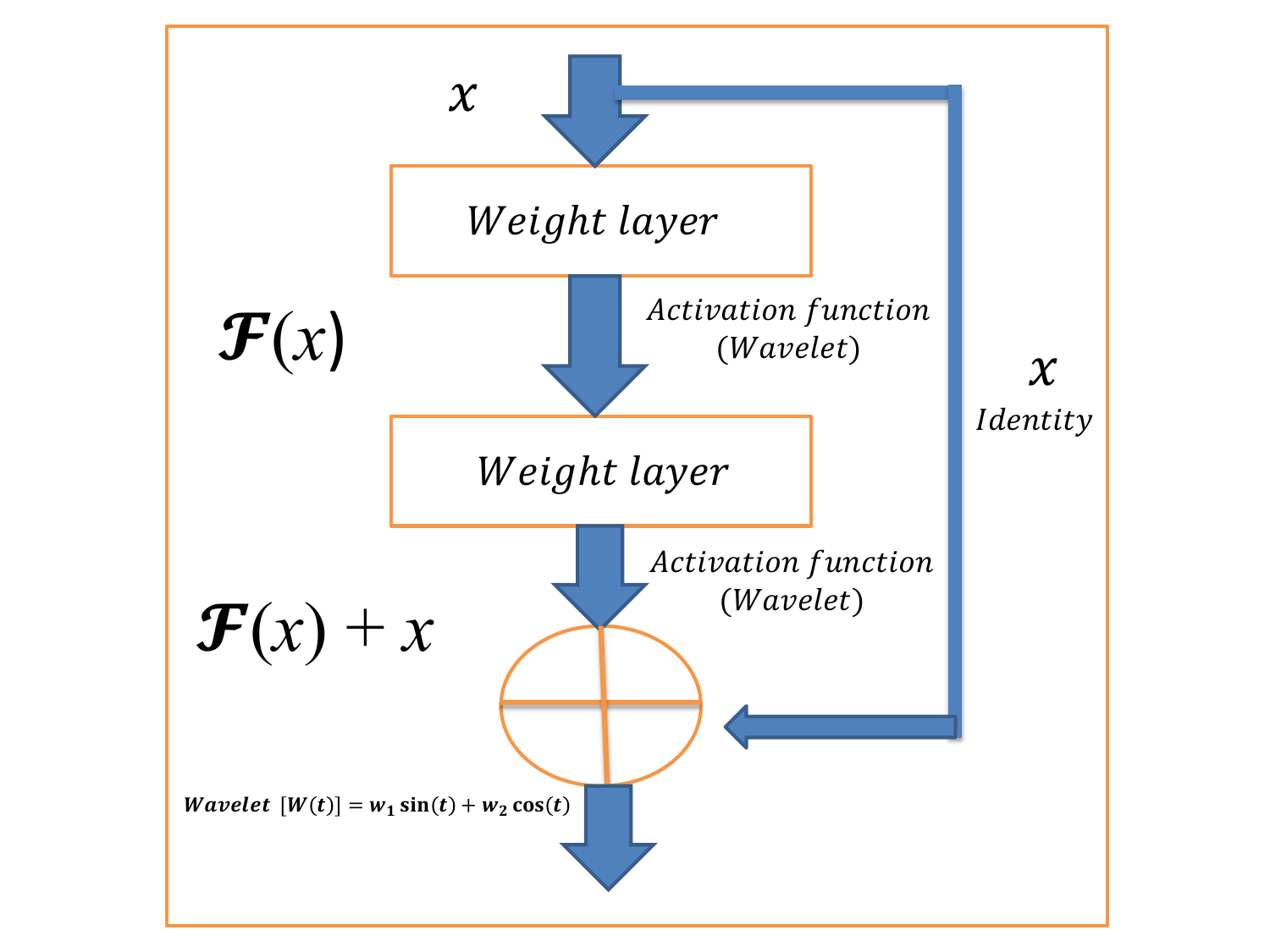}  % Replace with your actual image file name
    \caption{Residual learning: a building block.}
    \label{fig:res_block}
\end{figure}

Following the residual learning paradigm first introduced by He et al.~\cite{he2016deep}, the residual block employed in this work is illustrated in Figure.~\ref{fig:res_block}. Mathematically, a residual block can be written as
\begin{equation}
y = \mathcal{F}\left(x,\{W_i\}\right) + x,
\label{eq:residual_block}
\end{equation}
where \(x\) and \(y\) denote the input and output of the block, respectively, and \(\mathcal{F}(x,\{W_i\})\) represents the residual mapping parameterized by trainable weights.

As shown in Figure~\ref{fig:res_block}, the shortcut connection directly propagates the input \(x\) to the output, while the residual block consists of two linear transformations followed by wavelet-based activation functions. In this work, the residual mapping is defined as
\begin{equation}
\mathcal{F}(x) = W_2 \, W\!\left(W_1 x\right),
\label{eq:residual_formula}
\end{equation}
where \(W_1\) and \(W_2\) are weight matrices, and \(W(\cdot)\) denotes the wavelet-inspired activation function given by
$$
W(t) = w_1 \sin(t) + w_2 \cos(t).
$$
This formulation allows the network to combine the benefits of residual learning with oscillatory feature representation, which is particularly effective for capturing unsteady and periodic dynamics in flow reconstruction problems.

\subsection{Min-Max Normalization}
Normalization plays a crucial role in the data preprocessing phase of deep learning, aiming to modify the data range to improve processing efficiency. Various normalization methods, such as Min-Max normalization, Z-score normalization, and decimal scaling normalization \citep{patro2015normalization}, are widely used today. This research opts for the Min-Max normalization technique because it effectively removes scale differences between different features by compressing the data into a defined range. The Min-Max normalization is expressed as follows:

$$
X^*=\frac{X-X_{\min }}{X_{\max }-X_{\min }},
$$

In this context, $X^*$ denotes the normalized spatial coordinates $\left(x^*, y^*, z^*\right)$ in conjunction with the temporal coordinate $t^*$. The terms $X_{\text {min }}$ and $X_{\text {max }}$ represent the minimum and maximum values of $X$, respectively. Min-Max normalization is a method that scales data to fit within a specified range of $[0,1]$. This procedure involves adjusting the data by subtracting the minimum value and subsequently dividing by the range, which is the difference between the maximum and minimum values. By scaling the data to a fixed range, it facilitates the comparison of various features.

In our innovative wavelet physics-informed residual neural networks (W-PIRNNs), we use the custom wavelet function $W(t) = w_1\sin(t) + w_2\cos(t)$ as an activation function and integrate residual blocks. In multilayer perceptrons (MLPs), increasing network depth often leads to vanishing and exploding gradients. The implementation of residual connections mitigates the optimization difficulties associated with deep PINNs, leading to a more gradual loss progression and improved convergence during training. In conventional physics-informed neural networks (PINNs), a simpler model, particularly a shallow network, becomes insufficient for tackling a very complex problem. At that moment, we must enhance the networks' depth to attain satisfactory results in that instance. This clarifies the reasoning for employing ResNets \citep{he2016deep}. Our approach significantly mitigates the vanishing gradient phenomenon, and its performance is exceptional.

We initially provide a nonlinear wavelet-inspired activation function designed to approximate Fourier representations of broad target signals, thereby offering robust approximation capabilities across extensive neural networks. This activation is driven by its potential to capture oscillatory and multiscale characteristics observed in deep learning challenges. We examine its physical interpretation and its importance in vibrational phenomena, particularly in unsteady flow dynamics, including vortex shedding in the flow around a circular cylinder. Subsequently, we introduce the proposed wavelet–physics-informed residual neural network (W-PIRNN) framework.

\subsection{Wavelet activation function}
Simple harmonic motion (SHM), a fundamental oscillatory phenomenon, can be characterized by a mass–spring system on a frictionless surface, where the displacement \( y(t) \) of a mass \( m \) connected to an ideal spring follows to Hooke’s law \( \mathbf{F} = -k y(t) \), with \( k>0 \) representing the spring constant; the application of Newton’s second law results in the governing equation.
\begin{equation}
m \, y''(t) + k \, y(t) = 0,
\end{equation}
where \( y(t) \) represents the displacement of the mass from equilibrium at time \( t \), \( y''(t) \) is the corresponding acceleration, \( m \) is the mass of the object, and \( k \) characterizes the stiffness of the spring.

We employ the notation $y^{\prime \prime}$ to denote the second derivative of $y$ with respect to $t$. With $\omega=\sqrt{\mathrm{k} / \mathbf{m}}$, this second-order ordinary differential equation is transformed into

\begin{equation}
y^{\prime\prime}(t) + \omega^2 y(t) = 0
\label{eq:harmonic_oscillator}
\end{equation}

The general solution of equation (\ref{eq:harmonic_oscillator}) is given by

$$
y(t)=w_1 \cos \omega t+w_2 \sin \omega t
$$

Where, $w_1$ and $w_2$ are constants.

In the aforementioned expression for $y(t)$, the parameter $\omega$ is specified, although $w_1$ and $w_2$ may assume any real values. To ascertain the specific solution of the equation, it is necessary to apply two initial conditions due to the presence of the two unknown constants \( w_1 \) and \( w_2 \). If we are provided with $y(0)$ and $y^{\prime}(0)$, the initial position and velocity of the mass, the solution to the physical problem is unique and expressed as

$$
y(t)=y(0) \cos \omega t+\frac{y^{\prime}(0)}{\omega} \sin \omega t
$$

One can easily verify that there exist constants $A>0$ and $\varphi \in \mathbb{R}$ such that

$$
w_1 \cos \omega t+w_2 \sin \omega t=A \cos (\omega t-\varphi)
$$

Because of the physical interpretation given above, one calls $A=\sqrt{w_1^2+w_2^2}$ the "amplitude" of the motion, $\omega$ its "natural frequency," $\varphi$ its "phase" (uniquely determined up to an integer multiple of $2 \pi$ ), and $2 \pi / \omega$ the "period" of the motion.

\begin{figure}[htbp]
\centering
\includegraphics[width=0.7\textwidth]{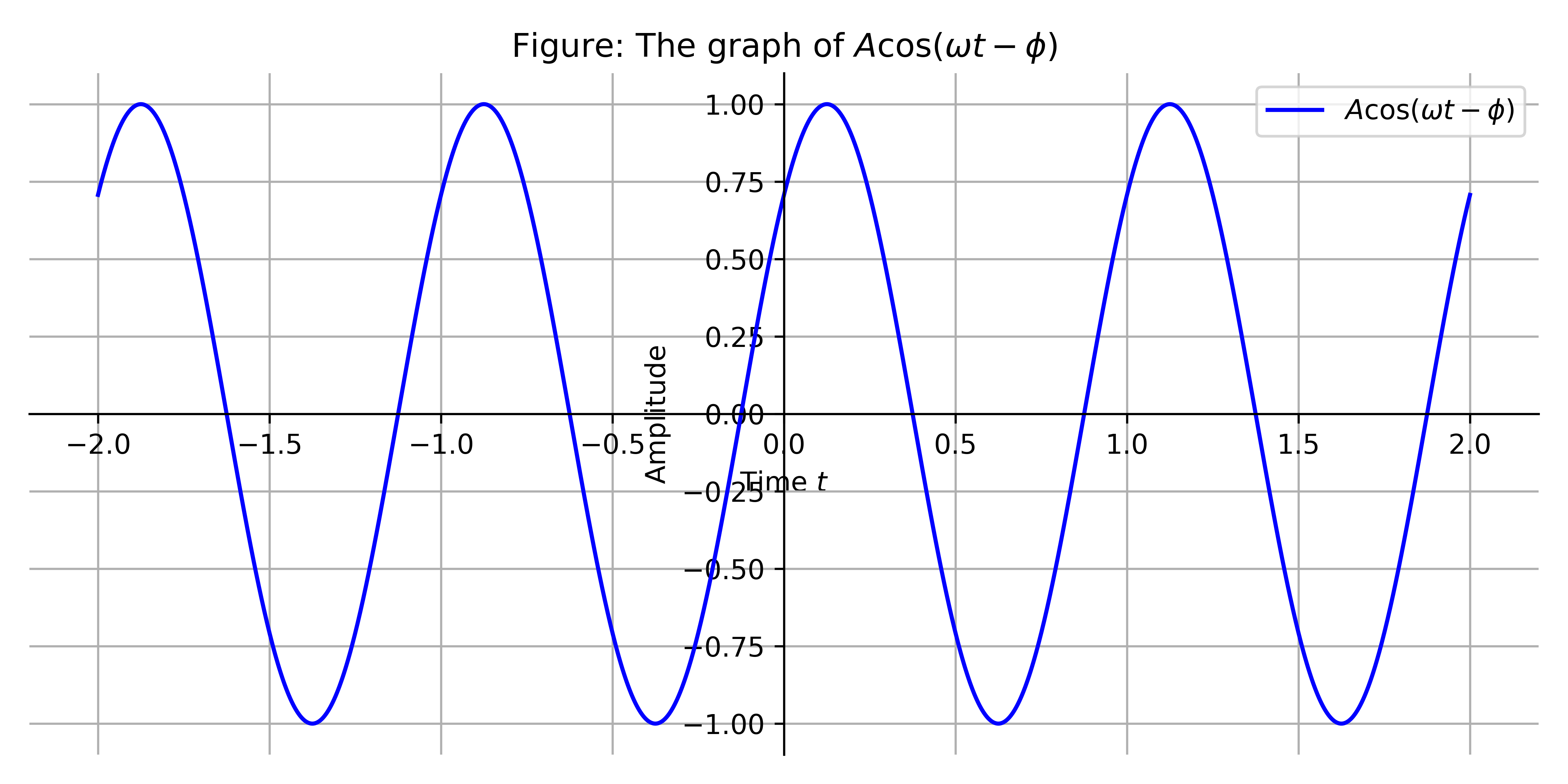} % Replace with your graph filename
\caption{The graph of \( A \cos(\omega t - \phi) \)}
\label{fig:cos_wave}
\end{figure}

The standard graph of the function $A \cos (\omega t - \varphi)$, depicted in Figure~\ref{fig:cos_wave}, demonstrates a wavelike pattern derived from the translation and scaling (or compression) of the conventional graph of $\cos t$.

We present two observations about our analysis of simple harmonic motion. The initial point is that the mathematical characterization of the most fundamental oscillatory system, specifically simple harmonic motion, entails

the most basic trigonometric functions $\cos t$ and $\sin t$. It will be important in what follows to recall the connection between these functions and complex numbers, as given in Euler's identity $e^{i t}=\cos t+i \sin t$. The second observation is that simple harmonic motion is determined as a function of time by two initial conditions, one determining the position, and the other the velocity (specified, for example, at time $t=0$ ). This property is shared by more general oscillatory systems, as we shall see below.

In our scenario, the wavelet activation function $\omega=1$ indicates that the natural frequency is the identity. The use of ReLU activation in PINNs may lead to suboptimal performance, since its efficacy is significantly dependent on precise derivative evaluation, but ReLU possesses a discontinuous derivative \citep{haghighat2021physics, de2021assessing}. Recent research employs snusoidal activation in particular contexts to replicate the periodic characteristics of solutions to partial differential equations \citep{li2020fourier, jagtap2020adaptive, song2022versatile}. Nonetheless, it necessitates robust prior understanding of the solution's behavior and is constrained in its applicability. To address this issue, \cite{zhao2023pinnsformer} introduced a novel and straightforward activation function, referred to as wavelet, defined as follows:

$$
W(t) = w_1 \sin (t)+w_2 \cos (t)
$$

$w_1$ and $w_2$ are designated as learnable parameters. The rationale for wavelet activation parallels the Real Fourier Transform: all signals, whether periodic or aperiodic, may be decomposed into an integral of sines and cosines of diverse frequencies.  Wavelets can approximate arbitrary functions with sufficient accuracy.

The novel wavelet function is infinitely differentiable, so there are no complications during automated differentiation (AD). For specific values of $w_1$ and $w_2$, $W(t)$ exhibits characteristics akin to several types of waves. Specifically, when $w_1=1$ and $w_2=0$, it resembles a sine wave, whereas when $w_1=0$ and $w_2=1$, it functions as a cosine wave. Between the values of $w_1$ and $w_2$, it exhibits characteristics of mixed waves.

\begin{figure}[htbp]
    \centering
    \includegraphics[width=0.9\linewidth]{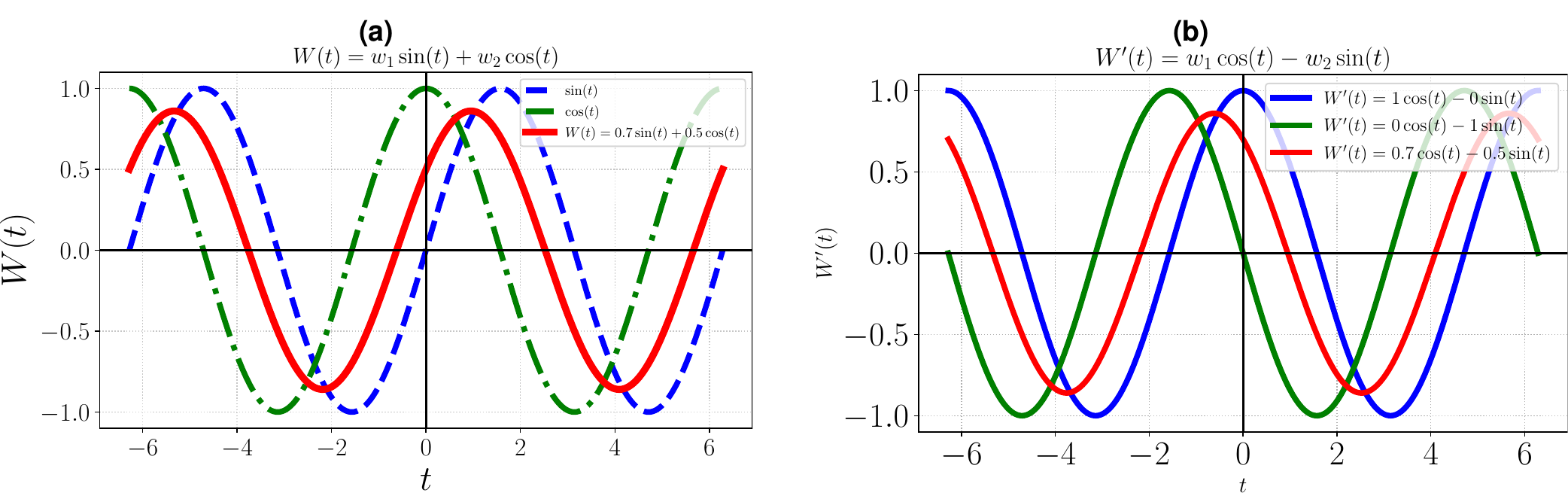}
    \caption{Plots of (a) the wavelet activation function \( W(t) = w_1\sin(t) + w_2\cos(t) \) and (b) its derivative \( W'(t) = w_1\cos(t) - w_2\sin(t) \), evaluated at three different parameter combinations: \( (w_1, w_2) = (1, 0) \), \( (0, 1) \), and \( (0.7, 0.5) \).}
    \label{fig:wavelet_side_by_side}
\end{figure}

The wavelet activation function $W(t) = w_1 \sin(t) + w_2 \cos(t)$ and its derivative $W'(t) = w_1 \cos(t) -w_2 \sin(t))$ will be shown in Figure ~\ref{fig:wavelet_side_by_side} in a variety of waveforms, such as sine, cosine, and mixed waves.

Prior to presenting the theorem, our wavelet activation function also adhered to the principles of universal approximation.

\begin{theorem}[Universal Approximation Theorem] \citep{hornik1991approximation} \label{thm:universal_approx}
  Let $\sigma$ be a continuous, bounded, and non-constant activation function. Then for any continuous function $\mathrm{f}$ defined on a compact subset $K \subset \mathbb{R}^n$ and for any $\epsilon>0$, there exists a feedforward neural network with a single hidden layer and a finite number of neurons such that the network's output $\hat{\mathrm{f}}$ approximates $\mathrm{f}$ within $\epsilon$, i.e.,

$$
|\mathrm{f}({x})-\hat{\mathrm{f}}({x})|<\epsilon \quad \forall {x} \in K
$$

In other words, this theorem states that a feedforward NN, with at least one hidden layer containing a sufficient number of neurons, can approximate any continuous function to any desired degree of accuracy.
\end{theorem}

\begin{theorem}\citep{zhao2023pinnsformer, zhaopinnsformer}
Let $\mathcal{N}$ be a two-hidden-layer neural network of infinite width, utilizing a wavelet activation function; then, $\mathcal{N}$ serves as a universal approximator for any real-valued target function $\mathrm{f}$.
\end{theorem}
Proof: The proof is based on the Real Fourier Transform (Fourier Integral Transform). For any specified input $x$ and its associated real-valued goal $\mathrm{f}(x)$, there exists the Fourier Integral:

$$
\mathrm{f}(x)=\int_{-\infty}^{\infty} A(\omega) \cos (\omega x) d \omega+\int_{-\infty}^{\infty} B(\omega) \sin (\omega x) d \omega
$$

Where $A(\omega)$ and $B(\omega)$ denote the coefficients of sine and cosine, respectively.  Secondly, the integral can be approximated by an infinite sum using Riemann sum approximation:

$$
\mathrm{f}(x) \approx \sum_{n=1}^N\left[A\left(\omega_n\right) \cos \left(\omega_n x\right) + B\left(\omega_n\right) \sin \left(\omega_n x\right)\right] \equiv {W}_2\left(\text { Wavelet }\left({W}_1 x\right)\right)
$$

Where ${W}_1$ and ${W}_2$ denote the weights of the first and second hidden layers of $\mathcal{N}$, respectively. Given that ${W}_1$ and ${W}_2$ possess infinite width, we can partition the piecewise summing into infinite small intervals, hence rendering the approximation increasingly precise in relation to the actual integral. Consequently, $\mathcal{N}$ serves as a universal approximator for any specified $\mathrm{f}$. In practice, most PDE solutions involve only a limited number of principal frequencies. Using a neural network with a finite number of parameters would yield accurate approximations of the exact solutions.

\textbf{Note:} The wavelet activation function $W(t) = w_1 \sin(t) + w_2 \cos(t)$ constitutes a linear combination of continuous trigonometric functions defined on $\mathbb{R}$. Consequently, $W(t)$ is continuous for every $w_1, w_2, t \in \mathbb{R}$.  Furthermore, $W(t)$ is not a constant function unless both $w_1 = 0$ and $w_2 = 0$, which constitutes a trivial example. To demonstrate that $W(t)$ is bounded, we reformulate it utilizing the identity
$$
W(t) = A \sin(t + \phi),
$$
where $A = \sqrt{w_1^2 + w_2^2}$ and $\phi = \tan^{-1}\left(\frac{w_2}{w_1}\right)$, assuming $w_1 \neq 0$. Given that $\sin(t + \phi) \in [-1, 1]$ for every $t \in \mathbb{R}$, it follows that $|W(t)| = |A \sin(t + \phi)| \leq A = \sqrt{w_1^2 + w_2^2}$. Thus, $W(t)$ is bounded. By the Universal Approximation ~\ref{thm:universal_approx} \citep{hornik1991approximation}, we can say that the wavelet activation function can be used in a feedforward neural network with at least one hidden layer and enough neurons to model any continuous function on a compact domain because it is continuous, bounded, and not constant.

\subsection{Proposed wavelet-physics informed residual neural networks (W-PIRNNs)}
In recent years, flow reconstruction using deep learning approaches has attracted significant interest in the fluid dynamics community. This methodology focuses on reconstructing high-resolution flow fields from limited or low-resolution data. Particle Image Velocimetry (PIV) is a widely used technique for obtaining experimental flow data, but it often yields limited observations. Traditional machine learning methods typically require large datasets for efficient training, making them less suitable for engineering scenarios where high-quality data is limited or expensive to obtain. To address this limitation, the Physics-Informed Neural Networks (PINNs) framework, introduced in 2019 \citep{raissi2019physics}, has demonstrated robustness by incorporating governing physical laws directly into the learning process. This enables efficient flow reconstruction even from restricted datasets. However, when applied to complex flow problems, traditional PINNs sometimes require very deep architectures, leading to challenges such as vanishing and exploding gradients and overfitting. To address these challenges, we present an innovative architecture, termed wavelet-physics-informed residual neural networks (W-PIRNNs). This architecture combines a custom wavelet-based activation function with residual learning via skip connections. The incorporation of custom wavelet activations effectively captures localized flow characteristics, while residual blocks alleviate training instabilities. Our proposed technique not only attains substantially lower training loss and relative $L_2$ error in fewer epochs for low-supervised learning but also demonstrates remarkable performance in terms of overall computational cost. The W-PIRNN architecture is computationally efficient and readily accessible, offering a practical solution for flow reconstruction in sparse data scenarios. The algorithm of our W-PIRNNs, presented in \textbf{Algorithm 1}.

\begin{algorithm}[H]
\caption{ Proposed Wavelet-Physics Informed Residual Neural Networks (W-PIRNNs)}
\label{alg:wavelet_pinn}
\begin{algorithmic}[1]

\REQUIRE Limited velocity measurement data  $\mathcal{D}_v = \{ (x_d, y_d, t_d), (u_d, v_d) \}_{d=1}^{\mathcal{N}_d}$
\REQUIRE Hyperparameters: Reynolds number \(Re\), loss weight \(\lambda\), learning rate \(\eta\), number of epochs \(\mathbf{N}_{epochs}\), batch size \(\mathbf{N}_b\)

\STATE \textbf{Step 1: Preparing Data and Collocation Points}
\STATE To cover the entire spatiotemporal domain, create a set of $\mathcal{N}_f$ collocation points $\{x_i, y_i, t_i\}_{i=1}^{\mathcal{N}_f}$ using Latin Hypercube Sampling (LHS).
\STATE Use Min-Max scaling to normalize all temporal and spatial inputs ($x, y, t$) to the range $[-1, 1]$.

\STATE \textbf{Step 2: W-PIRNNs Architecture and Initialization}
\STATE Build an input layer, $L$ residual blocks, and an output layer to create a deep residual neural network $\mathbb{f}(x, y, t; \theta)$.
\STATE After each linear layer, use a trainable **Wavelet Activation** function $W(t) = w_1 \sin(t) + w_2 \cos(t)$, where $w_1$ and $w_2$ are trainable parameters $\theta$.
\STATE Initialize all network parameters $\theta$ utilizing Xavier Normal initialization.
\STATE Elucidate the forward propagation within the residual core:
\FOR{$l = 1$ to $\mathcal{L}$}
    \STATE Calculate the residual mapping: $\mathcal{F}^{(l)}(x) = W(\text{Linear}^{(l)}_2(W(\text{Linear}^{(l)}_1(x))))$
    \STATE Implement skip connection:  $x \leftarrow x + \mathcal{F}^{(l)}(x)$
\ENDFOR
\STATE Establish a terminal linear output layer to forecast the flow fields:  $[u, v, p] = \text{Linear}_{\text{out}}(x)$

\STATE \textbf{Step 3: Define Composite Loss Function}
\STATE The loss function $\mathcal{L}_{total}$ is the weighted aggregation of the data loss and the physics loss.  The total loss function is expressed as $\mathcal{L}_{total} = \mathcal{L}_{data} + \lambda \mathcal{L}_{physics}$.
\STATE
\COMMENT{Data loss $\mathbf{L}_{data}$ is defined as the Mean Squared Error (MSE) of the sparse velocity measurements:}
\STATE $\mathcal{L}_{\text{data}} = \frac{1}{\mathcal{N}_d} \sum_{i=1}^{\mathcal{N}_d} \left[ (u(x_d^i, y_d^i, t_d^i) - u_d^i )^2 + (v(x_d^i, y_d^i, t_d^i) - v_d^i )^2 \right]$
\STATE
\COMMENT{Physics loss $\mathcal{L}_{residual}$ enforces the Navier-Stokes equations on the collocation points.}
\STATE Initially, delineate the equation residuals $e_1, e_2, e_3$ calculated using automatic differentiation:
\STATE $e_1 := u_t + (uu_x + vu_y) + p_x - Re^{-1}(u_{xx} + u_{yy})$
\STATE $e_2 := v_t + (uv_x + vv_y) + p_y - Re^{-1}(v_{xx} + v_{yy})$
\STATE $e_3 := u_x + v_y$
\STATE Subsequently, calculate the Mean Squared Error (MSE) of these residuals: $\mathcal{L}_{\text{residual}} = \frac{1}{\mathcal{N}_f} \sum_{j=1}^{\mathcal{N}_f} \left[ (e_1^j)^2 + (e_2^j)^2 + (e_3^j)^2 \right]$

\STATE \textbf{Step 4: Optimization and Training}
\STATE Configure the Adam optimizer with a learning rate $\eta$.
\STATE Optionally, implement a ReduceLROnPlateau learning rate scheduler for adaptive modification.
\FOR{epoch = 1 to $\mathbf{N}_{epochs}$}
    \STATE Shuffle the collocation points and the training data $\mathcal{D}_v$.
    \FOR{every mini-batch of size $\mathbf{N}_b$}
        \STATE Obtain a mini-batch from $\mathcal{D}_v$ and a mini-batch from the collocation points.
        \STATE Execute a forward pass to obtain predictions for both mini-batches.
        \STATE Calculate the aggregate loss $\mathcal{L}_{total}$ by combining $\mathcal{L}_{data}$ and $\mathcal{L}_{residual}$.
        \STATE Perform backpropagation of the gradients $\nabla_{\theta} \mathcal{L}_{total}$ and adjust the model parameters $\theta$ using the Adam optimization step.
    \ENDFOR
\ENDFOR

\STATE \textbf{Step 5: Prediction and Reconstruction}
\STATE The trained W-PIRNN, $\mathrm{f}(x, y, t; \theta^*)$, is now a continuous function representing the flow dynamics.
\STATE Use the model to reconstruct high-resolution velocity ($u, v$) and pressure ($p$) fields at any spatio-temporal coordinate in the domain.
\STATE Reconstruct derived quantities from the network's output derivatives, such as the fluid **streamlines** and the **vorticity field** ($\omega = v_x - u_y$).

\end{algorithmic}
\end{algorithm}

\section{Numerical experiments}
\label{sec:methodology}
\subsection{Numerical dataset for both training and validation at Re=100}
\label{sec:numerical_dataset}
\begin{figure}[htbp]
    \centering
    \includegraphics[width=0.8\textwidth]{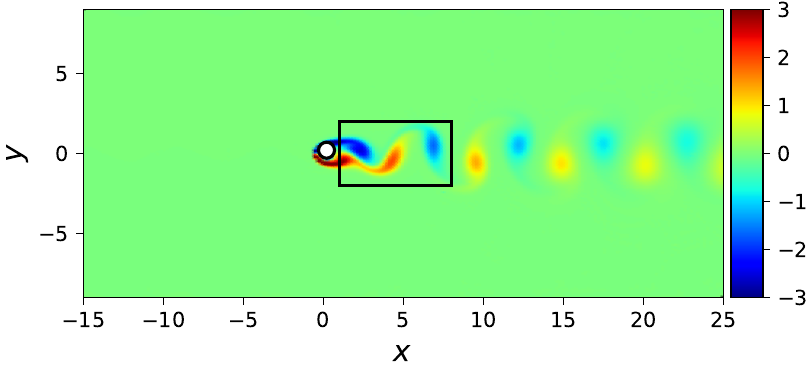}
    \caption{The vorticity contour of the flow past a circular cylinder with dynamic vortex shedding at Re=100.} \citep{raissi2019physics}
    \label{fig:vorticity_Re100}
\end{figure}

The resolution of the incompressible flow problem around a circular cylinder is a fundamental concern in fluid mechanics. This research uses data on two-dimensional wake flow around a circular cylinder at a low Reynolds number, $\operatorname{Re}=u_{\infty} D / \nu$. A non-dimensional free stream velocity of $u_{\infty}=1$, a cylinder diameter of $D=1$, and a kinematic viscosity of $\nu=0.01$ are assumed. The system attains a periodic steady flow state exhibiting an asymmetrical Kármán vortex street in the wake, \citep{raissi2019physics} as illustrated in Figure~\ref{fig:vorticity_Re100}. The grid and velocity are nondimensionalized using the free stream velocity $u_{\infty}$ and the diameter of the cylinder $D$. The boundary conditions consist of a uniform free flow velocity on the left, a zero-pressure outlet on the right boundary situated 25 diameters downstream, and symmetric conditions $[-15,25] \times[-8,8]$ at the upper and lower boundaries of the domain. The dataset was acquired using direct numerical simulation of the Navier-Stokes equations.

For the purpose of simplification, a rectangular region downstream of the cylinder was selected, and grids of 100 equidistant points along the x-axis and 50 equidistant points along the y-axis were established inside the spatial domain of $[1,8] \times[-2,2]$. The initial data were gathered on a grid of 5000 points within the specified timeframe, from 0 to 19.9, with an interval of 0.1. To create the original training set, 1 million data points comprising just velocity information were extracted from the dataset at various sparsity levels, including $100\%$, $20\%$, $5\%$, $1\%$, $0.16\%$, $0.5\%$, and $0.05\%$. Furthermore, an additional $10^6$ two-dimensional equation points (collocation points) were sampled using Latin hypercube sampling (LHS) to ensure a uniform distribution across the multidimensional parameter space \citep{xu2023practical}. The training set consisted of the processed data points and collocation points combined. The training set, which exclusively includes velocity data at different sparsity levels, such as $100\%$, $20\%$, $5\%$, $1\%$, $0.16\%$, and $0.05\%$. To ensure that the anticipated results gradually align with the values in the original data, the Navier-Stokes constraints are incorporated into the loss function to help the model in effectively predicting pressure gradients, vorticity, and streamlines. The original dataset, including both velocity and actual pressure data, was used as the validation set.

\begin{figure}[htbp]
\centering
\includegraphics[width=0.85\linewidth]{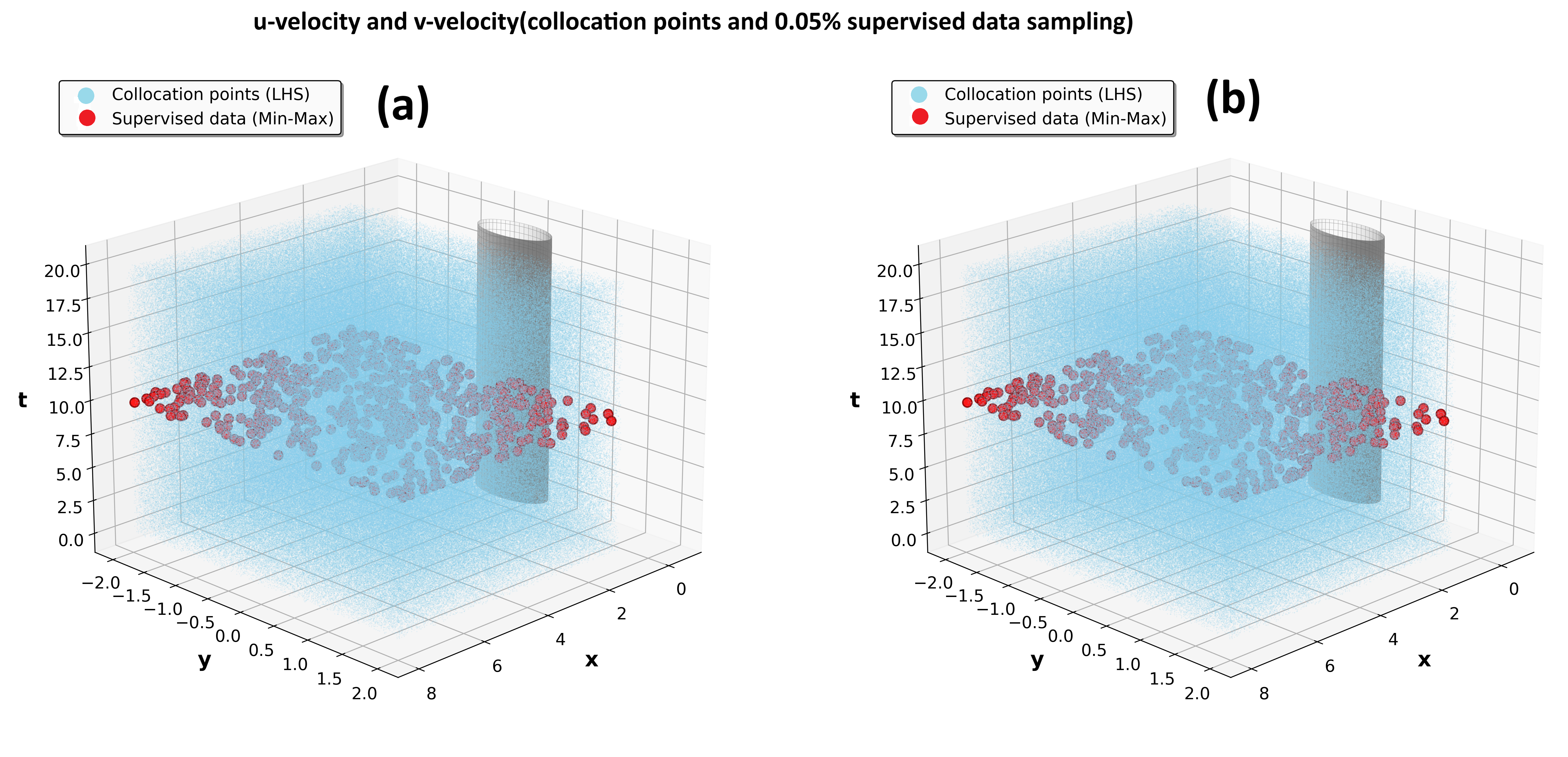}
\caption{Collocation points and $0.05\%$ supervised data sampling distribution for the flow past a circular cylinder at $\mathrm{Re}=100$: (a) $u$-velocity and (b) $v$-velocity.}
\label{fig:uv_collocation_supervised_Re100}
\end{figure}

\begin{figure}[htbp]
\centering
\includegraphics[width=0.85\linewidth]{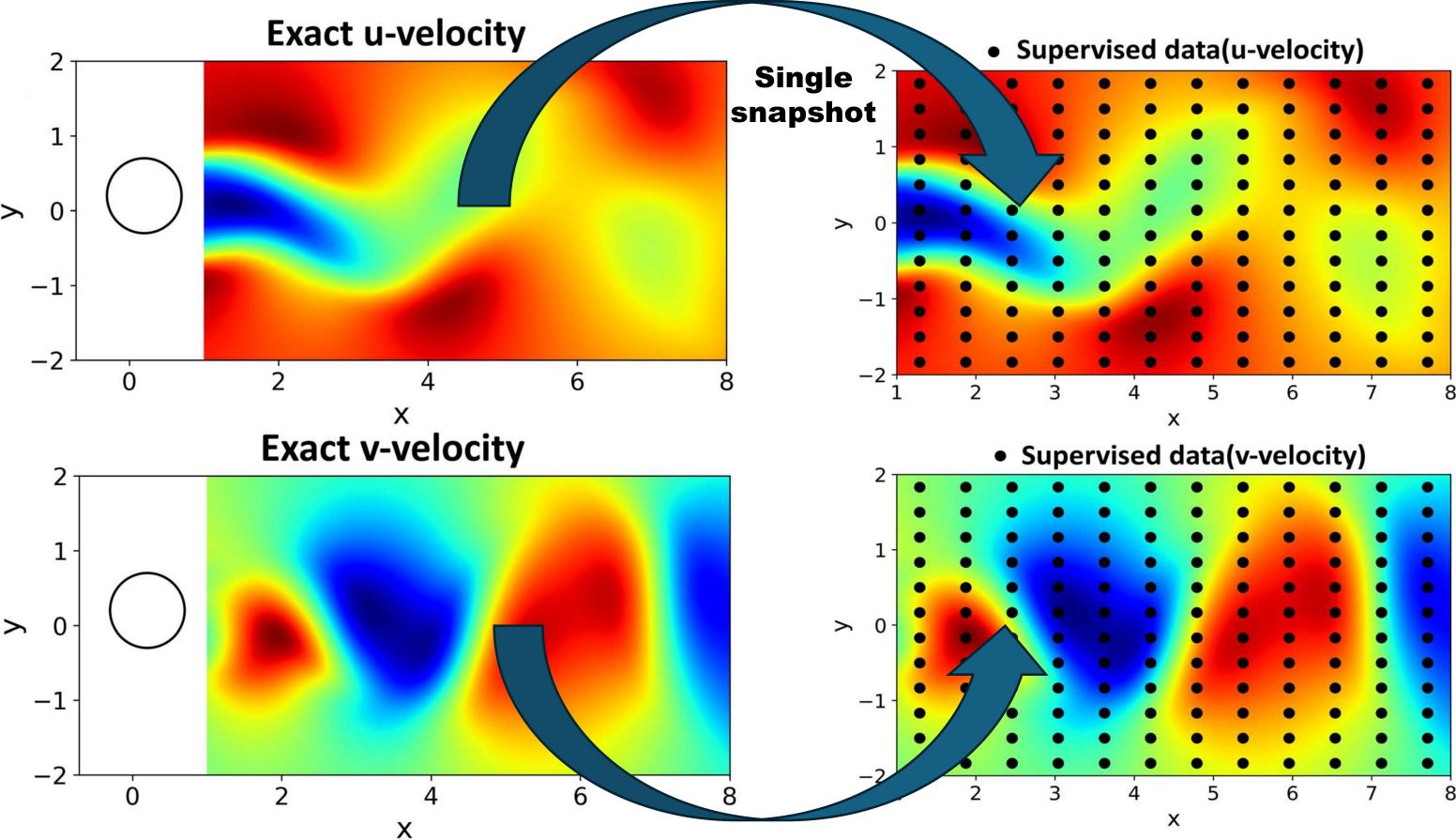}
\caption{Supervised data selection of the $u$-velocity and $v$-velocity on the single temporal snapshot for the flow past a circular cylinder at Re=100, top row for u-velocity and bottom row for v-velocity, respectively.}
\label{fig:uv_supervised_snapshot}
\end{figure}

\begin{figure}[htbp]
    \centering
    \includegraphics[width=\textwidth]{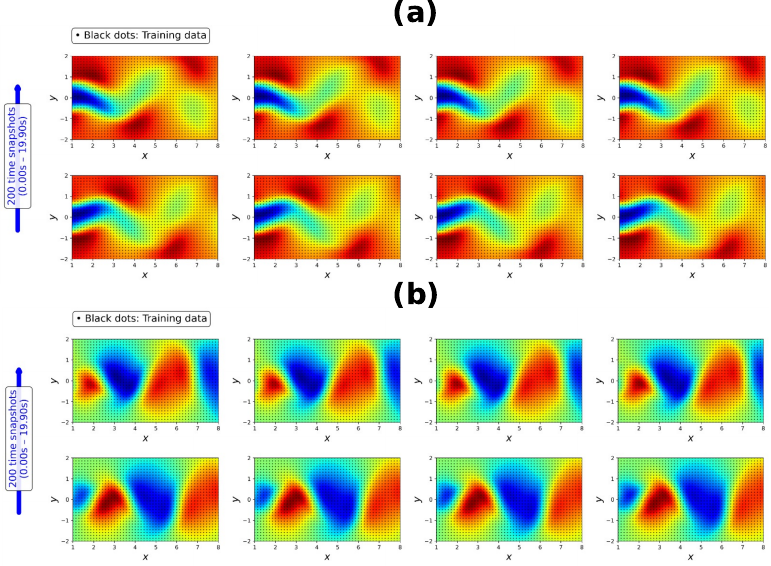}
    \caption{
   General procedure of supervised data selection for the flow past a circular cylinder at Re=100: (a) and (b) are the supervised data selection procedures of the u-velocity and v-velocity, respectively.
    }
    \label{fig:temporal_supervised_sampling}
\end{figure}

For the cylinder wake configuration, velocity data are extracted over the temporal interval $t \in [0.00, 19.90]\,\mathrm{s}$ with a uniform sampling interval of $0.10\,\mathrm{s}$, resulting in a total of $200$ temporal snapshots. At each snapshot, both the streamwise ($u$) and cross-stream ($v$) velocity components are utilized as supervised training data under different sparsity levels, as illustrated in Figure~\ref{fig:temporal_supervised_sampling},~\ref{fig:uv_supervised_snapshot}. In a single snapshot, the process by which we obtain supervised data is clearly shown in Figure~\ref{fig:uv_supervised_snapshot}. In addition to the supervised measurements, $10^6$ collocation points are employed to enforce the governing equations. The spatio-temporal three-dimensional distributions of the collocation points and the $0.05\%$ supervised dataset are shown in Figure~\ref{fig:uv_collocation_supervised_Re100}, similar to the other supervised sparsity datasets.

\subsection{Flow reconstruction from sparse data for a circular cylinder at Re=100}
\begin{figure}[htbp]
    \centering
    \includegraphics[width=1.0\textwidth]{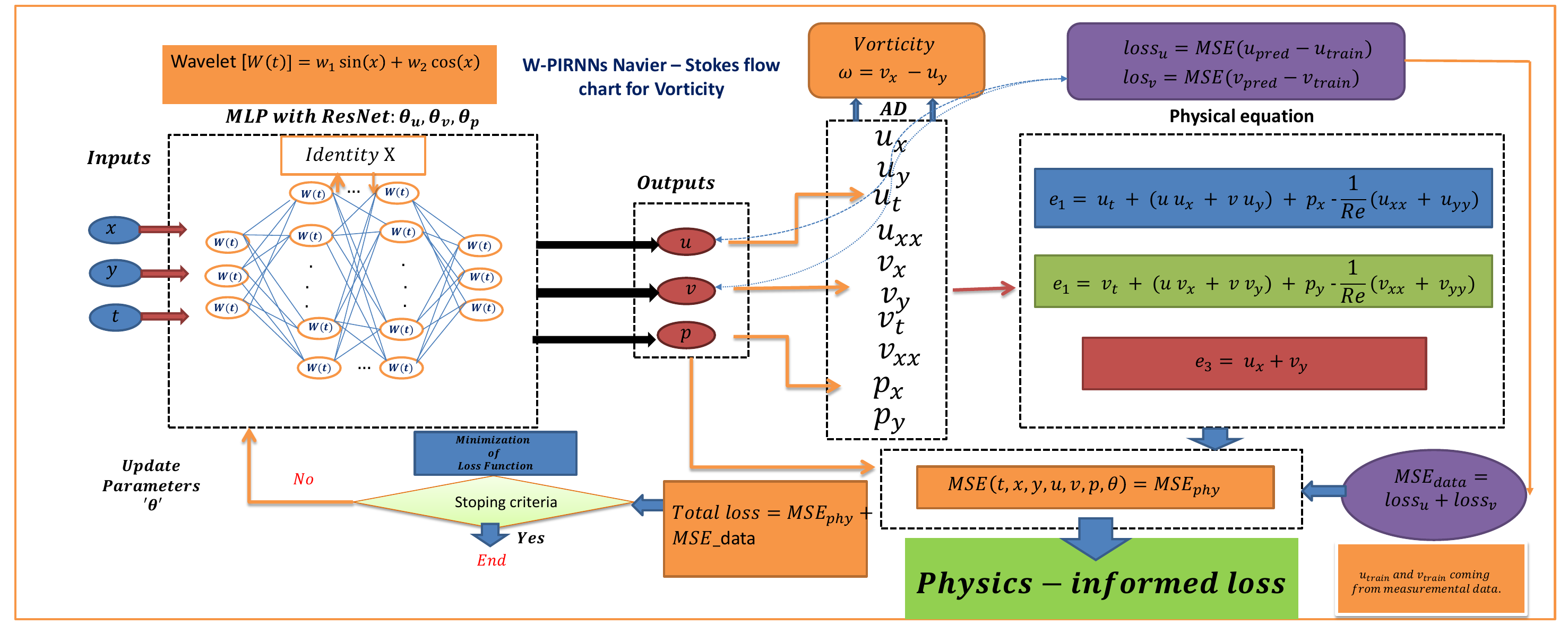} % Replace with actual file name
    \caption{The schematic flowchart of the proposed W-PIRNNs for the flow reconstruction problem.}
    \label{fig:wpirnns_vorticity}
\end{figure}

The main goal of this study is to reconstruct the flow field using extremely sparse supervised velocity data. Specifically, the aim is to recover high-resolution flow velocity and pressure fields, together with wake vorticity and streamline structures around a circular cylinder. The overall flowchart of the proposed network is illustrated in Figure~\ref{fig:wpirnns_vorticity}. Despite using limited velocity data for training, the proposed framework is systematically validated across datasets with supervision levels of $100\%$, $20\%$, $5\%$, and $1\%$. In addition, the reconstruction capability is further examined under extremely sparse training conditions, with velocity data sparsity levels of $0.16\%$ and $0.05\%$. Across all sparsity regimes, the proposed method consistently outperforms the conventional PINN~\citep{xu2023practical} and the physics-informed convolutional network based on feature fusion (FFPICN)~\citep{liu2025physics}. For validation and comparison, results obtained from PINNs~\citep{xu2023practical} and FFPICN~\citep{liu2025physics} are considered. To further assess the robustness of the proposed framework, additional comparisons are carried out with PINNs employing adaptive activation functions (Adaptive PINNs) and PINNs using sinusoidal activations (SIREN PINNs). The corresponding quantitative results are summarized in Table~\ref{tab:relative_error_comparison1}. This extended comparison enables an evaluation of the proposed approach against commonly adopted activation strategies intended to improve trainability and representational capacity. The results indicate that, even with these enhancements, the proposed W-PIRNN framework consistently achieves improved loss balance and lower relative $L_2$ errors in the reconstruction of both velocity components. Overall, W-PIRNNs demonstrate robust agreement and superior performance relative to all comparison methods. To the best of our knowledge, these existing approaches have not previously been applied to flow reconstruction problems involving velocity, pressure, vorticity, and streamlines under extremely sparse supervision. Moreover, the proposed W-PIRNNs attain this improved performance with reduced computational cost, achieving convergence within $2000$ training epochs. While results are examined over multiple epochs, the performance achieved within the $2000$-epoch window is particularly notable across all compared methodologies.

To reconstruct the complete flow field from minimal measurements, we consider a deep learning-based approach that leverages sparse velocity measurements. Specifically, we employ only a limited number of sparse velocity data points from experimental measurements (DNS) to reconstruct the full-field velocity, pressure, vorticity, and streamline.

Our proposed network is designed to reconstruct the velocity components, pressure, vorticity, and streamlines. Vorticity and streamlines are computed from the predicted velocity using our network. We utilize Min-Max normalization to preprocess the sparse velocity measurement data, ensuring sufficient representation of the input domain. Furthermore, we use Latin Hypercube Sampling (LHS) to apply 1 million collocation points across the domain.

The proposed approach incorporates the governing principles of fluid flow and limited observational data into the learning process. By explicitly enforcing the incompressible Navier--Stokes equations, the model maintains physical consistency while improving reconstruction quality. This enables accurate reconstruction of important flow features, such as the Kármán vortex street, even under extremely sparse data conditions of up to $0.05\%$.

The governing equations of the 2D incompressible Navier–Stokes equations, written in residual form, are:

\begin{equation}
\begin{aligned}
e_1 &= u_t + u u_x + v u_y + p_x - \frac{1}{\mathrm{Re}} (u_{xx} + u_{yy}) \\
e_2 &= v_t + u v_x + v v_y + p_y - \frac{1}{\mathrm{Re}} (v_{xx} + v_{yy}) \\
e_3 &= u_x + v_y
\end{aligned}
\label{eq:ns_residuals}
\end{equation}

In this context, \( u \) and \( v \) represent the streamwise and crosswise velocity components, respectively, \( p \) signifies the pressure, and \( \mathrm{Re} \) indicates the Reynolds number. These residual equations are enforced at the collocation points via automatic differentiation within the physics-informed loss function.

We present the wavelet-physics-informed residual neural network (W-PIRNNs) framework for flow reconstruction. The network architecture consists of 4 hidden layers, each with 160 neurons. The hidden layers use a wavelet-inspired activation function, which is defined as
$$
W(t) = w_1 \sin(t) + w_2 \cos(t),
$$
where the coefficients $w_1$ and $w_2$ are trainable parameters.

The Adam optimization technique is used to train network parameters, along with an adaptive learning rate strategy implemented by the \texttt{ReduceLROnPlateau} scheduler. When available, GPU acceleration is used to increase computing efficiency during training. To ensure stable and efficient convergence, the network weights are initialized using the Xavier initialization strategy \citep{glorot2010understanding}.

\paragraph{Loss Functions and Training Strategy:}

The network inputs comprise the spatial coordinates \((x, y)\) and the temporal coordinate \(t\). The outputs consist of the velocity components \(u, v\) and the pressure \(p\). While the network does not explicitly reconstruct vorticity \(\omega\), it is obtained from the velocity outputs via automatic differentiation. In a similar manner, we replicated these streamlines utilizing this velocity.

The total loss function used for training consists of two components: a data loss term and a physics-informed loss term (the residual). The total loss is defined as:

$$
\mathcal{L}_{\text{total}} = \mathcal{L}_{\text{data}} + \lambda \mathcal{L}_{\text{residual}}
$$

where \(\lambda = 1\) in our setup.

The data loss is characterized by the mean squared error between predicted and actual velocity values:

$$
\mathcal{L}_{\text{data}} = \frac{1}{\mathcal{N}_d} \sum_{i=1}^{\mathcal{N}_d} \left[ \left( u_{\text{pred}}^{(i)} - u^{(i)} \right)^2 + \left( v_{\text{pred}}^{(i)} - v^{(i)} \right)^2 \right]
$$

The physics loss enforces the residuals of the incompressible Navier–Stokes equations:

$$
\mathcal{L}_{\text{residual}} = \frac{1}{\mathcal{N}_f} \sum_{i=1}^{\mathcal{N}_f} \left[ e_1^{(i)^2} + e_2^{(i)^2} + e_3^{(i)^2} \right]
$$

where \(\mathcal{N}_f = 10^6\) denotes the number of collocation points, sampled using Latin Hypercube Sampling (LHS), and the residuals are defined as (\ref{eq:ns_residuals}).

These partial derivatives are computed using automatic differentiation (AD), a technique that ensures accurate gradient computation during training.

\paragraph{Relative $L_2$ Error Metrics:}

To assess the quality of the reconstruction, we compute relative $L_2$ errors for both velocity components:

$$
\text{Relative $L_2$ Error $(L_2(u))$} = 
\frac{ 
\sqrt{ \sum_{i=1}^{\mathcal{N}} \left( u_{\text{true}}^{(i)} - u_{\text{pred}}^{(i)} \right)^2 } 
}{
\sqrt{ \sum_{i=1}^{\mathcal{N}} \left( u_{\text{true}}^{(i)} \right)^2 }
}
$$

$$
\text{Relative $L_2$ Error $(L_2(v))$} = 
\frac{ 
\sqrt{ \sum_{i=1}^{\mathcal{N}} \left( v_{\text{true}}^{(i)} - v_{\text{pred}}^{(i)} \right)^2 } 
}{
\sqrt{ \sum_{i=1}^{\mathcal{N}} \left( v_{\text{true}}^{(i)} \right)^2 }
}
$$

\begin{table}[htbp]
\centering
\caption{Hyperparameter details of the proposed W-PIRNNs model for flow past a circular cylinder at Re=100.}
\label{tab:hyperparameters}
\renewcommand{\arraystretch}{1.2}
\begin{tabular}{lc}
\toprule
\textbf{Hyperparameter} & \textbf{Value} \\
\midrule
Epochs & 2000 \\
Optimizer & Adam \\
Learning Rate Schedule & ReduceLROnPlateau (patience = 2) \\
Initial Learning Rate & $1 \times 10^{-4}$ \\
Loss Function & Data + PDE residual (MSE) \\
Activation Function & Wavelet \\
Hidden Layers & 4 \\
Neurons per Layer & 160 \\
\bottomrule
\end{tabular}
\end{table}

\paragraph{Training Duration and Observations:}

Although satisfactory convergence is observed within the first 1000 epochs, we continue training up to 1500 and 2000 epochs to ensure stability and observe marginal improvements in accuracy. The model demonstrates the ability to reconstruct high-resolution flow quantities from sparse, low-resolution velocity data with high fidelity.

In our model training, we initially set the learning rate as \( \text{lr} = 0.0001 \). Furthermore, an adaptive learning rate strategy was employed, as discussed previously, using a scheduler to adjust the learning rate dynamically during training. We also show all the hyperparameter selections of our model in detail in Table~\ref{tab:hyperparameters}. This approach ensures efficient model convergence.

The results demonstrate that the overall loss continually diminishes and converges steadily to a stable minimum. The adaptive learning rate technique substantially enhanced the stability and efficiency of the optimization process.  

% Subsequently, we will present the results for several epochs: $1000$, $1500$, and $2000$, as well as differing percentages of sparse data supervision.

% \begin{figure}[htbp]
%     \centering
%     \includegraphics[width=\textwidth]{Wavelet_ResNet_PINNs_data_distribution.pdf}
%     \caption{
%     Three-dimensional (3D) visualization of the 1 million collocation points, together with supervised data at different sparsity levels, for flow past a circular cylinder at Re = 100. Figures (a) and (b) are the 100\% supervised data u-velocity and v-velocity sampling distributions, respectively. Figures (c) and (d) are the 20\% supervised data u-velocity and v-velocity sampling distributions, respectively. Figures (e) and (f) are the 5\% supervised data u-velocity and v-velocity sampling distributions, respectively. Figures (g) and (h) are the 1\% supervised data u-velocity and v-velocity sampling distributions, respectively. Figures (i) and (j) are the 0.16\% supervised data u-velocity and v-velocity sampling distribution, respectively. Figures (k) and (l) are the 0.05\% supervised data u-velocity and v-velocity sampling distributions, respectively. In all figures, supervised velocity and collocation points are represented by the red and purple colors, respectively.
%     }
%     \label{fig:sampling_visualization}
% \end{figure}

\subsubsection{Reconstructing flow fields using different percentages of sparse data (e.g., 100\%, 20\%, 5\%, and 1\%)}
\begin{figure}[htbp]
    \centering
    \includegraphics[width=\textwidth]{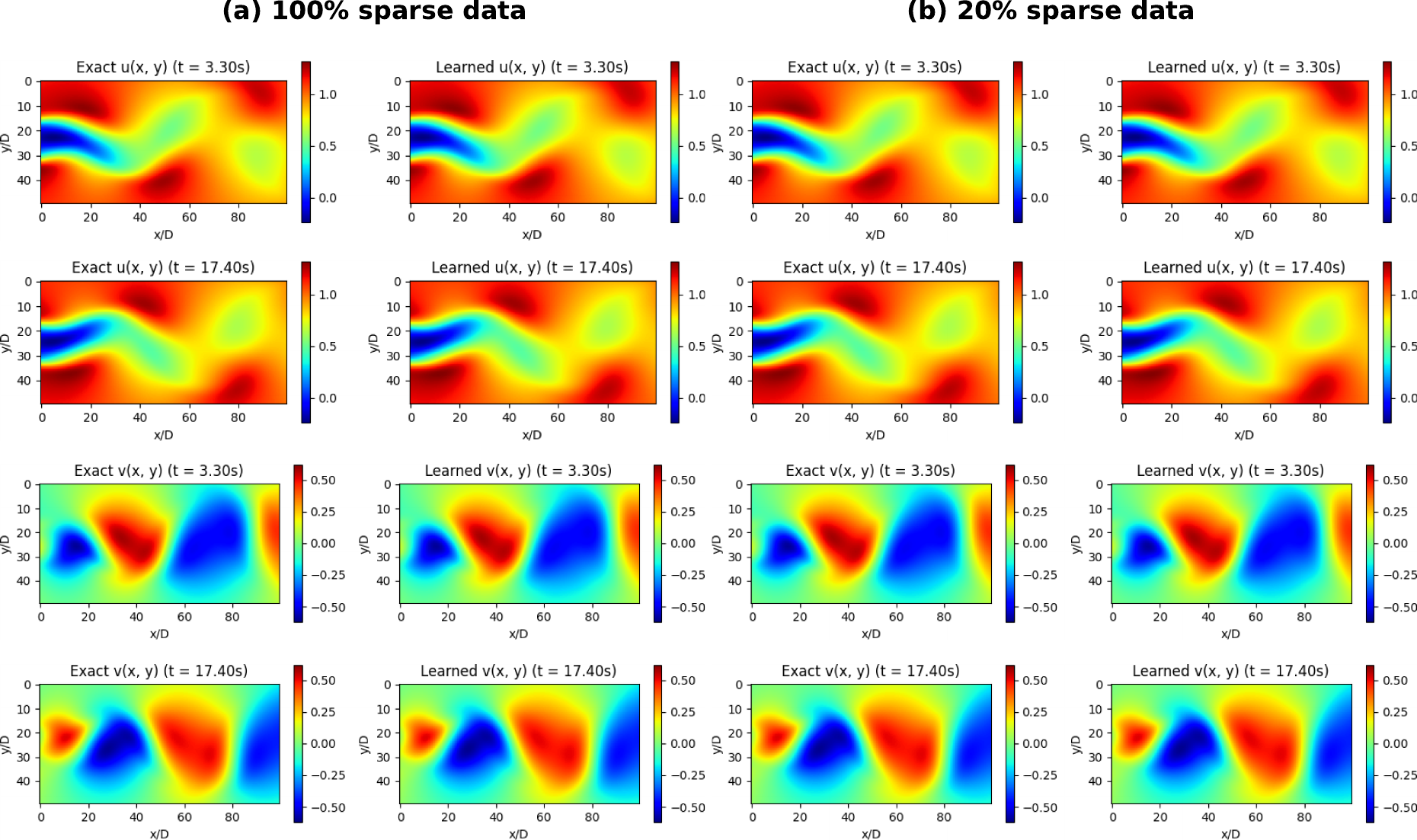} % Replace with your actual filename
    \caption{Comparison of u-velocity and v-velocity results at two different temporal snapshots at t=3.30s and t=17.40s for the flow past a circular cylinder at Re=100: the left column (a) is the exact (CFD benchmark) \citep{raissi2019physics}, and the right column (a) is the learned (proposed W-PIRNNs) results for 100\% supervised velocity data, and the left column (b) is the exact (CFD benchmark) \citep{raissi2019physics}, and the right column (b) is the learned (proposed W-PIRNNs) results for 20\% supervised velocity data, respectively.}
    \label{fig:sparse_velocity_results}
\end{figure}

\begin{figure}[htbp]
    \centering
    \includegraphics[width=\textwidth]{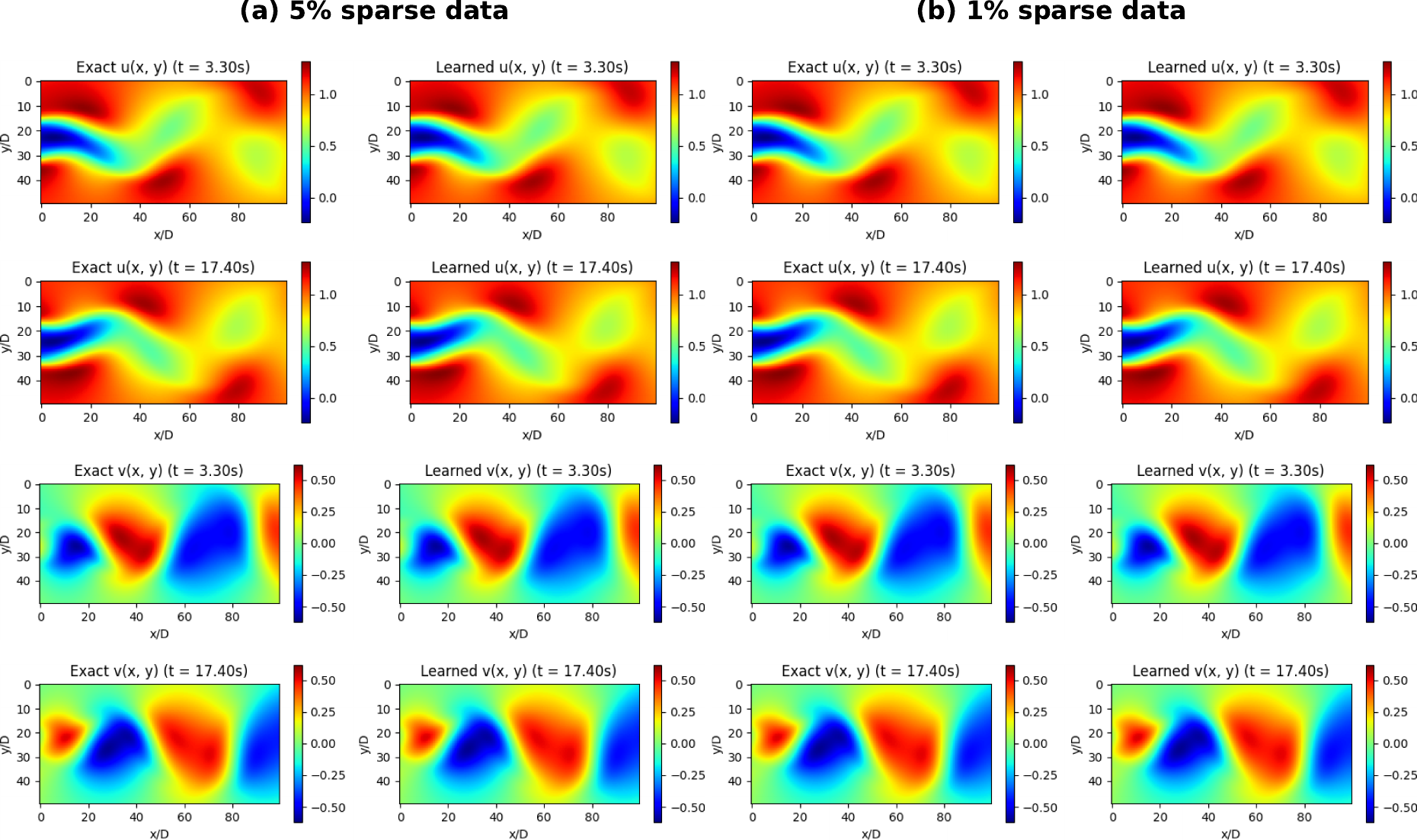} % Replace with your actual filename
    \caption{Comparison of u-velocity and v-velocity results at two different temporal snapshots at t=3.30s and t=17.40s for the flow past a circular cylinder at Re=100: the left column (a) is the exact~\citep{raissi2019physics} (CFD benchmark), and the right column (a) is the learned (proposed W-PIRNNs) results for 5\% supervised velocity data; the left column (b) is the exact~\citep{raissi2019physics} (CFD benchmark), and the right column (b) is the learned (proposed W-PIRNNs) results for 1\% supervised velocity data.}
    \label{fig:sparse_velocity_lowdata}
\end{figure}

\begin{figure}[htbp]
    \centering
    \includegraphics[width=\textwidth]{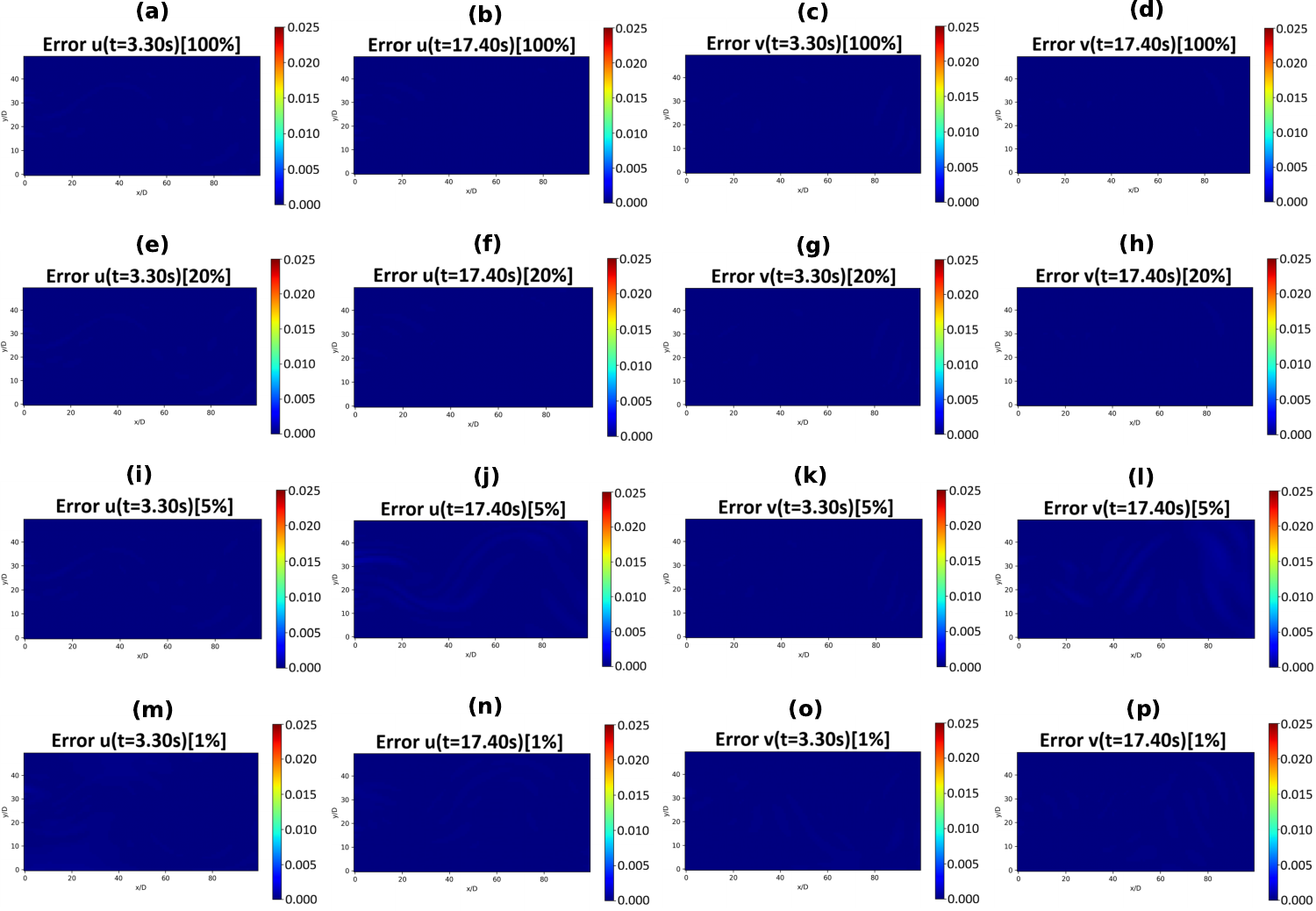}
    \caption{
    Absolute error of the u-velocity and v-velocity for the flow past a circular cylinder at Re=100, evaluated at two different temporal snapshots, t=3.30s and t=17.40s: (a)-(d) are the errors of u-velocity and v-velocity for 100\% supervised data, (e)-(h) are the errors of u-velocity and v-velocity for 20\% supervised data, (i)-(l) are the errors of u-velocity and v-velocity for 5\% supervised data, and (m)-(p) are the errors of u-velocity and v-velocity for 1\% supervised data.
    }
    \label{fig:absolute_error_sparse_velocity}
\end{figure}

\begin{figure}[htbp]
    \centering
    \includegraphics[width=\textwidth]{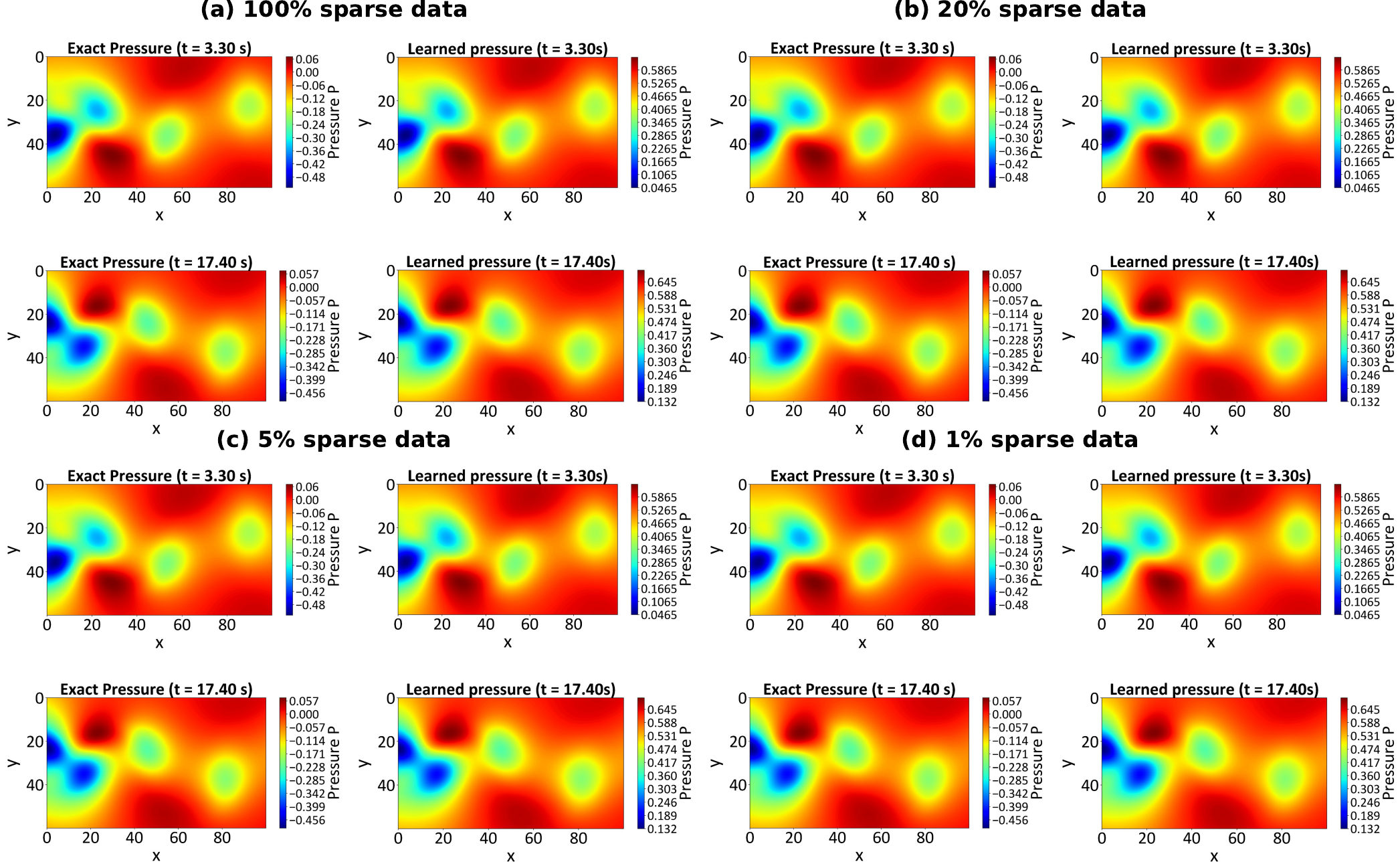} % Replace with your actual PDF filename
    \caption{Comparison between exact pressure~\cite{raissi2019physics} (CFD benchmark) [left column] and learned pressure (proposed W-PIRNNs) [right column] results at two different temporal snapshots at t=3.30s and t=17.40s for the flow past a circular cylinder at Re=100: (a) results for 100\% supervised velocity data, (b) results for 20\% supervised velocity data, (c) results for 5\% supervised velocity data, and (d) results for 1\% supervised velocity data, respectively.}
    \label{fig:pressure_sparse_all}
\end{figure}

\begin{figure}[htbp]
    \centering
    \includegraphics[width=\textwidth]{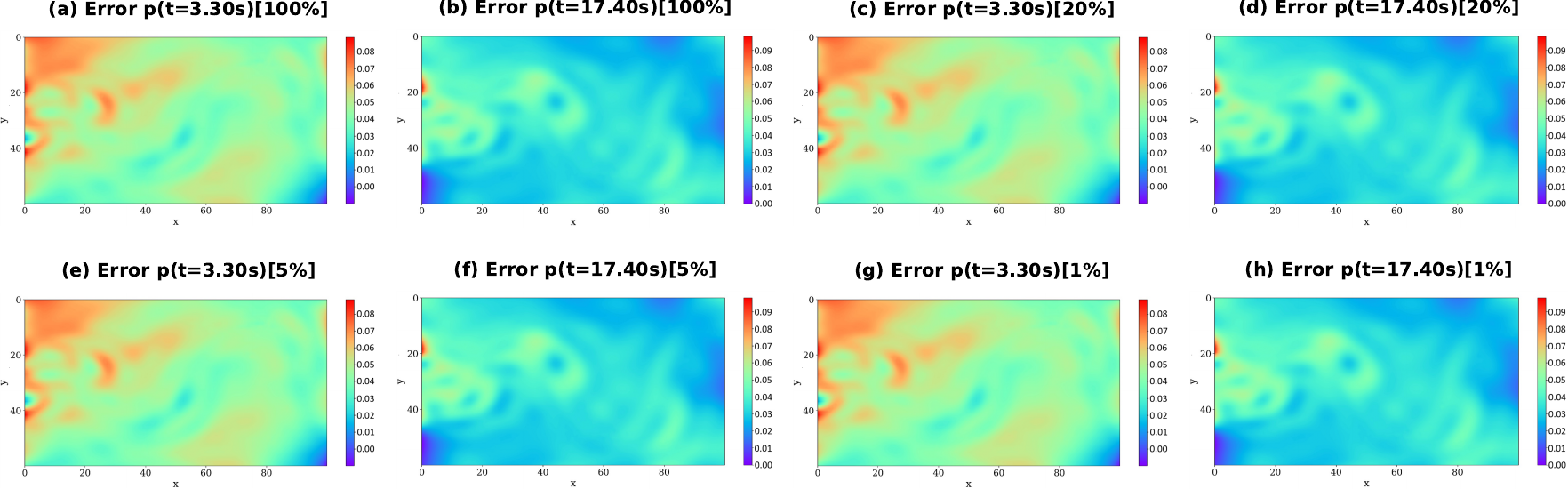}
    \caption{
   The absolute error of the reconstructed pressure for the flow past a circular cylinder at Re=100 is shown at two different temporal snapshots, t=3.30s and t=17.40s; the pairs (a,b), (c,d), (e,f), and (g,h) correspond to the 100\%, 20\%, 5\%, and 1\% supervised data, respectively. 
    }
    \label{fig:pressure_absolute_error_sparse}
\end{figure}

\begin{figure}[htbp]
    \centering
    \includegraphics[width=1.0\linewidth]{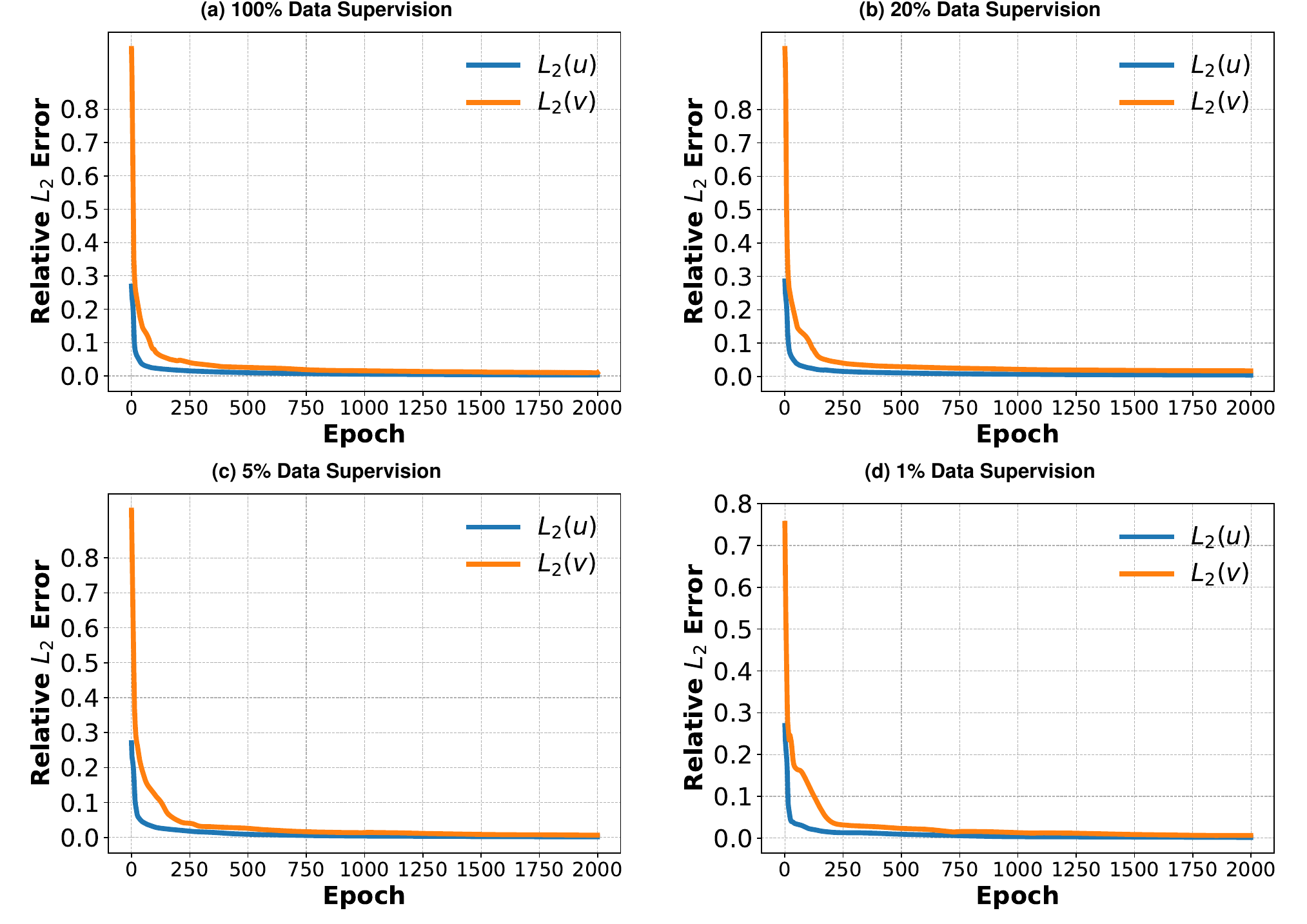}
    \caption{Convergence history of relative $L_2$ error of u-velocity and v-velocity with respect to training epochs under different sparsity levels of data for flow past a circular cylinder at Re=100; (a), (b), (c), and (d) are the 100\%, 20\%, 5\%, and 1\% data, respectively.
    }
    \label{fig:wpirnns_rel_error_grid}
\end{figure}

\begin{figure}[htbp]
    \centering
    \includegraphics[width=1.0\linewidth]{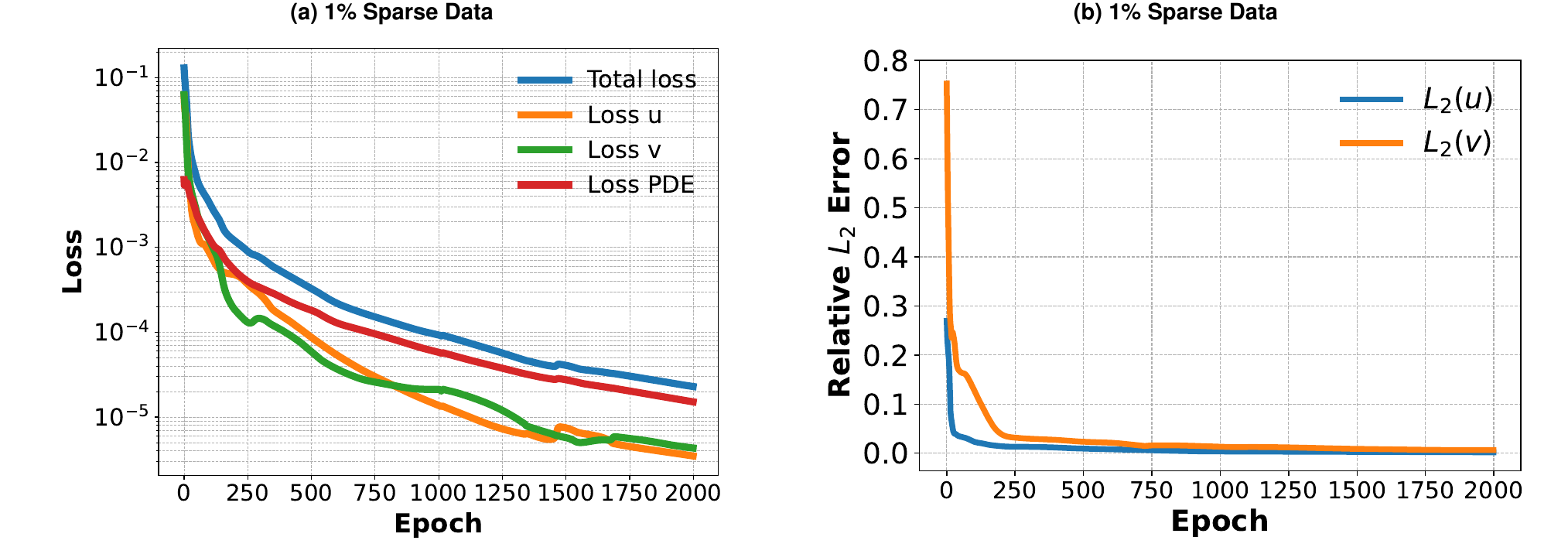}
    \caption{Results of flow past a circular cylinder at Re=100 with 1\% supervised data: (a) shows the different loss component progress with respect to epochs, and (b) demonstrates how the relative $L_2$ error progresses with respect to epochs during training.}
    \label{fig:loss_relerror_sparse1}
\end{figure}

The primary objective of this study is to develop an accurate and robust framework for flow-field reconstruction under varying levels of data sparsity. Accordingly, the proposed wavelet–physics-informed neural networks (W-PIRNNs) are evaluated using four sparsity configurations of the velocity measurements: $100\%$, $20\%$, $5\%$, and $1\%$. For all sparsity levels considered, W-PIRNNs consistently outperform conventional physics-informed neural networks (PINNs) \citep{raissi2019physics} and physics-informed convolutional networks with feature fusion (FFPICN) \cite{liu2025physics}, as demonstrated through both qualitative flow visualizations and quantitative error assessments. To assess the proposed formulation's ability to capture unsteady flow dynamics, W-PIRNNs are trained using velocity data for each sparsity level. The reconstructed velocity and pressure fields are evaluated at two representative temporal snapshots: an early-stage snapshot at $t = 3.30\,\mathrm{s}$ and a later-stage snapshot at $t = 17.40\,\mathrm{s}$. These time snapshots are selected to assess the reconstruction accuracy during both transient development and near-wake flow regimes. Validation is performed by comparing the predicted velocity and pressure fields against reference solutions using sparse observational data. Figures~\ref{fig:sparse_velocity_results} and~\ref{fig:sparse_velocity_lowdata} present the reconstructed streamwise ($u$) and cross-stream ($v$) velocity components obtained from W-PIRNNs for all sparsity levels at the selected time snapshots. The corresponding absolute error distributions for the velocity predictions at $100\%$, $20\%$, $5\%$, and $1\%$ data availability, evaluated at $t = 3.30\,\mathrm{s}$ and $t = 17.40\,\mathrm{s}$, are shown in Figure~\ref{fig:absolute_error_sparse_velocity}. Pressure field reconstructions at the same temporal snapshots are provided in Figure~\ref{fig:pressure_sparse_all}, while the absolute pressure error distributions across different sparsity configurations are illustrated in Figure~\ref{fig:pressure_absolute_error_sparse}. The results demonstrate that the proposed W-PIRNNs maintain stable and accurate reconstruction performance across all levels of data sparsity. In the fully supervised case with $100\%$ velocity data, the model attains a total loss of $1.57 \times 10^{-5}$, indicating strong convergence behavior. Notably, even under severely limited data conditions, the reconstructed velocity and pressure fields remain in close agreement with the reference and experimental results. For completeness, the training loss histories for $1\%$ dataset and relative $L_2$ error curves corresponding to all sparsity levels are shown in Figure~\ref{fig:wpirnns_rel_error_grid}, \ref{fig:loss_relerror_sparse1}.

\begin{table}[htbp]
\centering
\caption{Comparison of the relative $L_2$ error of PINNs~\citep{xu2023practical}, FFPICN~\cite{liu2025physics}, and proposed W-PIRNNs at Re=100 for the flow past a circular cylinder, u-velocity and v-velocity, using two different supervised data sets: 100\% and 1\%, respectively.}
\label{tab:relative_error_comparison}
\renewcommand{\arraystretch}{1.2}
\begin{tabular}{lccc}
\toprule
\textbf{Method} & \textbf{Data Level} & \( L_2(u) \) & \( L_2(v) \) \\
\midrule
\multirow{2}{*}{\centering PINNs~\cite{xu2023practical}} 
  & 100\% & 0.0052 & 0.0485 \\
  & 1\%   & 0.0093 & 0.0847 \\
\midrule
\multirow{2}{*}{\centering FFPICN~\citep{liu2025physics}} 
  & 100\% & 0.0013 & 0.0047 \\
  & 1\%   & 0.0041 & 0.0083 \\
\midrule
\multirow{2}{*}{\centering W-PIRNNs (Proposed)} 
  & 100\% & 0.0010 & 0.0045 \\
  & 1\%   & 0.0018 & 0.0065 \\
\bottomrule
\end{tabular}
\end{table}

\begin{figure}[htbp]
    \centering
    \textbf{1\% sparse data}\\[0.5em]  % Bold centered title above image
    \includegraphics[width=\textwidth]{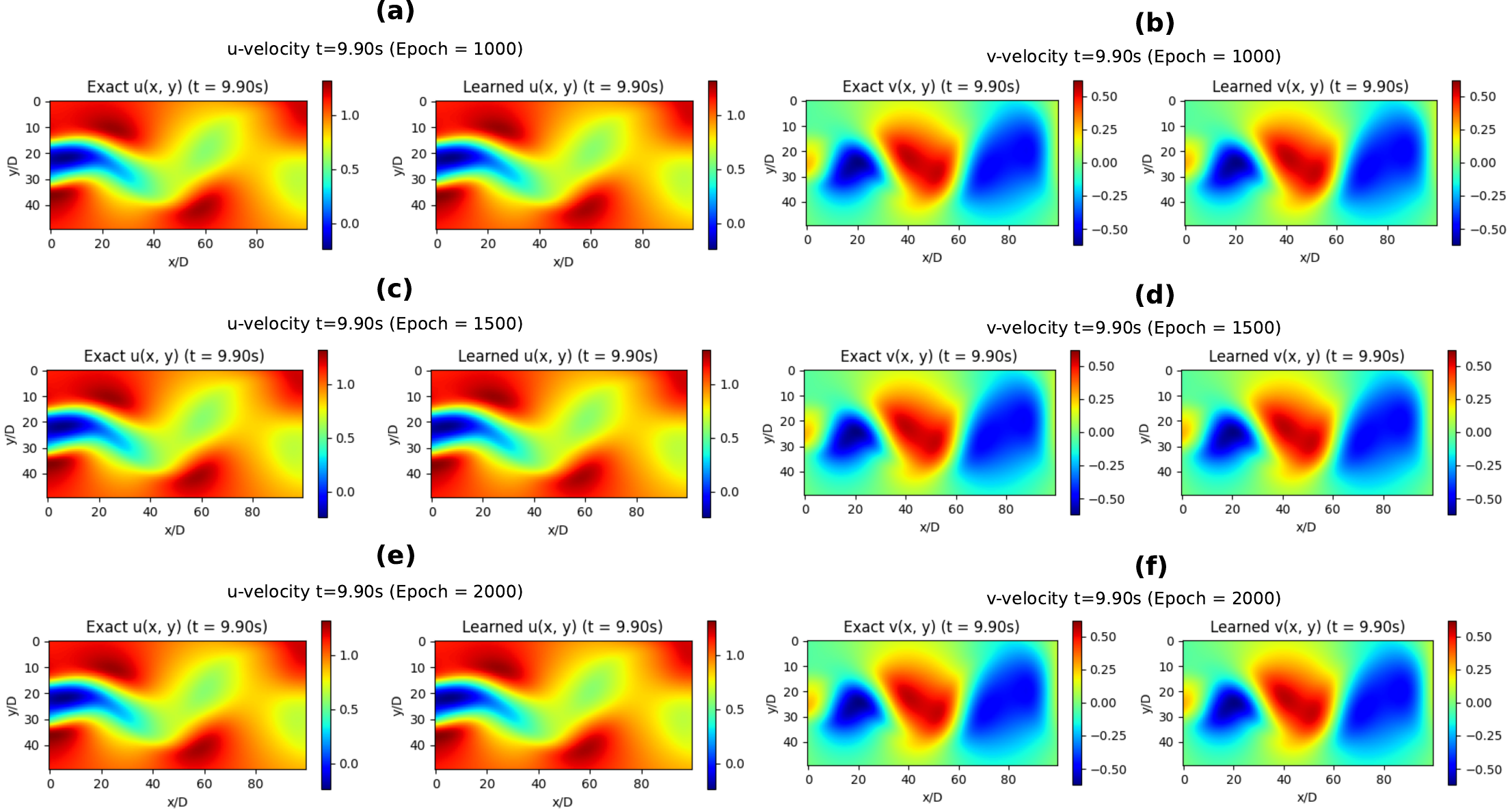}  % <-- replace with your actual filename
    \caption{
        Comparison of the exact~\citep{raissi2019physics} (CFD benchmark) [left column] and learned (reconstructed) [right column] u-velocity and v-velocity results for the flow past a circular cylinder at Re=100 at t=9.90s: (a), (c), and (e) show the u-velocity at training epochs 1000, 1500, and 20000, respectively, while (b), (d), and (f) show the corresponding v-velocity field.
    }
    \label{fig:uv_velocity_epoch}
\end{figure}

\begin{figure}[htbp]
    \centering
    \includegraphics[width=0.9\textwidth]{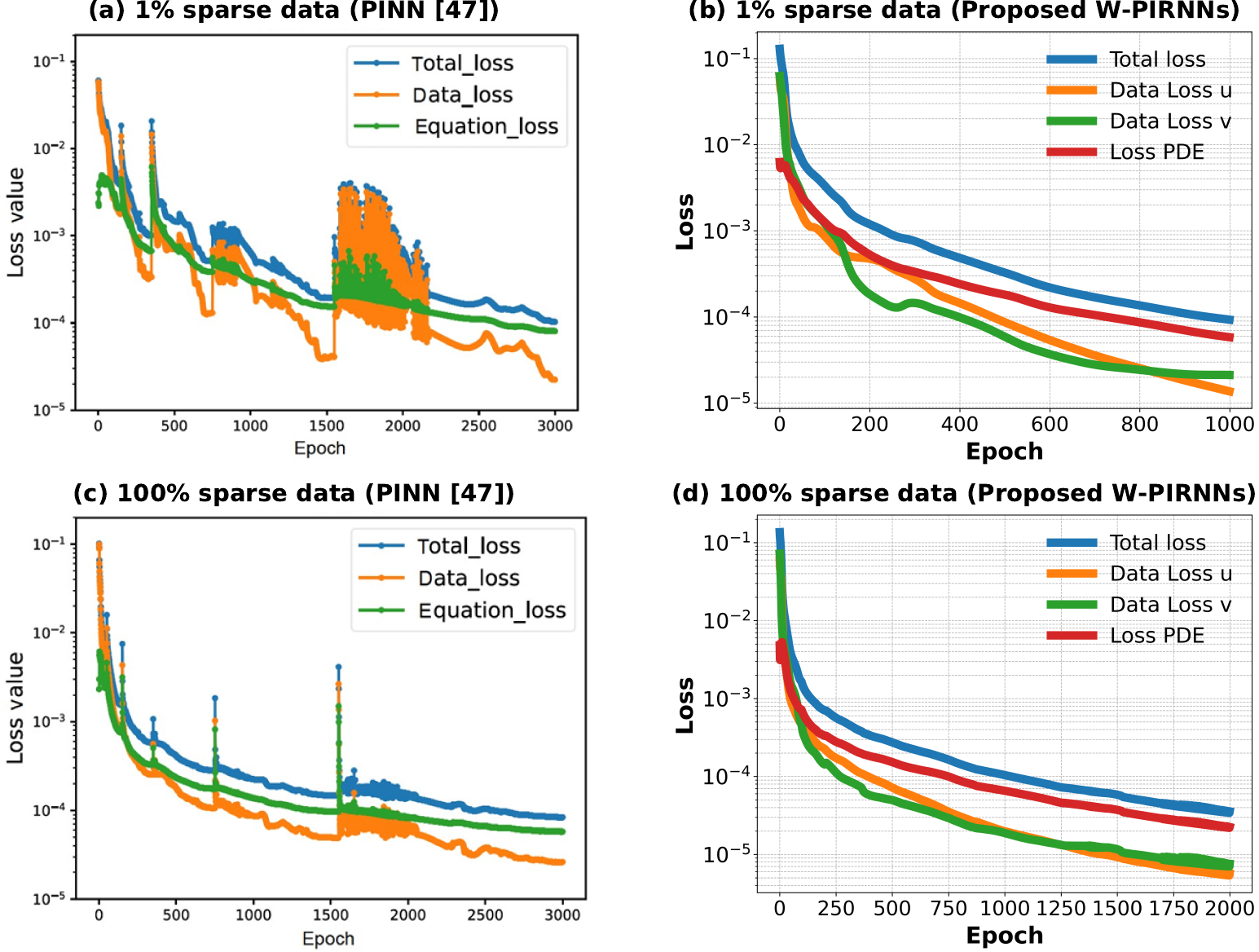} % Replace with your actual filename

    % Manual labels overlaid on image (adjust position as needed)
    % \vspace{-0.5cm}
    % \hspace*{-13.5cm}(a)\hspace{6.6cm}(b)\hspace{6.5cm}(c)\hspace{2.2cm}(d)
    
    \caption{Comparison between PINNs~\citep{xu2023practical} (left column) and the proposed W-PIRNNs (right column) method of the different loss components with respect to epoch during training for a flow past a circular cylinder at Re=100, pairs (a, b) and (b, d) results for 1\% and 100\% supervised data, respectively.}
    \label{fig:loss_comparison1}
\end{figure}

\begin{figure}[htbp]
    \centering
    \includegraphics[width=0.8\textwidth]{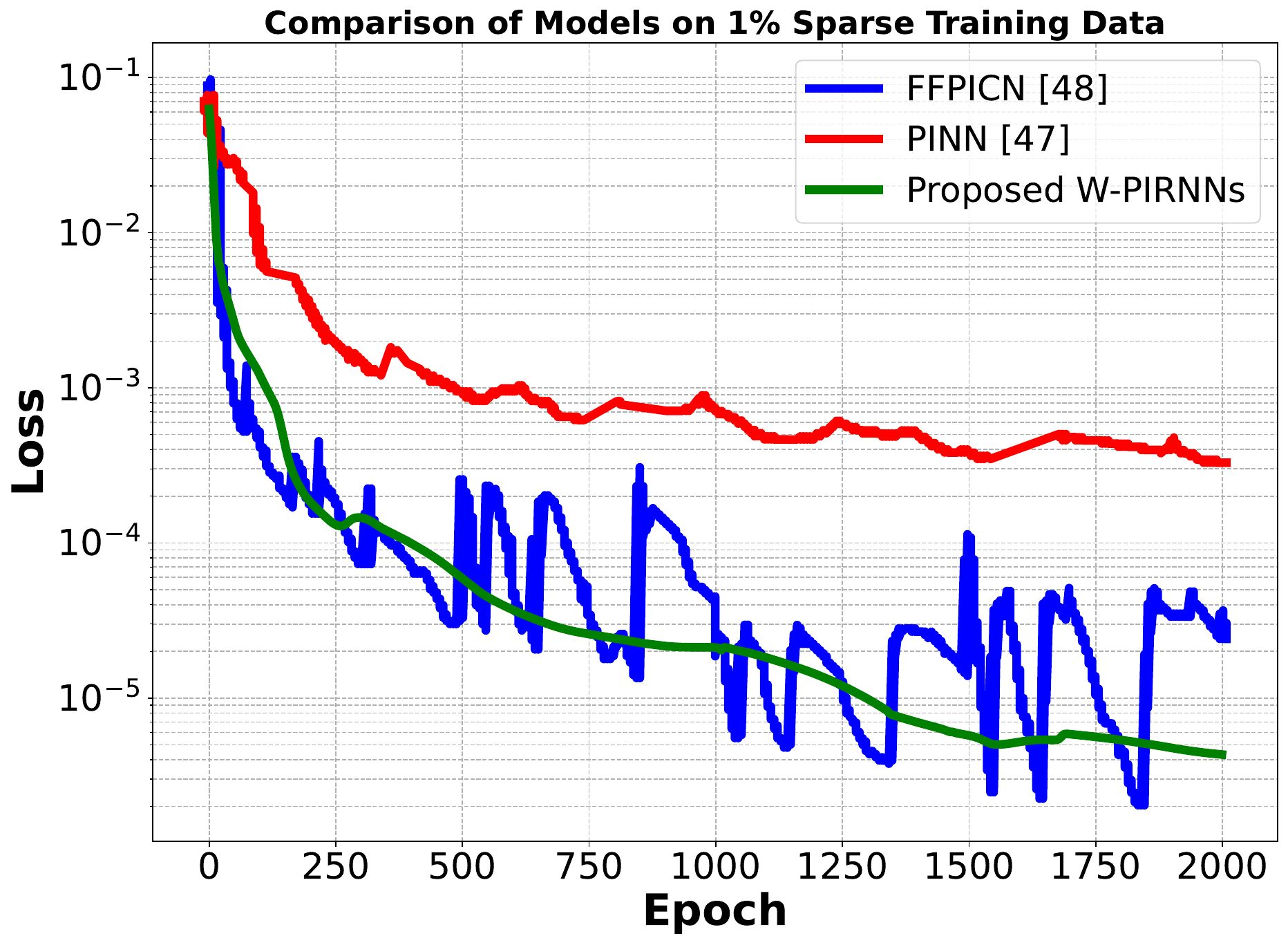}

    \caption{
        Comparison PINNs \citep{xu2023practical}, FFPICN \citep{liu2025physics}, and the proposed method (W-PIRNNs) of the total training loss progress with respect to epochs during training for the flow past a circular cylinder at Re=100 under 1\% supervised data.
    }
    \label{fig:loss_comparison_1percent}
\end{figure}

\begin{figure}[htbp]
    \centering
    \includegraphics[width=1.0\linewidth]{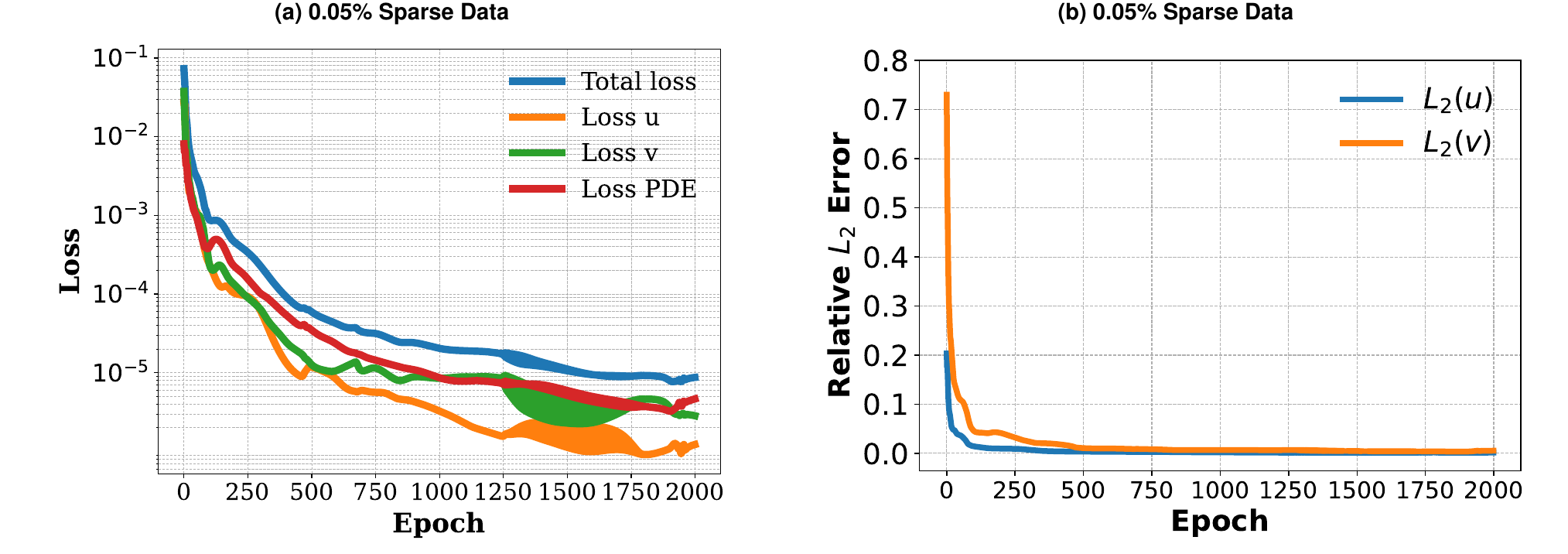}
    
    \caption{
       For the flow past a circular cylinder at Re=100 for 0.05\% supervised data, (a) shows the different loss component behaviors with respect to epochs, and (b) shows the relative $L_2$ error of both u-velocity and v-velocity progress with respect to epochs during training.
    }
    \label{fig:loss_error_plots}
\end{figure}

\begin{figure}[htbp]
    \centering
    \textbf{0.16\% sparse data}\par\vspace{0.3em}
    \includegraphics[width=0.90\textwidth]{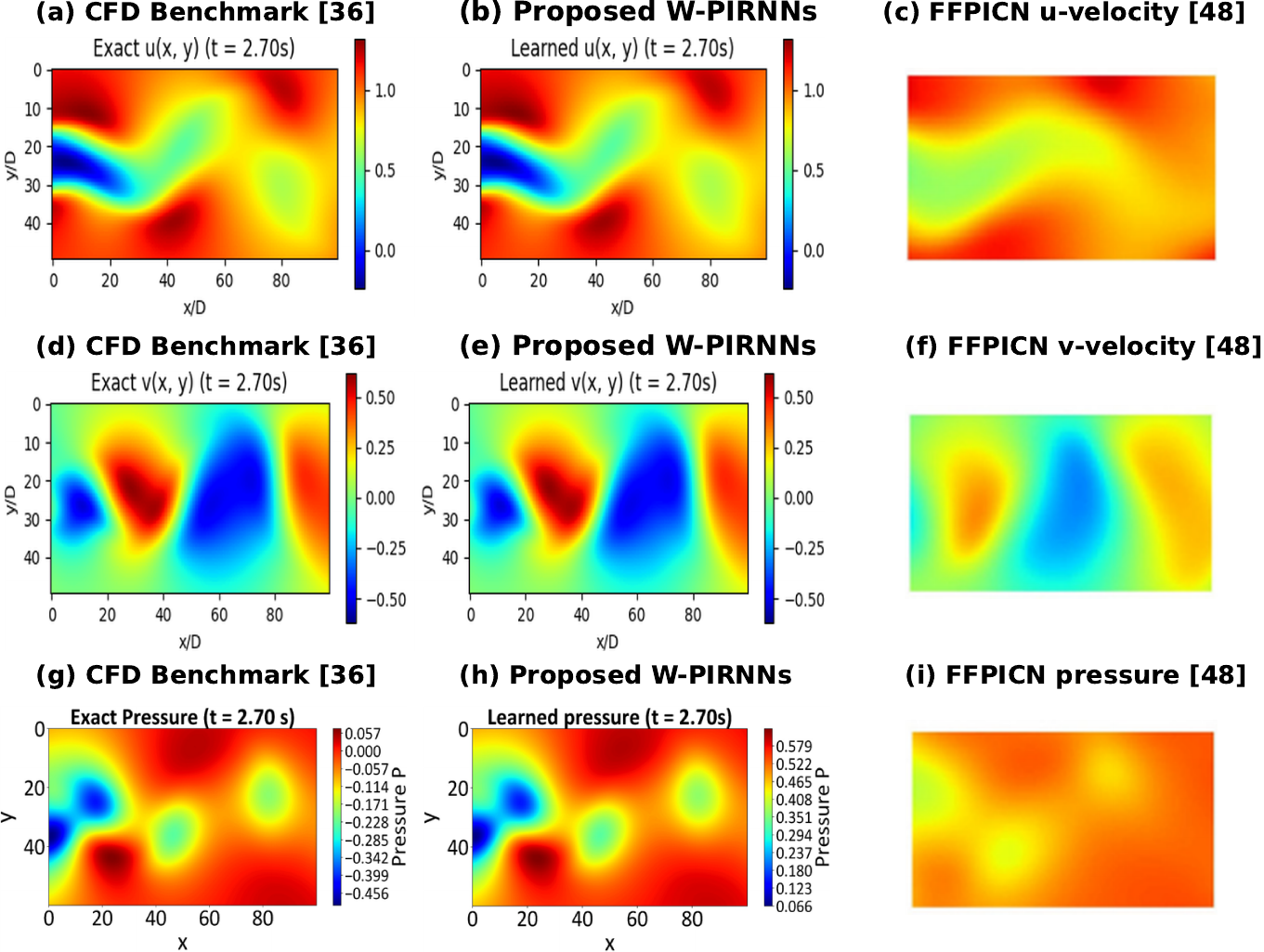} % Replace with actual filename
    \caption{Comparison of the exact~\citep{raissi2019physics} (CFD Benchmark) [first column], learned (proposed W-PIRNNs) [second column], and FFPICN~\citep{liu2025physics} [third column] of u-velocity, v-velocity, and pressure results of the flow past a circular cylinder at Re=100 for 0.16\% supervised data at the temporal snapshot t=2.70s: (a), (d), and (g) are the exact~\citep{raissi2019physics} (CFD benchmark); (b), (e), and (h) are the learned (proposed W-PIRNNs); and (c), (f), and (i) are the FFPICN~\citep{liu2025physics} results, respectively.}
    \label{fig:comparison-016}
\end{figure}

\begin{figure}[htbp]
    \centering
    \textbf{0.05\% sparse data} \\[1ex] % <-- Adds bold label at the top
    \includegraphics[width=0.95\textwidth]{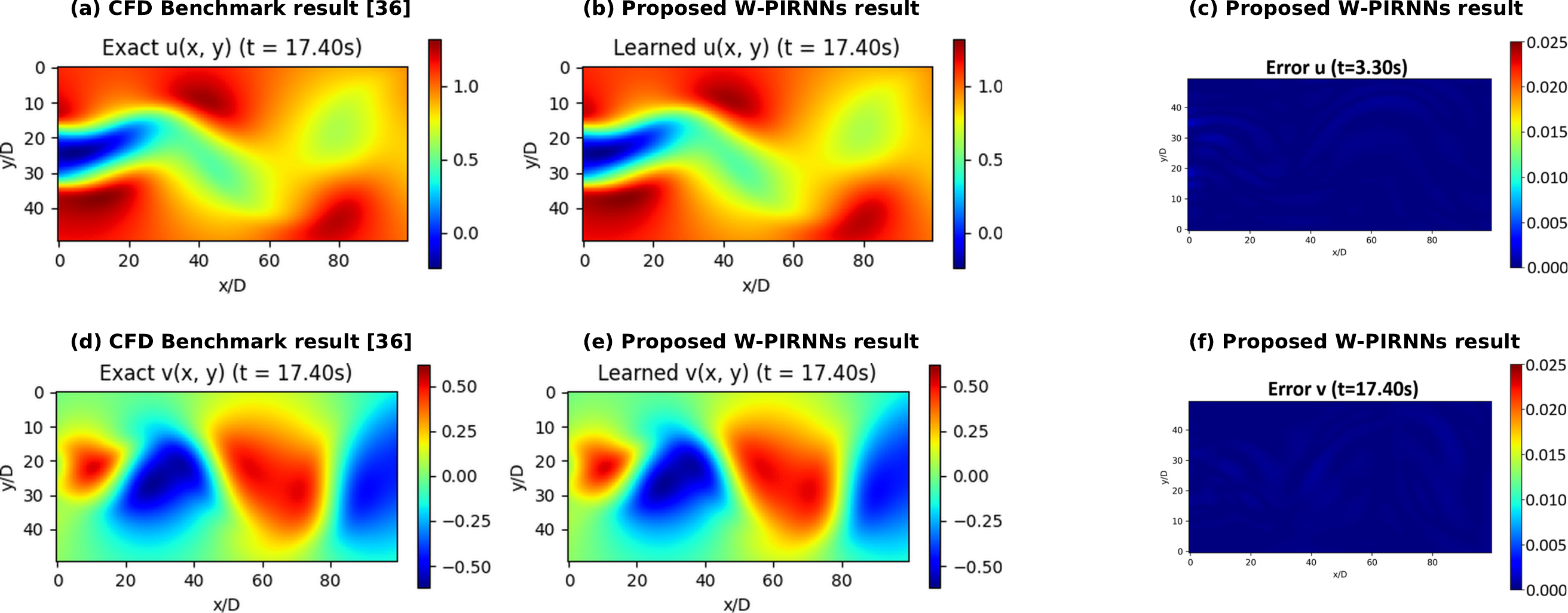}
    \caption{
        Comparison of the exact~\citep{raissi2019physics} (CFD benchmark) and learned (proposed W-PIRNNs) velocity fields for flow past a circular cylinder at $\mathrm{Re}=100$ at $t=17.40\,\mathrm{s}$ using $0.05\%$ supervised data, (a) and (b) show $u$-velocity, (d) and (e) show $v$-velocity results, respectively, and pair (c,f) are the corresponding absolute error fields.
    }
    \label{fig:velocity_comparison}
\end{figure}

\begin{figure}[htbp]
    \centering
    \textbf{0.05\% sparse data} \\[1ex]
    \includegraphics[width=1.0\textwidth]{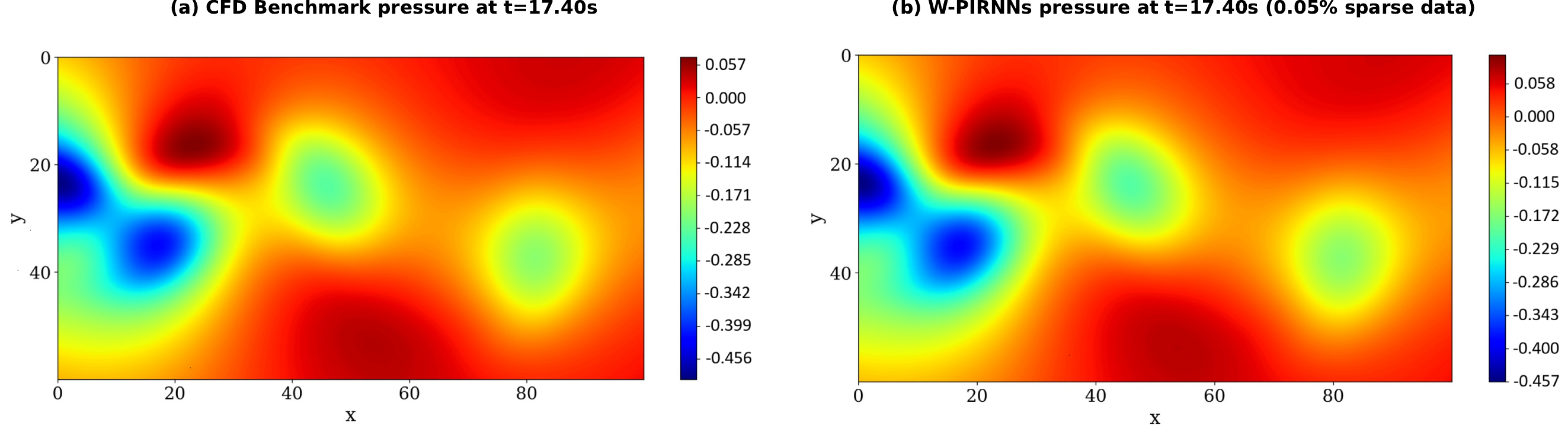}
    \caption{
    Comparison of the pressure at the temporal snapshot t=17.40s for the flow past a circular cylinder at Re=100 for 0.05\% supervised data: (a) is the CFD benchmark~\citep{raissi2019physics} pressure, and (b) is the proposed W-PIRNN reconstructed pressure at t=17.40 s.
    }
    \label{fig:pressure_comparison}
\end{figure}

\begin{figure}[htbp]
    \centering
    \textbf{0.05\% sparse data} \\[1ex]
    \includegraphics[width=0.9\linewidth]{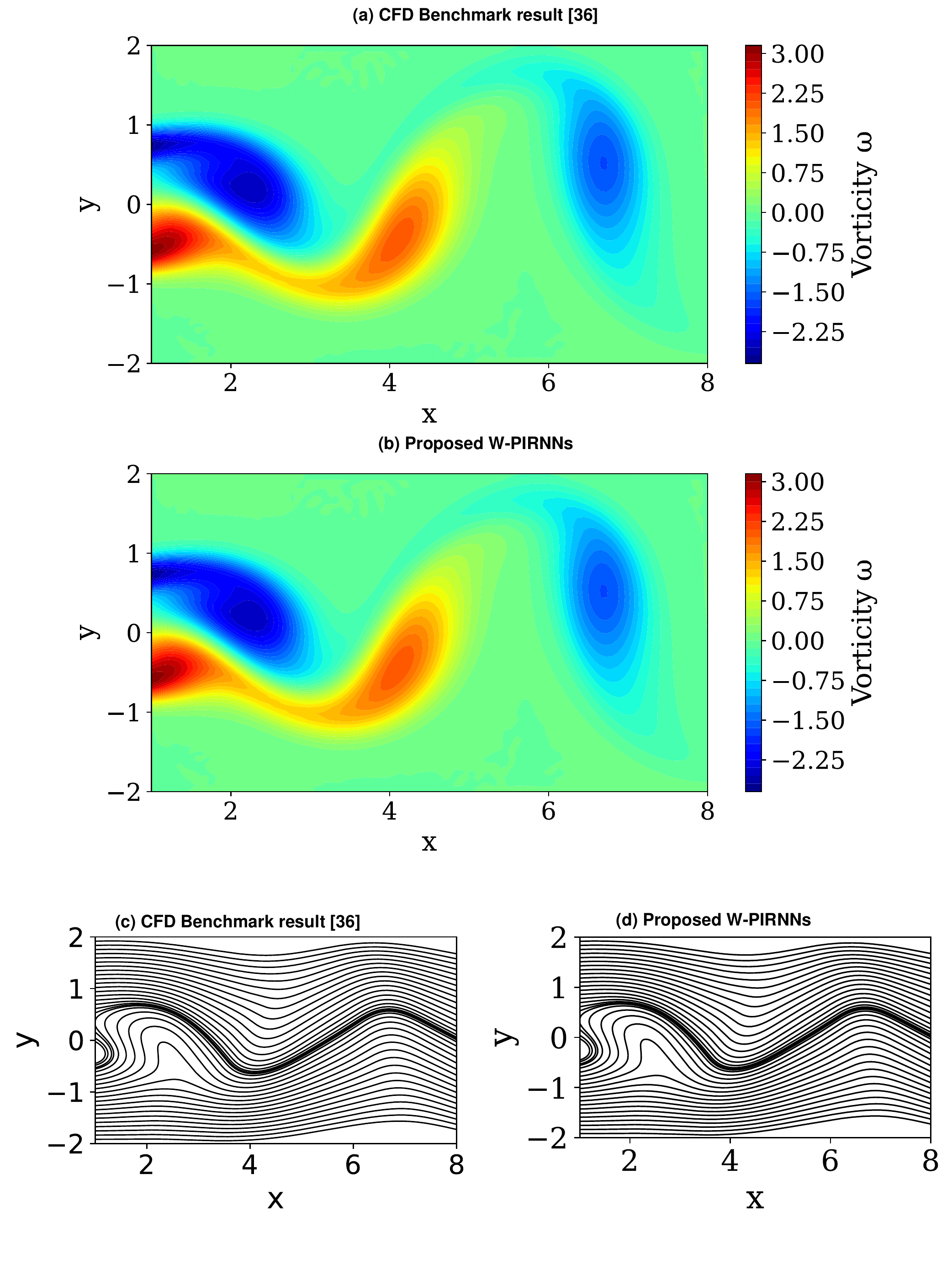}

    \caption{Comparison of the vorticity contour and streamlines of the flow past a circular cylinder at Re=100, CFD benchmark~\citep{raissi2019physics}, and reconstructed (proposed W-PIRNNs) at the temporal snapshot t=17.40s: (a) is the CFD benchmark~\citep{raissi2019physics} vorticity contour, (b) is the reconstructed (proposed W-PIRNNs) vorticity contour, and (c) and (d) are CFD benchmark streamlines~\citep{raissi2019physics} and proposed W-PIRNN streamlines, respectively.}
    \label{fig:comparison_vorticity_streamline}
\end{figure}

\begin{figure}[htbp]
    \centering
    \textbf{1\% sparse data} \\[1ex]
    \includegraphics[width=\textwidth]{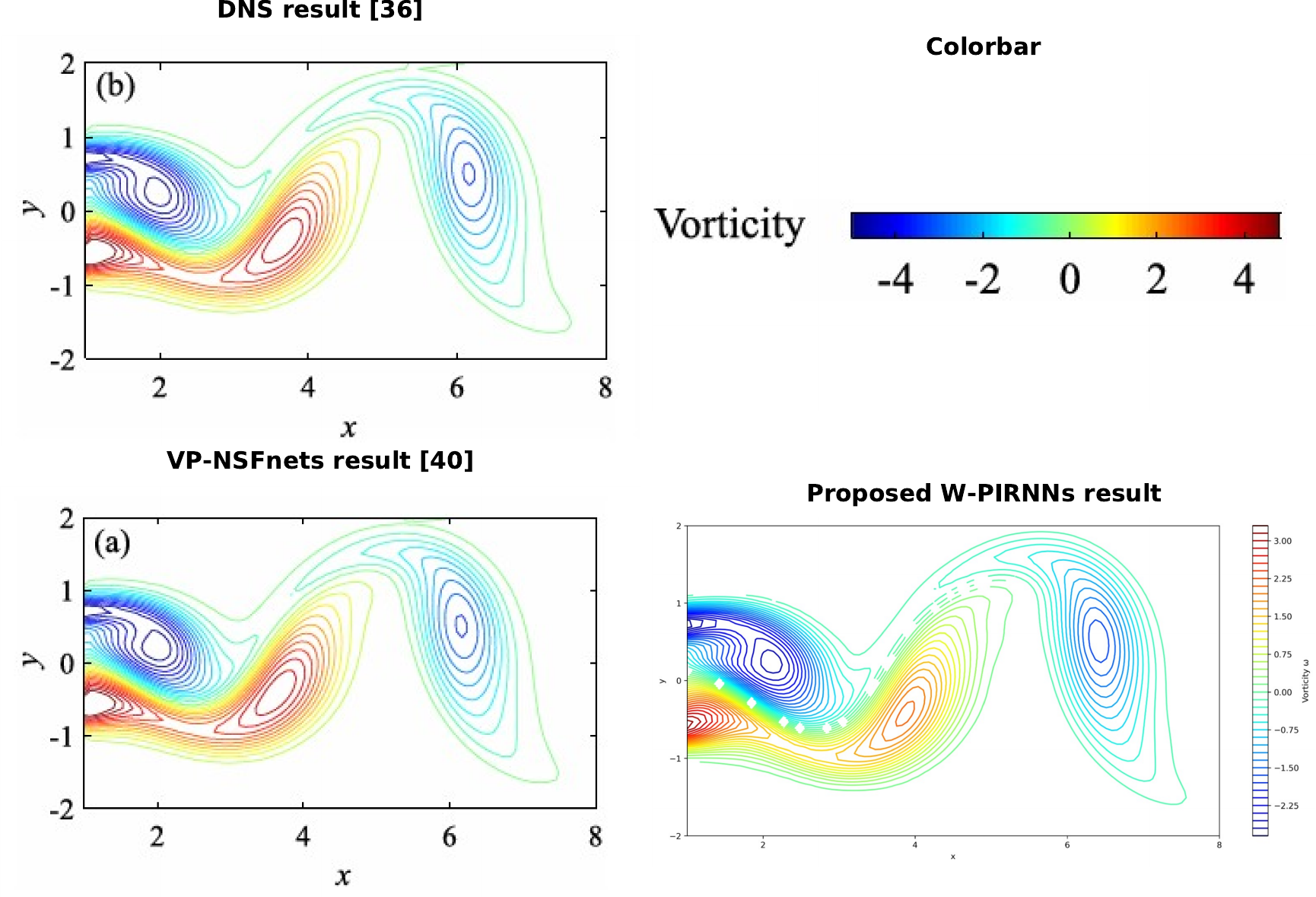}
    \caption{
        Comparison of vorticity contours at \( t = 4.0 \,\text{s} \) for the flow past a circular cylinder at Re=100: The first row presents the DNS outcomes \citep{raissi2019physics}. The second row left VP-NSFnet \citep{jin2021nsfnets} result (with fixed weights \( \alpha = \beta = 1 \)) and the right proposed W-PIRNNs result under $1\%$ sparse data supervision. 
    }
    \label{fig:vorticity_comparison}
\end{figure}

\begin{figure}[htbp]
    \centering
    \textbf{1\% sparse data} \\[1ex]
    \includegraphics[width=1.0\linewidth]{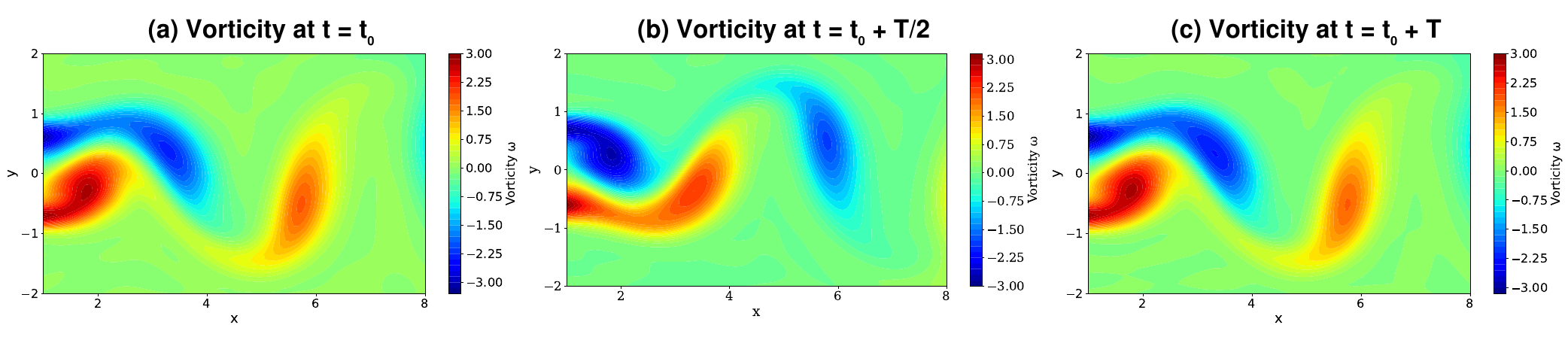}
    \caption{The full period (T) of vortex shedding reconstruction for the flow past a circular cylinder at Re=100 using 1\% supervised data is illustrated in vorticity contours (a), (b), and (c) at times $t = t_0$, $t = t_0 + T/2$, and $t = t_0 + T$, respectively.}
    \label{fig:vorticity_cycle}
\end{figure}

\begin{figure}[htbp]
    \centering
    % \textbf{1\% sparse data} \\[1ex]
    \includegraphics[width=\linewidth]{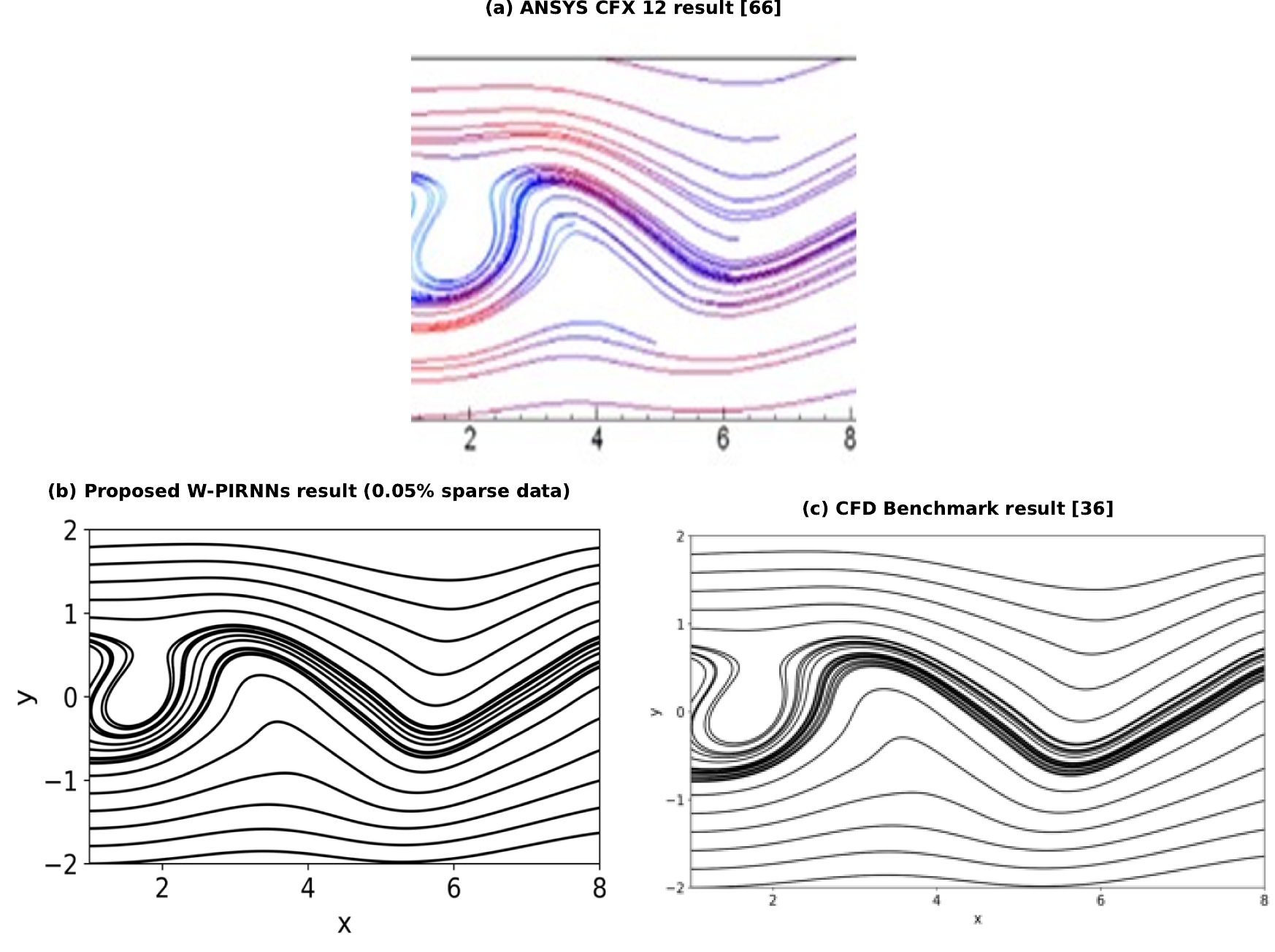}
    \caption{
        Streamline comparisons of the flow past a circular cylinder at Re=100 for 1\% supervised data: (a), (b), and (c) are the ANSYS CFX 12 result by Laroussi et al. (2014) \citep{laroussi2014triggering}, the proposed W-PIRNNs reconstructed streamlines, and the CFD benchmark result \citep{raissi2019physics}, respectively.
    }
    \label{fig:streamline_comparison}
\end{figure}

\begin{figure}[htbp]
    \centering
    \textbf{1\% sparse data} \\[1ex]
    
    \includegraphics[width=1.0\linewidth]{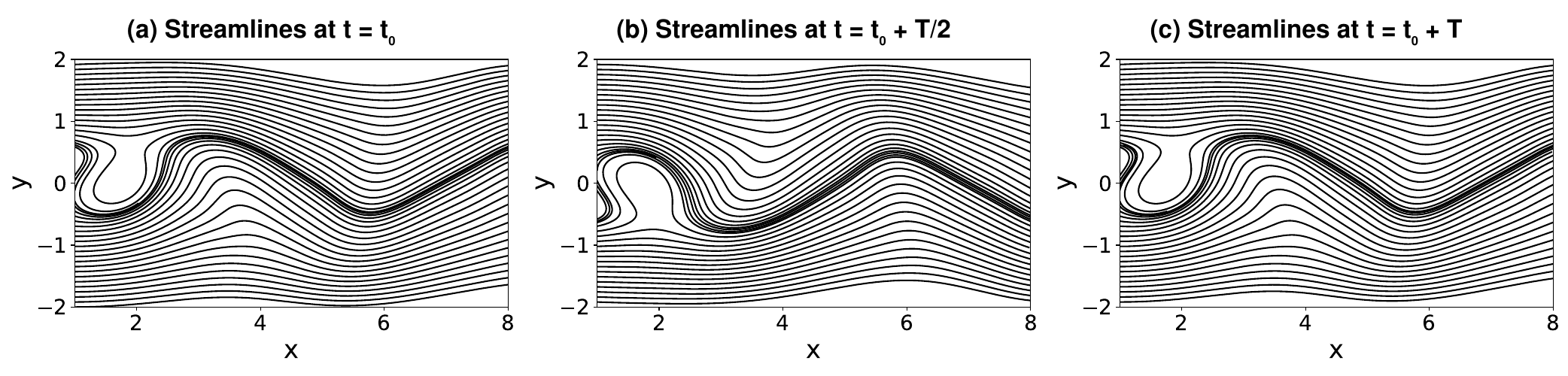}
    \caption{The whole period (T) streamlines reconstruction of the flow past a circular cylinder at Re=100 using 1\% supervised data; (a), (b), and (c) are the streamlines at time $t = t_0$, $t = t_0 + T/2$, and $t = t_0 + T$, respectively.}
    \label{fig:streamline_cycle}
\end{figure}

The comparison of findings at two temporal snapshots, $t=3.30s$ and $t=17.40s$, reveals exceptional agreement; specifically, our proposed results closely match the ground truth results. We present the findings at both the initial and final time steps to demonstrate that our method accurately captures the dynamic nature of unsteady phenomena. Moreover, our W-PIRNNs produced quantitatively superior outcomes relative to the conventional PINN and FFPICN findings within $2000$ epochs, whereas the vanilla PINN and FFPICN required $3000$ epochs, further establishing the superiority of our results over both methodologies. In the setting of $1\%$ sparse data supervision, the total loss is $0.0000239$, with velocity losses of $0.00000454$ for $u$ and $0.00000478$ for $v$, resulting in relative $L_2$ errors of $0.0018$ for $u$ and $0.0065$ for $v$. In traditional PINNs~\citep{xu2023practical}, the relative error of $u$ is $0.0087$ after $3000$ epochs, whereas in FFPICN~\citep{liu2025physics}, the relative error of $u$ is $0.0046$ and the relative $L_2$ error of $v$ is $0.0083$; we achieve better results in fewer epochs. Additionally, we
include our model's performance on $100\%$ and $1\%$ training data, measured by the relative $L_2$ error, compared to both the PINNs~\citep{xu2023practical} and FFPICN~\citep{liu2025physics} models in Table~\ref{tab:relative_error_comparison}.

Our model achieves superior performance on test data because we do not provide training data for pressure prediction; yet, our proposed approach accurately reconstructs these pressures Figure~\ref{fig:pressure_sparse_all}. Figure~\ref{fig:uv_velocity_epoch} displays the u-velocity and v-velocity results at the temporal snapshot $t=9.90s$ for three distinct epochs: $1000$, $1500$, and $2000$, utilizing $1\%$ sparse velocity data. In all cases, our predicted outcomes align completely with the experimental findings. To achieve enhanced accuracy and stable convergence of loss and relative $L_2$ error, which approaches zero, we present Figure~\ref{fig:loss_relerror_sparse1} for this purpose, and we standardize the epoch at $2000$ across all cases. With $1\%$ supervised data, we achieve high precision in flow field reconstruction. That's why we use these supervised datasets to validate other physical phenomena in the near wake region (i.e., vorticity and streamlines). Additionally, we incorporate results for the supplementary physical parameters, vorticity ($\omega$) and streamlines. Our finding of vorticity reconstruction is comparable with DNS results \citep{raissi2019physics} and VP-NSFnet results~\citep{jin2021nsfnets}, which are presented in the Figure~\ref{fig:vorticity_comparison} and in the Figure~\ref{fig:streamline_comparison} present the streamlines comparison against ANSYS CFX 12 result~\citep{laroussi2014triggering} and benchmark result (DNS)~\citep{raissi2019physics}. We evaluate our model's performance at epochs $1000$, $1500$, and $2000$. For three epochs, our model additionally captures the lower epoch of $1000$, yielding notable results. 

The performance of the proposed algorithm is further examined from the perspective of different loss components, including the data loss, physics loss, and total loss, and is compared with the corresponding PINN results~\citep{xu2023practical}. Figure~\ref{fig:wpirnns_rel_error_grid} presents the relative $L_2$ errors of both u and v velocity convergence history with respect to epochs of the proposed W-PIRNNs for data sparsity levels of $100\%$, $20\%$, $5\%$, and $1\%$. In addition, Figure~\ref{fig:loss_comparison1} illustrates the evolution of the different loss functions for both PINNs and W-PIRNNs under $1\%$ and $100\%$ sparse data supervision. For the $1\%$ sparse data case, the loss associated with the proposed W-PIRNNs converges within approximately $1000$ epochs. In contrast, the PINN loss exhibits strong fluctuations across all sparsity levels of supervised velocity data. The losses obtained using W-PIRNNs remain smooth and stable across a reliable training process and stable convergence behavior. In particular, under $1\%$ sparse supervision, the PINN~\citep{xu2023practical} loss shows significant oscillations around the $2000$th epoch and becomes only moderately stable after $3000$ epochs. Meanwhile, the proposed W-PIRNNs maintain stable convergence throughout the first $2000$ epochs, with consistently lower losses across all components. These observations demonstrate that the proposed method exhibits greater training stability than the vanilla PINN~\citep{xu2023practical}.Figure~\ref{fig:loss_relerror_sparse1} shows the loss and relative error behavior for the $1\%$ sparse velocity supervision case. The results indicate that the proposed method reaches a stable minimum loss by the $2000$th epoch, characterized by a smooth and monotonic decay. 

The adaptability of the proposed approach is further evaluated against the recently developed high-resolution flow reconstruction framework based on physics-informed convolutional networks with feature fusion (FFPICN) \citep{liu2025physics}. In contrast, the PINN~\citep{xu2023practical} results remain significantly less accurate, while the FFPICN~\citep{liu2025physics} loss exhibits noticeable fluctuations throughout training Figure~\ref{fig:loss_comparison_1percent}, leading to a final loss that differs substantially from that of the proposed approach. For a fair comparison, the results of PINN~\citep{xu2023practical} and FFPICN~\citep{liu2025physics} are considered over $2000$ epochs; however, even when extended to $3000$ epochs, their losses do not reach the level achieved by W-PIRNNs at $2000$ epochs. These results confirm that the proposed algorithm is effective for flow reconstruction under extremely sparse data conditions ($1\%$). In the following, we further investigate the performance of the proposed approach for an even smaller dataset of $0.05\%$. With $1\%$ supervised data, the vorticity field reconstructed using W-PIRNNs shows close agreement with the DNS results~\citep{raissi2019physics} and is comparable to the VP-NSFnets~\citep{jin2021nsfnets} predictions, as illustrated in Figure~\ref{fig:vorticity_comparison}. The corresponding streamline reconstructions are presented in Figure~\ref{fig:streamline_comparison}, where similar consistency with existing methods and experimental observations is observed. Moreover, the reconstructed vorticity and streamline contours preserve the underlying flow physics, successfully capturing the von Kármán periodic vortex shedding, as shown in Figures~\ref{fig:vorticity_cycle} and~\ref{fig:streamline_cycle}.

\subsubsection{Flow field reconstruction with extremely sparse data ($0.16\%$ and $0.05\%$ supervision)}
A more detailed analysis is performed to examine the performance of the proposed W-PIRNNs under extremely sparse velocity data conditions. In this study, supervision levels of $0.16\%$ and $0.05\%$ are considered for reconstructing the velocity, pressure, vorticity ($\omega$), and streamline fields, as shown in Figures~\ref{fig:loss_error_plots}, \ref{fig:comparison-016}, \ref{fig:velocity_comparison}, \ref{fig:pressure_comparison}, and \ref{fig:comparison_vorticity_streamline}. Using the predicted velocity fields, the corresponding vorticity and streamline quantities are computed. The results demonstrate that the proposed W-PIRNNs are able to accurately reconstruct all flow fields even under such limited data availability. For this extremely sparse setting, the total loss is $8.74 \times 10^{-6}$, with data losses of $1.23 \times 10^{-6}$ and $2.85 \times 10^{-6}$ for the $u$ and $v$ velocity components, respectively. The corresponding relative $L_2$ errors are $0.0015$ for $u$ and $0.0054$ for $v$. Figure~\ref{fig:loss_error_plots} illustrates the variation of individual loss components together with the relative $L_2$ errors of $u$ and $v$ during training. Although a mild degree of overfitting is observed, the reconstructed flow field remains in close agreement with the reference solution at the temporal snapshot $t = 17.40\,\mathrm{s}$. Figure~\ref{fig:comparison_vorticity_streamline} further presents a direct comparison of the vorticity and streamline fields obtained from the experimental~\citep{raissi2019physics} (DNS) data and the proposed W-PIRNNs. The close correspondence between the two confirms that the essential flow physics is well preserved by the proposed approach, even at extremely low data sparsity levels. To the best of our knowledge, neither the conventional PINN nor the FFPICN framework is able to successfully reconstruct the flow field at sparsity levels of $0.16\%$ and $0.08\%$. For the $0.16\%$ sparse velocity data case, Figure~\ref{fig:loss_error_plots} compares the reconstructed velocity and pressure fields obtained using W-PIRNNs with those produced by FFPICN \citep{liu2025physics}. The results clearly indicate that the proposed method accurately reconstructs both velocity and pressure at the temporal snapshot $t = 2.70\,\mathrm{s}$. A quantitative summary of the reconstruction accuracy for supervised velocity training at $0.16\%$ and $0.05\%$ data availability is provided in Table~\ref{tab:relative_error_comparison1}, with comparisons to FFPICN included for the $0.16\%$ case. Furthermore, the results obtained using $0.05\%$ supervised data demonstrate that the proposed methodology maintains a high level of accuracy when compared with the experimental measurements (DNS). These findings confirm the robustness of the proposed approach under extremely sparse data conditions. Notably, within $1000$ training epochs, our proposed algorithm reaches the lowest relative $L_2$ error.

\begin{table}[htbp]
\centering
\caption{
Comparison of FFPICN~\citep{liu2025physics}, Adaptive PINNs~\citep{jagtap2020adaptive}, SIREN PINNs, and the proposed W-PIRNN of the relative $L_2$ errors of the $u$-velocity and $v$-velocity of the flow past a circular cylinder at Re=100 for 0.16\% and 0.05\% supervised data sets. For all methods, the case with $0.16\%$ supervised data is trained for $1000$ epochs, while the case with $0.05\%$ supervised data is trained for $2000$ epochs. The FFPICN result is taken from this paper~\citep{liu2025physics}.
}
\label{tab:relative_error_comparison1}
\renewcommand{\arraystretch}{1.2}
\begin{tabular}{lccc}
\toprule
\textbf{Method} & \textbf{Data Level} & $L_2(u)$ & $L_2(v)$ \\
\midrule

\multirow{1}{*}{FFPICN~\citep{liu2025physics}} 
& 0.16\% & 0.2492 & 0.7491 \\
& 0.05\% & --- & --- \\

\midrule

\multirow{2}{*}{Adaptive PINNs~\citep{jagtap2020adaptive}} 
& 0.16\% & 0.0050 & 0.0090 \\
& 0.05\% & 0.0045 & 0.0088 \\

\midrule

\multirow{2}{*}{SIREN PINNs} 
& 0.16\% & 0.0067 & 0.0104 \\
& 0.05\% & 0.0051 & 0.0089 \\

\midrule

\multirow{2}{*}{\textbf{W-PIRNNs (Proposed)}} 
& 0.16\% & \textbf{0.0019} & \textbf{0.0066} \\
& 0.05\% & \textbf{0.0015} & \textbf{0.0054} \\

\bottomrule
\end{tabular}
\end{table}

Figure~\ref{fig:vorticity_comparison} displays our predicted vorticity results at the temporal instance of $t=4.00s$, which qualitatively correspond with the findings from experimental (DNS) \citep{raissi2019physics} and VP-NSFnet \citep{jin2021nsfnets}. Our reconstructed streamlines further validate the experimental result and ANSYS CFX 12 \citep{larosa2023halting}, as illustrated in Figure~\ref{fig:streamline_comparison}, and are noticeable. Figures~\ref{fig:vorticity_cycle} and \ref{fig:streamline_cycle} demonstrate that we are accurately capturing the complete duration of the Von Kármán street, similar to the initial timestep $t = t_0$ shedding. At time $t = t_0 + T/2$, we observe opposite shedding (where $T$ is the duration of the whole period), and at the timestep $t = t_0 + T$, we get similar shedding at time $t_0$. Similarly, we observe a periodicity in the streamlines, indicating that our proposed method comprehensively captures the problem's physics at a Reynolds number Re=$100$.

\subsection{Validation on a more complex flow field at Re=3900}
To validate turbulent flow, we use the 2D circular cylinder data \citep{xu2023spatiotemporal} at Reynolds number Re=3900. Due to the restrictions of present measuring methods and instrument accuracy, flow field reconstruction data is typically sparse in the spatial domain. However, at high sample frequencies, the observed data can become dense in the temporal domain. The $k-\epsilon$ model \citep{yan2023exploring} simulates wake flow past a 2D circular cylinder at Reynolds number $R e=3900$. The flow data $u, v, p$ from 36 sparsely distributed points during a 42.9 s period are utilized as labeled training data. Each of the 100 snapshots contains 36 data points, so the total number of data points, $\mathcal{N}_d$, is 3600. The total number of collocation points $\mathcal{N}_r$ is 1000000, and they are sampled using the Latin hypercube sampling (LHS) method throughout the spatiotemporal domain. In this turbulent flow example, we used supervised velocity and pressure data; the total dataset comprised only $0.09\%$ of the data. This study's neural network architecture consists of $4$ hidden layers, each with $100$ neurons. An initial learning rate of $1 \times 10^{-3}$ is used, and a learning rate scheduler is included to improve convergence behavior during training. The model is optimized with the Adam optimizer and trained for $4000$ epochs. The proposed framework's performance depends on the network configuration implemented. The wavelet activation function, which mixes sine and cosine components, requires accurate hyperparameter tuning. Incorrect parameter selection can lead to high GPU memory utilization during training.

\begin{figure}[htbp]
    \centering
    \includegraphics[width=\textwidth]{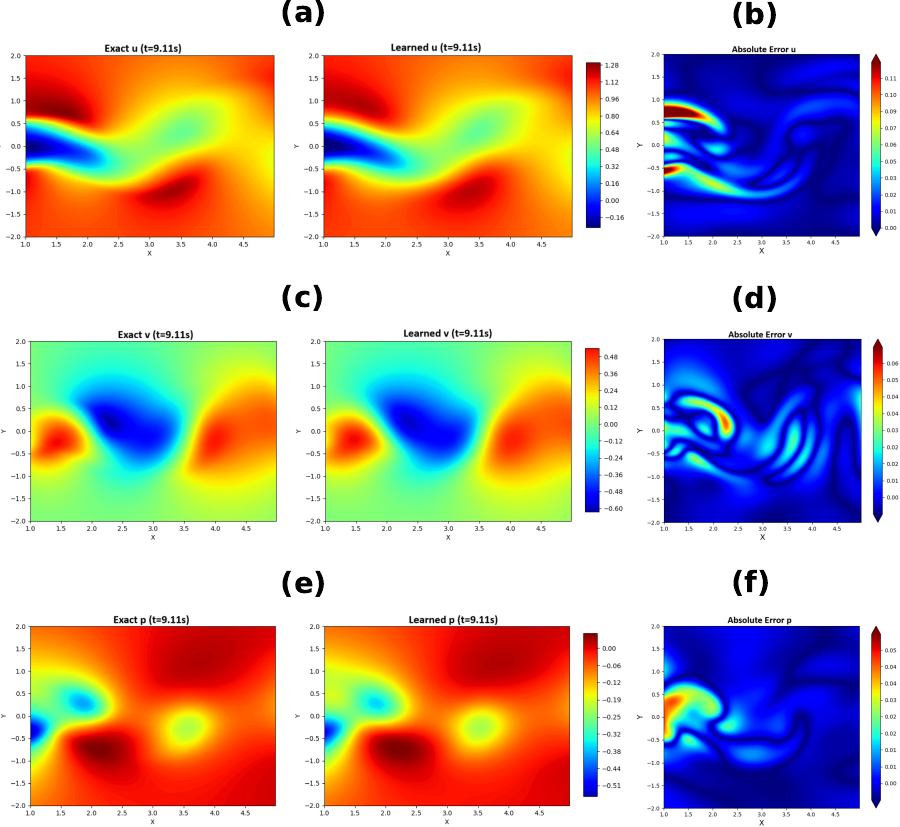}
    \caption{
    The complex flow reconstruction for flow past a circular cylinder at Reynolds number ($\mathrm{Re}=3900$)~\citep{xu2023spatiotemporal}: (a), (c), and (e) are the exact (CFD benchmark result) [left] and learned (proposed W-PIRNNs result) [right] $u$-velocity, $v$-velocity, and pressure, respectively, at $t = 9.11\,\mathrm{s}$, and (b), (d), and (f) show the corresponding absolute error at the same time snapshot.
    }
    \label{fig:Re3900_turbulent_validation}
\end{figure}

A comparison was performed between CFD benchmark flow-field findings and our proposed W-PIRNN predictions. Figure~\ref{fig:Re3900_turbulent_validation} illustrates that the proposed W-PIRNNs framework efficiently reconstructed the original flow field and captured the vortex shedding pattern from a sparse observation of merely 36 points. The relative $L_2$ error $(= \frac{\lVert \hat{U} - U \rVert}{\lVert U \rVert})$ has been defined to quantitatively evaluate the difference between the predicted flow field and the original flow field over the full temporal domain, where, $\|\hat{U}-U\|$ represents the $L_2$ norm of the prediction error for the quantities of interest $\{u, v, p\}$ at a specific time, while $\|U\|$ denotes the $L_2$ norm of the original quantity at the same time. The resultant average relative $L_2$ error values for $u$, $v$, and $p$ were $0.017$, $0.022$, and $0.090$, respectively.

\clearpage
\section{Methodological validation}
\label{sec:Methodological validation}
Our proposed W-PIRNNs are not only restricted to reconstructing the flow field, but they are also robust enough to solve forward and inverse partial differential equations. For validation purposes, we use Burger's equation and the Schrodinger equation. Burger's equation is solved in both forward and inverse settings. In contrast, the Schrödinger equation is solved in the forward setting to ensure the method is robust enough to enforce the periodic boundary condition.
\subsection{Burgers' equation: Forward problem}

The goal in forward settings is to approximate the solution of a partial differential equation (PDE) given with full governing equation and its initial and boundary conditions. This study uses Burgers' equation as a primary example.

W-PIRNNs implement the PDE restrictions through a residual-based neural architecture utilizing a wavelet activation function, that is, 
$$
W(t) = w_1 \sin(t) + w_2 \cos(t)
$$
 
This activation function increases the model's effectiveness in allowing it to efficiently capture high-frequency components of the solution. The network parameters are optimized by minimizing a composite loss function that combines physics-informed residual loss with supervised data loss whenever needed. In one spatial dimension, the Burgers equation with Dirichlet boundary conditions is written as,

\begin{equation}
\begin{aligned}
& \frac{\partial u}{\partial t} + u \frac{\partial u}{\partial x} - \frac{0.01}{\pi} \frac{\partial^2 u}{\partial x^2} = 0, \quad && \text{for } x \in [-1, 1], \ t \in [0, 1], \\
& u(x, 0) = -\sin(\pi x), \quad \\
& u(-1, t) = u(1, t) = 0. \quad
\end{aligned}
\label{equ:Burgers_forward}
% \tag{a.1}
\end{equation}

The residual function of equation~\ref{equ:Burgers_forward} is defined by:

\begin{equation}
\mathrm{f}(x, t) = \frac{\partial u}{\partial t} + u \frac{\partial u}{\partial x} - \frac{0.01}{\pi} \frac{\partial^2 u}{\partial x^2}
% \tag{a.2}
\end{equation}

\begin{figure}[htbp]
    \centering
    \includegraphics[width=1.0\textwidth]{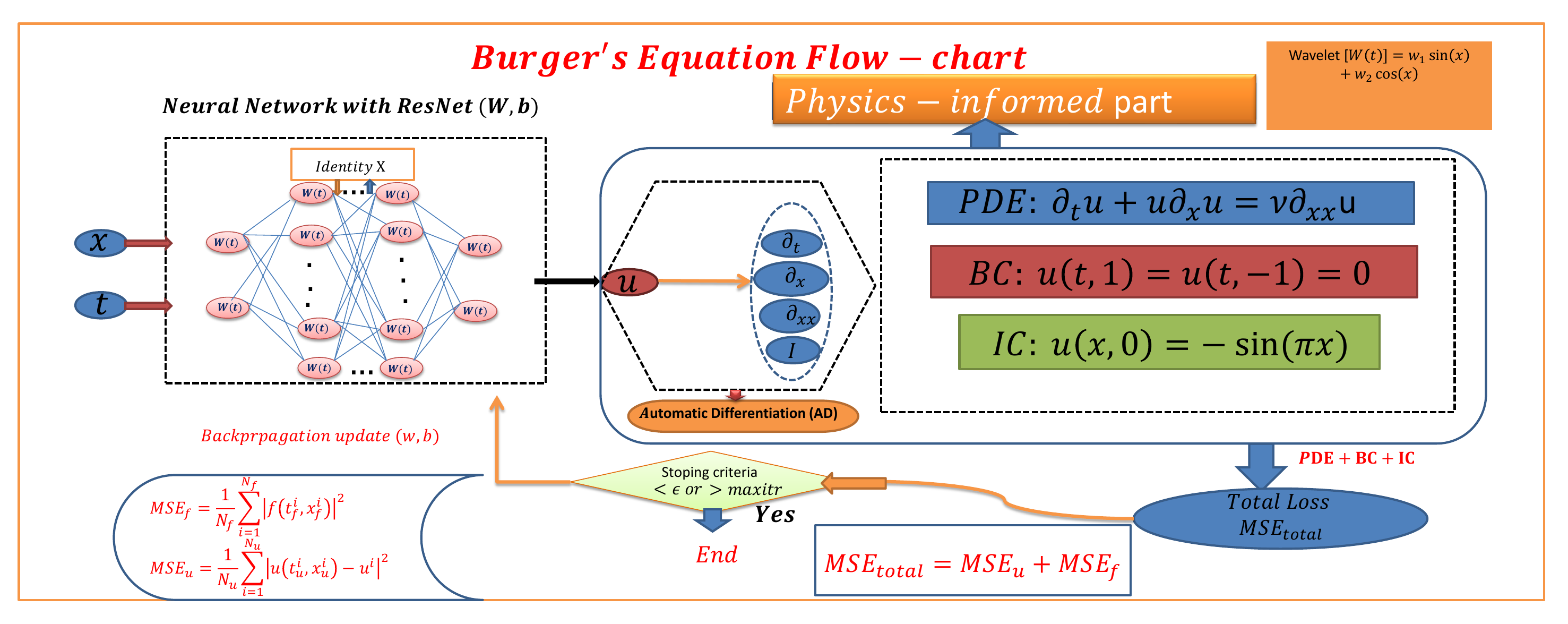}
    \caption{This flowchart illustrates the solution of Burgers' equation via the proposed method.}
    \label{fig:burger_flowchart}
\end{figure}

We use a wavelet-physics-informed residual neural network (W-PIRNN) as a function approximator to approximate the unknown solution \(u(x, t) \), allowing for the learning of complicated spatio-temporal patterns inherent to the governing equation. The more details for our W-PIRNNs are illustrated in Figure~\ref{fig:burger_flowchart}.

The parameters (weight and bias) for the $u(x, t)$ and $\mathrm{f}(x, t)$ neural network approximations can be obtained by minimizing the mean squared error loss, which is defined as follows,

\begin{equation}
\mathcal{L}_{total} = \mathcal{L}_{ic/bc} + \mathcal{L}_\mathrm{residual}
% \tag{a.2}
\end{equation}

where

\begin{equation}
\mathcal{L}_{ic/bc} = \frac{1}{\mathcal{N}_u} \sum_{i=1}^{\mathcal{N}_u} \left| u(x_u^i,t_u^i) - u^i \right|^2
% \tag{a.3}
\end{equation}

and

\begin{equation}
\mathcal{L}_{residual} = \frac{1}{\mathcal{N}_\mathrm{f}} \sum_{i=1}^{\mathcal{N}_\mathrm{f}} \left| \mathrm{f}(x_\mathrm{f}^i,t_\mathrm{f}^i) \right|^2
% \tag{a.4}
\end{equation}

Here, $\left\{x_u^i,t_u^i, u^i\right\}_{i=1}^{\mathcal{N}_u}$ represents the initial and boundary training data for $u(x, t)$, whereas $\left\{x_\mathrm{f}^i,t_\mathrm{f}^i\right\}_{i=1}^{\mathcal{N}_\mathrm{f}}$ represents the collocation points for $\mathrm{f}(x, t)$. The loss $\mathcal{L}_{ic/bc}$ refers to the initial and boundary conditions, while $\mathcal{L}_{residual}$ denotes the residual loss or physics loss. To solve burgers' equation using proposed W-PIRNNs, we take 7 hidden layers, with 20 neurons per layer. In our proposed algorithm in the multilayer perceptron (MLP), we employ residual blocks with skip connections, using $W(t) = w_1 \sin(t) + w_2 \cos(t)$ as an activation function for each neuron. In these examples, we take a total of 100 randomly distributed initial and boundary data $(\mathcal{N}_u)$ and a total of 10000 collocation points $(\mathcal{N}_f)$ in the interior domain of the problem. For better sampling of the collocation points, we use the Latin hypercube sampling strategy. For the model training, we use Adam~\citep{adam2014method} with an initial learning rate (lr=0.001) for the first few epochs, then switch to the second-order optimizer L-BFGS~\citep{liu1989limited}.  The flowchart of our proposed network in Figure~\ref{fig:burger_flowchart} is attached.

\begin{figure}[htbp]
    \centering
    \includegraphics[width=0.75\textwidth]{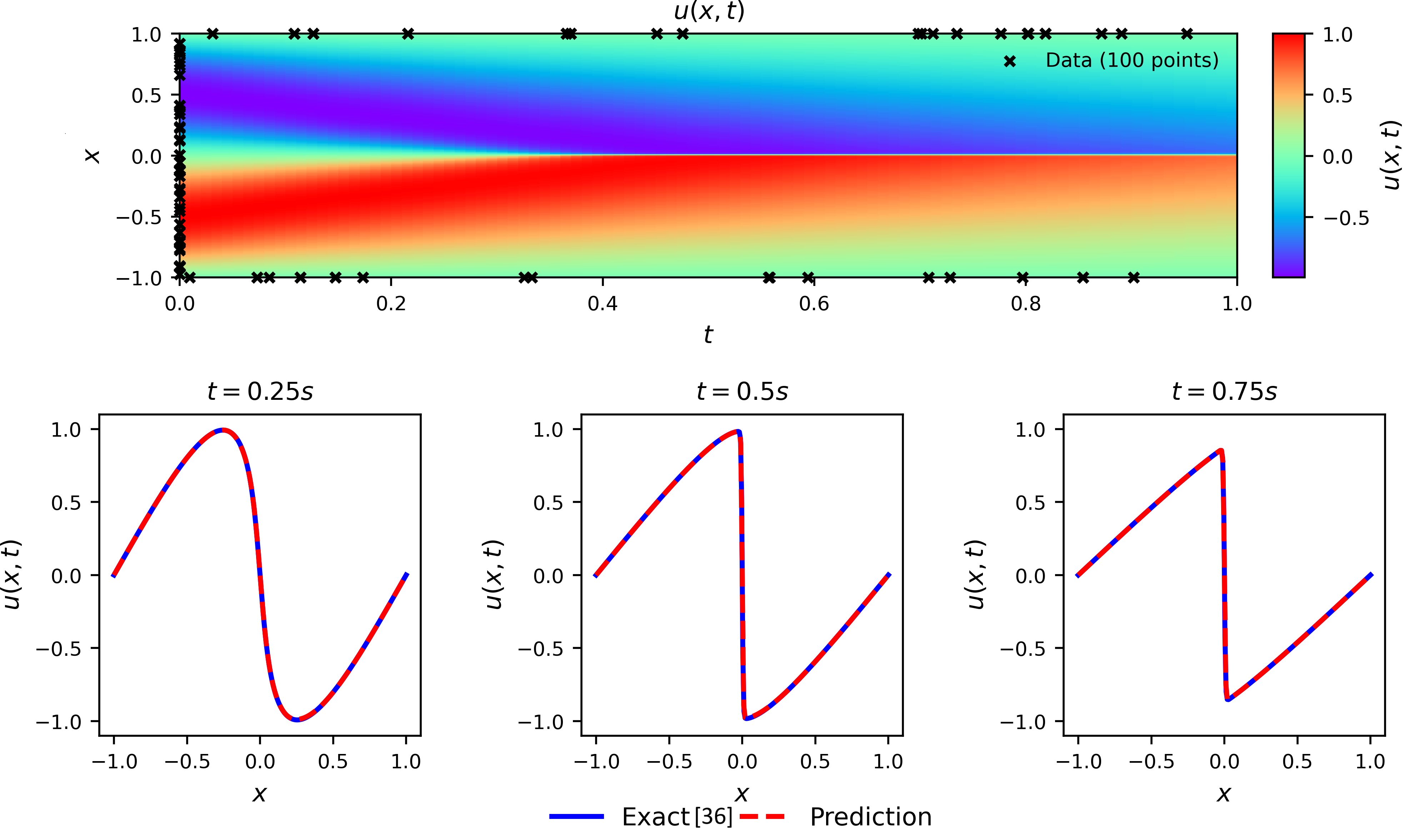} % Replace with your actual image filename

    \caption{Burger's forward problem: The top row is our predicted solution u(x,t) along with initial and boundary training data $(\mathcal{N}_u=100)$ and collocation points $(\mathcal{N}_f)=10000$. The bottom row shows the comparison between the exact~\citep{raissi2019physics} and predicted results for three temporal snapshots at t=0.25s, t=0.50s, and t=0.75s, respectively. }
    \label{fig:burgers_forward}
\end{figure}

The Burgers' equation is widely recognized as an excellent illustration of a hyperbolic conservation law (as $v \rightarrow 0$). The residual function $\mathrm{f}(x, t)$ is derived by using the automatic differentiation. In our study, we substitute $u(x, t)$ with a neural network approximation $u(x, t ; {W}, {b})$ and the residual $\mathrm{f}(x, t)$ with a neural network approximation $\mathrm{f}(x, t ; {W}, {b})$ calculated using automatic differentiation. Thus, the resultant pair $u(x, t ; {W}, {b})$ and $\mathrm{f}(x, t ; {W}, {b})$ must adhere to the Burgers' equation irrespective of the selection of the weight ${W}$ and bias ${b}$ parameters. During training, using the datasets $x_u^i,t_u^i$ and $x_f^i,t_f^i$, we aim to identify the optimal parameters ${W}^*$ and ${b}^*$ to achieve an optimal fit to both the training data and the residuals of the differential equation. Throughout this process, our aim is to minimize these losses as close as possible to zero.
The exact solution to this problem is available, and we collect the dataset from these paper~\citep{raissi2019physics}. By using these exact solutions and our model prediction, the relative $L_2$ error we achieved was $1.23 \times 10^{-4}$ by using $7$ hidden layers with 20 neurons per layer, which is comparable to PINNs results, where their architecture is $9$ hidden layers with $20$ neurons per layer. For further details, see Table~\ref{tab:burger_forward}. We compare the exact and predicted solutions at specific time intervals: $t = 0.25, 0.50,$ and $0.75$ that also closely align with the exact solution. In Figure~\ref{fig:burgers_forward}, the top row displays our proposed algorithm's predicted solution $u(x,t)$, while the bottom row presents three different time snapshots: t=0.25s, t=0.50s, and t=0.75s, along with a comparison of the exact and predicted solutions. Using a limited set of initial and boundary conditions, the wavelet-physics-informed residual neural network can accurately capture the complex nonlinear dynamics of the Burgers’ equation.

\begin{table}[htbp]
\centering
\caption{Burgers' forward problem: Network configuration and relative $L_2$ error comparisons. In these examples, both models have the same total number of initial and boundary training data $(\mathcal{N}_u=100)$ and collocation points $(\mathcal{N}_f=10000)$.}
\label{tab:architecture_burgers}
\renewcommand{\arraystretch}{1.2}
\begin{tabular}{@{}lcccc@{}}
\toprule
\textbf{Method} 
& \textbf{Hidden Layers} 
& \textbf{Neurons/Layer} 
& \textbf{Activation Function} 
& \textbf{Relative $L_2$ Error} \\
\midrule
PINN~\citep{raissi2019physics} 
& 9 
& 20 
& Tanh 
& $6.7 \times 10^{-4}$ \\

Proposed W-PIRNNs 
& \textbf{7} 
& 20 
& Wavelet 
& $\mathbf{1.23 \times 10^{-4}}$ \\
\bottomrule
\end{tabular}
\label{tab:burger_forward}
\end{table}

\subsection{Burgers' equation: Inverse problem}

In inverse problems, the goal is to identify unknown parameters, source terms, or missing field values using known governing equations. W-PIRNNs excel at these tasks because of their physics-integrated learning framework.

To begin, we look at the Burgers' equation. This equation appears in several areas of applied mathematics, including fluid mechanics, nonlinear acoustics, gas dynamics, and traffic flow. It is a fundamental partial differential equation that can be derived from the Navier-Stokes equations for the velocity field by removing the pressure gradient factor. Burgers' equation can generate shocks at low viscosity, which are frequently hard to resolve using traditional numerical approaches. In one space dimension, the equation is written as

\begin{equation}
\frac{\partial u}{\partial t} + \lambda_1 u \frac{\partial u}{\partial x} - \lambda_2 \frac{\partial^2 u}{\partial x^2} = 0,
% \tag{b.1}
\label{eq:burgers_forward}
\end{equation}

Let us define the residual function associated with the governing equation as follows:

\begin{equation}
\mathrm{f}(x, t) := \frac{\partial u}{\partial t} + \lambda_1 u \frac{\partial u}{\partial x} - \lambda_2 \frac{\partial^2 u}{\partial x^2}
% \tag{b.2}
\end{equation}

Firstly, approximate the solution u(x,t) using a wavelet-physics-informed residual neural network, the same as the forward problem~\ref{fig:burger_flowchart}. Then the residual function \( \mathrm{f}(x, t) \) is approximated by using automatic differentiation. The common parameters (weight and bias) of the neural networks $u(x, t)$ and $\mathrm{f}(x, t)$, along with the parameters $\lambda=\left(\lambda_1, \lambda_2\right)$, are learned by minimizing the mean squared error loss,

\begin{align}
\mathcal{L}_{total} = \mathcal{L}_{data} + \mathcal{L}_{residual}
% \tag{b.3}
\end{align}

where

\begin{align}
\mathcal{L}_{data} = \frac{1}{\mathcal{N}} \sum_{i=1}^\mathcal{N} \left| u\left(t_u^i, x_u^i\right) - u^i \right|^2
% \tag{b.4}
\end{align}

and

\begin{align}
\mathcal{L}_{residual} = \frac{1}{\mathcal{N}} \sum_{i=1}^\mathcal{N} \left| \mathrm{f}\left(t_u^i, x_u^i\right) \right|^2
% \tag{b.5}
\end{align}

Here, $\left\{x_u^i,t_u^i, u^i\right\}_{i=1}^N$ represents the training dataset for $u(x, t)$.  $\mathcal{L}_{data}$ corresponds to the data loss, whereas $\mathcal{L}_{residual}$ corresponds to the residual loss, which enforces the physical phenomena of the problem.

\begin{figure}[htbp]
    \centering
    \includegraphics[width=0.8\textwidth]{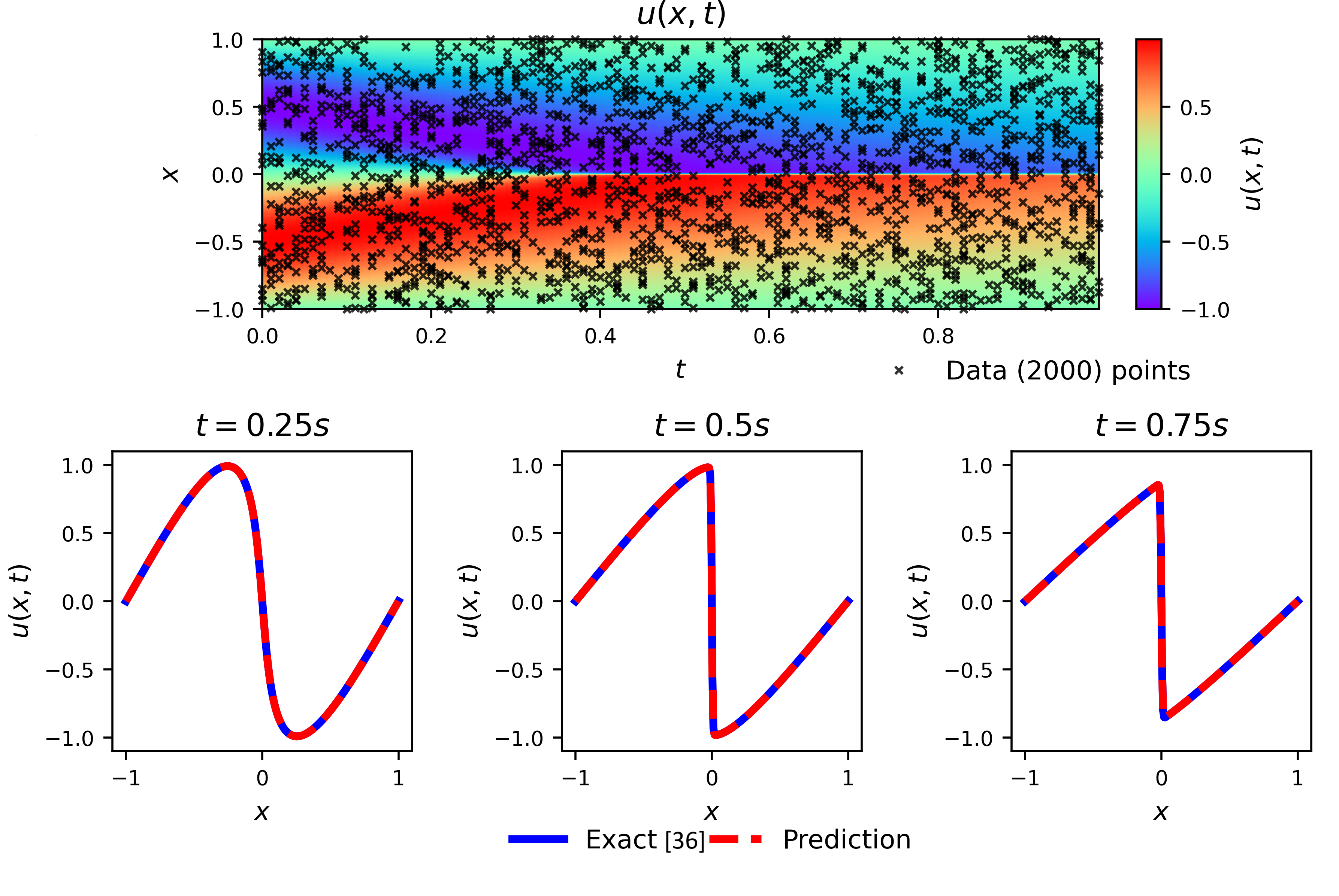}
    \caption{Burger's inverse problem: The top row shows the predicted solution u(x,t) along with training data $(\mathcal{N}=2000)$. The bottom row shows the comparison between the exact~\cite{raissi2019physics} and predicted solutions at three temporal snapshots: t=0.25s, t=0.50s, and t=0.75s, respectively.}
    \label{fig:burgers_inverse}
\end{figure}

To demonstrate the efficacy of our methodology, we have constructed a training dataset~\citep{raissi2019physics} by randomly generating $\mathcal{N}=2,000$ points throughout the complete spatio-temporal domain based on the predicted solution associated with $\lambda_1=1.0$ and $\lambda_2=0.01 / \pi$. The positions of the training points are depicted in the upper panel of Figure~\ref{fig:burgers_inverse}. This data loss $(\mathcal{L}_{data})$, along with the residual loss $(\mathcal{L}_{residual})$, is used to train a $7$-layer deep neural network with $20$ neurons per hidden layer and uses wavelet activation in each layer. Minimizing mean squared error loss $(\mathcal{L}_{total})$ using the Adam~\citep{adam2014method} optimizer with an initial learning rate (lr=0.001), the first few epochs, and then switched to the L-BFGS optimizer \citep{liu1989limited}. The network configuration details are shown in Table~\ref{tab:burgers_inverse}. During training, the network is adjusted to the entire solution $u(t, x)$ and the unknown parameters $\lambda = \left (\lambda_1, \lambda_2 \right)$ that characterize the underlying dynamics. The network can accurately identify the underlying partial differential equation in Table~\ref{tab:identified_PDE}.

\begin{table}[htbp]
\centering
\caption{Burger's inverse problem: Architecture comparison of PINNs~\citep{raissi2019physics} and proposed W-PIRNNs.}
\begin{tabular}{@{}lccc@{}}
\toprule
\textbf{Method} & \textbf{Hidden Layers} & \textbf{Neurons/Layer} & \textbf{Collocation Points} \\
\midrule
PINNs~\citep{raissi2019physics} & 8 & 20 & 2,000 \\
Proposed W-PIRNNs & \textbf{7} & 20 & 2,000 \\
\bottomrule
\end{tabular}
\label{tab:burgers_inverse}
\end{table}

Proposed W-PIRNNs accurately estimate the unknown parameters $\lambda_1 = 1.00002$ and $\lambda_2 = 0.0031810$, with absolute errors of 0.002\% and 0.065\% for clean data, respectively. We also compare the absolute error percentages of the identified parameters with PINNs~\citep{raissi2019physics}, which are also remarkable; the comparable results are shown in Table~\ref{tab:burgers_param_error_clean}. Using 2000 training data $(\mathcal{N})$ and the partial differential equation~\ref{eq:burgers_forward}, proposed wavelet-physics-informed residual neural networks (W-PIRNNs) efficiently solve the equation; the solution $u(x,t)$ is shown in Figure~\ref{fig:burgers_inverse}. Moreover, the proposed W-PIRNNs efficiently identified the governing equation; see Table~\ref{tab:identified_PDE} for details. The dataset of exact solutions to these equations was collected from~\citep {raissi2019physics}.

\begin{table}[htbp]
\centering
\caption{Burger's inverse problem: Comparison of PINNs~\citep{raissi2019physics} and the proposed W-PIRNNs, absolute errors (in \%) for the identified parameters \(\lambda_1\) and \(\lambda_2\) (clean data).}
\label{tab:burgers_param_error_clean}
\renewcommand{\arraystretch}{1.2}
\begin{tabular}{lcc}
\toprule
Method & \(\lambda_1\) Error (\%) & \(\lambda_2\) Error (\%) \\
\midrule
PINNs~\citep{raissi2019physics} & 0.141 & 1.902 \\
Proposed W-PIRNNs & \textbf{0.002} & \textbf{0.065} \\
\bottomrule
\end{tabular}
\end{table}

\begin{table}[htbp]
\centering
\caption{Burger's inverse problem: Comparison of the correct partial differential equations (PDEs) along with identified PDEs.}
\label{tab:identified_PDE}
\renewcommand{\arraystretch}{1.0} % Increase row height
{
\begin{tabular}{|c|c|}
\hline
\textbf{Correct PDE} & $u_t + uu_x - 0.0031831 u_{xx} = 0$ \\
\hline
\textbf{Identified PDE (clean data)} & $u_t + \mathbf{1.00002} u u_x - \mathbf{0.0031810} u_{xx} = 0$ \\
\hline
\end{tabular}
}
\end{table}

\subsection{Schrodinger equation: Forward problem}
This example demonstrates our method's ability to handle periodic boundary conditions, complex-valued solutions, and various nonlinearities in the governing partial differential equations. The one-dimensional nonlinear Schrödinger equation is a classical field equation used to study quantum-mechanical systems, such as nonlinear wave propagation in optical fibers and waveguides, Bose-Einstein condensates, and plasma waves. The nonlinear term in optics arises from a material's intensity-dependent index of refraction. Similarly, the nonlinear term for Bose-Einstein condensates arises because of mean-field interactions in an N-body system. The nonlinear Schrödinger equation, along with periodic boundary conditions, is given by

\begin{equation}
\begin{aligned}
& \mathrm{i}\, h_t + \frac{1}{2} h_{xx} + |h|^{2} h = 0,
\quad x \in [-5,5], \; t \in \left[0,\frac{\pi}{2}\right], \\[4pt]
& h(x,0) = 2\,\mathrm{sech}(x), \\[4pt]
& h(-5,t) = h(5,t), \\[4pt]
& h_x(-5,t) = h_x(5,t).
\end{aligned}
\label{eq:nls_system}
\end{equation}

where $h(x, t)$ is the complex-valued solution. Let $f(x, t)$ be the residual function associated with equation~\ref{eq:nls_system}, which is defined by

\begin{equation}
f := \mathrm{i}\, h_t + \frac{1}{2} h_{xx} + |h|^{2} h.
\label{eq:nls_residual}
\end{equation}

We approximate the complex-valued function \(h(x,t)\) using wavelet-physics-informed residual neural networks (W-PIRNNs). Where \(u(x,t)\) and \(v(x,t)\) denote the real and imaginary parts of \(h(x,t)\), respectively, such that the network output can be written as,

$$
h(x,t) = [\,u(x,t)\;\; v(x,t)\,].
$$

This construction naturally needs a multi-output wavelet-physics-informed residual neural network for the associated residual \(f(x,t)\). The parameters (weight and bias) between the networks' approximation \(h(x,t)\) and \(f(x,t)\) are identified through minimization of the mean squared error loss,

\begin{equation}
\mathcal{L}_{total} = \mathcal{L}_{ic} + \mathcal{L}_{bc} + \mathcal{L}_{residual}.
\label{eq:total_mse}
\end{equation}

where

\begin{equation}
\begin{aligned}
& \mathcal{L}_{ic}
= \frac{1}{\mathcal{N}_0} \sum_{i=1}^{\mathcal{N}_0}
\left| h\!\left(x_0^{\,i},0\right) - h_0^{\,i} \right|^{2}.
\end{aligned}
\label{eq:loss_ic}
\end{equation}

\begin{equation}
\begin{aligned}
& \mathcal{L}_{bc}
= \frac{1}{\mathcal{N}_b} \sum_{i=1}^{\mathcal{N}_b}
\Big(
\left| h\!\left(-5,t_b^{\,i}\right) - h\!\left(5,t_b^{\,i}\right) \right|^{2}
+
\left| h_x\!\left(-5,t_b^{\,i}\right) - h_x\!\left(5,t_b^{\,i}\right) \right|^{2}
\Big).
\end{aligned}
\label{eq:loss_bc}
\end{equation}

\begin{equation}
\begin{aligned}
& \mathcal{L}_{residual}
= \frac{1}{\mathcal{N}_f} \sum_{i=1}^{\mathcal{N}_f}
\left| f\!\left(x_f^{\,i},t_f^{\,i}\right) \right|^{2}.
\end{aligned}
\label{eq:loss_residual}
\end{equation}

Here, $\left\{x_0^i, h_0^i\right\}_{i=1}^{\mathcal{N}_0}$ is the initial data, $\left\{t_b^i\right\}_{i=1}^{\mathcal{N}_b}$ corresponds to the boundary collocation point, and $\left\{x_f^i,t_f^i\right\}_{i=1}^{\mathcal{N}_f}$ is the collocation points on the residual ($f(x, t)$). $\mathcal{L}_{ic}$ represents the loss on the initial training data, $\mathcal{L}_{bc}$ is the loss corresponding to the boundary collocation points, and $\mathcal{L}_{residual}$ is the residual loss (i.e., the loss trying to enforce the Schrodinger equation to model), respectively.

\begin{figure}[htbp]
    \centering
    \includegraphics[width=0.95\linewidth]{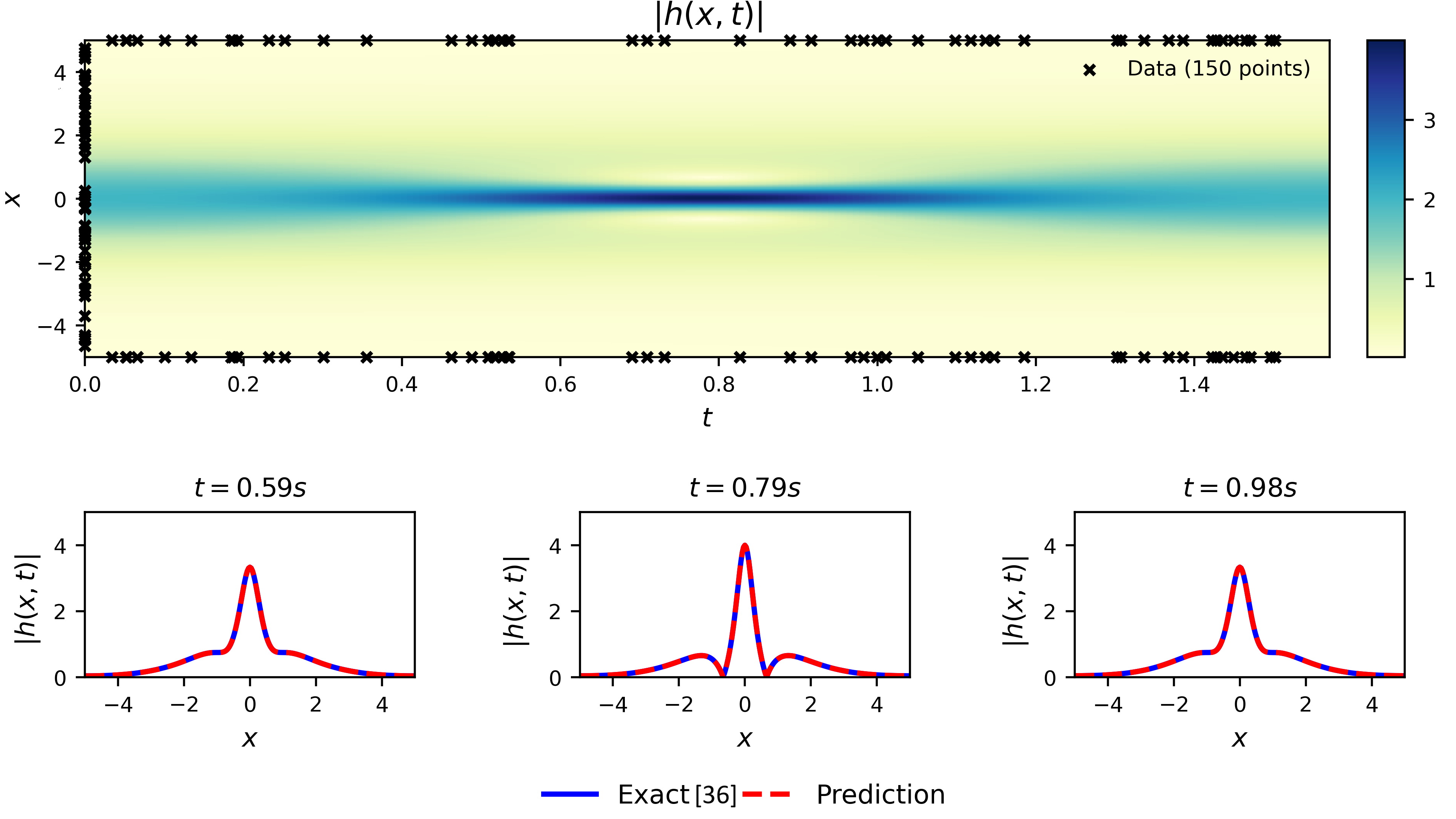}
    \caption{Schrodinger forward problem: The top row shows the predicted solution, |h(x,t)|, along with the initial and boundary data. The bottom row shows the comparison between the exact~\citep{raissi2019physics} and predicted solutions at three temporal snapshots: t=0.59s, t=0.79s, and t=0.98s.}
    \label{fig:schrodinger_results}
\end{figure}

This example shows the efficiency of the proposed method for enforcing periodic boundary conditions. To solve these equations, we fixed the W-PIRNNs framework with a 4-layer hidden network of 20 neurons per layer and used wavelet activations for all layers except the output layer. For network parameters (weight and bias), optimized using the Adam~\citep{adam2014method} optimizer with initial learning rate (lr=0.001) for the first few epochs, then switched to L-BFGS~\citep{liu1989limited} to obtain the best-optimized parameter values. The training set consists of a total of $\mathcal{N}_{ic}=50$ data points on $h(x, 0)$ randomly parsed from the complete high-resolution data set, as well as $\mathcal{N}_{bc}=50$ randomly sampled collocation points $\left\{t_b^i\right\}_{i=1}^{\mathcal{N}_{bc}}$ for enforcing the periodic boundaries. Moreover, we set $\mathcal{N}_{f} = 20,000$ randomly sampled collocation points to implement equation~\ref{eq:nls_system} within the solution domain. All randomly sampled point locations were generated using a space-filling Latin hypercube sampling strategy. In the top panel of Figure~\ref{fig:schrodinger_results}, the magnitude of the predicted solution $|h(x, t)|=\sqrt{u^2(x, t)+v^2(x, t)}$ is shown, along with the locations of the initial and boundary training data. Moreover, we validate our prediction by using relative $L_2$ error, which is measured at $1.02 \cdot 10^{-3}$; it is also comparable to PINNs\citep{raissi2019physics} in details, as shown in Table~\ref{tab:schrodinger_equ}. In the bottom panel of Figure~\ref{fig:schrodinger_results}, we compare the exact and predicted solutions at the three different time instants $t=0.59s, 0.79s, 0.98s$. The exact solution dataset was taken from this paper~\citep{raissi2019physics}.
\begin{table}[htbp]
\centering
\caption{Schrodinger forward problem: Comparison of the PINNs~\citep{raissi2019physics} and proposed W-PIRNNs architecture and relative $L_2$ error.}
\label{tab:architecture_schrodinger}
\renewcommand{\arraystretch}{1.2}
\begin{tabular}{@{}lcccc@{}}
\toprule
\textbf{Method} 
& \textbf{Hidden Layers} 
& \textbf{Neurons/Layer} 
& \textbf{Activation Function} 
& \textbf{Relative $L_2$ Error} \\
\midrule
PINNs~\citep{raissi2019physics} 
& 5 
& 100 
& Tanh 
& $\,1.97 \times 10^{-3} $ \\

Proposed W-PIRNNs 
& \textbf{4} 
& 100 
& Wavelet 
& $\, \mathbf{1.30 \times 10^{-3}} $ \\
\bottomrule
\end{tabular}
\label{tab:schrodinger_equ}
\end{table}

\newpage

\section{Conclusion}
\label{sec:conclusion}
This study introduced wavelet-physics-informed residual neural networks (W-PIRNNs), a novel deep learning architecture designed to accurately reconstruct complex flow fields, specifically pressure, streamline, and vorticity, using only sparse velocity data for supervision. In this supervision, we examined various levels of sparsity, i.e, $100\%$, $20\%$, $5\%$, $1\%$, $0.16\%$, $0.5\%$ and $0.05\%$. It is worthy to mention that, proposed algorithm can effectively reconstruct complex flow field by using as low as $0.05\%$ supervised data. We validate our results with the existing benchmark results. Additionally, we compared our results with existing results of PINNs~\citep{xu2023practical} and FFPICN~\citep{liu2025physics}.  Our results demonstrate better performance across all levels of data sparsity percentages. The main objective of the present study is to reconstruct the complex flow field with minimal sparse data; thus, we tested the performance of our proposed algorithm under extremely sparsity with $0.05\%$ supervised data, for Re=100, which effectively reconstructs dynamic vorticity, streamlines, pressure, and both u \& v velocity fields. Moreover, for the overall loss component, all supervised datasets smoothly reach to a stable minimum. We also validated our proposed algorithm to reconstruct the more complex flow field at Re=3900, using only 0.09\% sparse supervised velocity and pressure data. Overall, proposed algorithm shows an excellent performance to reconstruct complex flow field using extremely sparse dataset.

We also have shown that proposed algorithm effeciently solve the forward and inverse problems. We considered Burgers' equation for forward and inverse problems and the Schrödinger equation for forward problem to validate our proposed algorithm. Our proposed algorithm efficiently calculated the solution of Bugers' equation (for both forward and inverse problems) using 7 hidden with 20 neurons per layer, and Schrodinger equation (forward problem) using 
4 hidden layers with 100 neurons per layer, wheras the PINNs~\citep{raissi2019physics} results are calculated with more hidden layer for both forward and inverse problems still it produces more errors compared to our results. Also it can be noted that for Burgers' inverse problem, the proposed algorithm identified the parameters $\lambda_1$ and $\lambda_2$ with high precision, resulting in absolute errors of $0.004\%$ and $0.065\%$ for clean data, respectively.  
This illustrates the efficiency of our proposed algorithm in addressing both forward and inverse problems, which is noteworthy. 

\section*{Author Declaration}
The authors have no conflicts of interest to disclose.

\section*{Data Availability}
The data that support the findings of this study are available from the corresponding author upon reasonable request.

% \clearpage
\appendix
\section*{Appendix A. Ablation study on activation functions}
\label{app:activation_ablation}

To assess the robustness of the proposed framework with respect to the choice of activation function, we conduct a systematic ablation study by replacing the nonlinear activation while keeping the network architecture, optimizer, learning rate, batch size, and all other hyperparameters identical to those used in the main reconstruction experiments.

The test case considered is the two-dimensional incompressible flow past a circular cylinder at Reynolds number $\mathrm{Re}=100$. The study focuses exclusively on the wake region, where the flow exhibits strong nonlinear dynamics dominated by vortex shedding. Velocity data corresponding to both streamwise ($u$) and cross-stream ($v$) components are used. A total of $10^6$ spatio-temporal samples are available, from which only $0.5\%$ of the data are randomly selected and used as supervised measurements. The temporal window spans from $t=0.00\,\mathrm{s}$ to $t=19.90\,\mathrm{s}$.

We compare several activation functions that are widely adopted in deep learning and physics-informed learning literature, namely the sigmoid, hyperbolic tangent (tanh), SiLU, and sinusoidal representation networks (SIREN). These are benchmarked against the proposed custom wavelet activation, which corresponds to a single-frequency Fourier-based activation designed to enhance the representation of oscillatory flow features inherent to vortex-dominated wakes.

For all activation functions, the network is trained for up to $1000$ epochs. The evaluation metrics include the final training loss, the relative $L_2$ errors of the velocity components $u$ and $v$, and the total wall-clock training time measured in hours. The relative error is computed as
\setcounter{equation}{0}
\renewcommand{\theequation}{A.\arabic{equation}}

\begin{equation}
\text{Relative } L_2 \text{ Error}
= \frac{\| \mathbf{u}_{\text{pred}} - \mathbf{u}_{\text{ref}} \|_2}
       {\| \mathbf{u}_{\text{ref}} \|_2},
\end{equation}

where $\mathbf{u}_{\text{pred}}$ and $\mathbf{u}_{\text{ref}}$ denote the predicted and reference velocity fields, respectively.

Table~\ref{tab:activation_ablation} summarizes the quantitative comparison. Although the proposed wavelet activation incurs a slightly higher computational cost when training is extended to the full $1000$ epochs, it consistently achieves the lowest reconstruction errors. Notably, the wavelet-based model converges significantly faster and attains its best performance within approximately $500$ epochs, at which point the runtime becomes comparable to that of the baseline activations.

This trend is not limited to the $0.5\%$ supervision case reported here; similar early-convergence behavior is observed across all sparsity levels considered in this work. These results indicate that the proposed activation function improves optimization efficiency and expressive capability for oscillatory flow reconstruction, enabling accurate predictions with fewer training iterations and competitive computational cost.

\begin{table}[htbp]
\centering
\caption{
Comparison of different activation functions for the $\mathrm{Re}=100$ cylinder wake reconstruction using $0.5\%$ supervised data. The table reports the final training loss, the relative $L_2$ errors of the velocity components $u$ and $v$, and the total wall-clock runtime in hours. All models are trained with identical hyperparameters for a maximum of $1000$ epochs.
}
\label{tab:activation_ablation}
\renewcommand{\arraystretch}{1.2}
\begin{tabular}{lcccc}
\toprule
\textbf{Activation Function} 
& \textbf{Loss} 
& \textbf{${L_2}(u)$} 
& \textbf{${L_2}(v)$} 
& \textbf{Runtime (hrs)} \\
\midrule
Sigmoid
& $6.44 \times 10^{-3}$
& $0.0352$
& $0.1993$
& 3 \\

Tanh 
& $1.48 \times 10^{-4}$ 
& $0.0043$ 
& $0.0116$ 
& 3 \\

SiLU 
& $4.12 \times 10^{-5}$ 
& $0.0037$ 
& $0.0087$ 
& 4 \\

SIREN (Sinusoidal) 
& $4.35 \times 10^{-5}$ 
& $0.0039$ 
& $0.0089$ 
& 3 \\

Wavelet Activation (epoch = 500)
& $4.07 \times 10^{-5}$
& $\mathbf{0.0035}$
& $\mathbf{0.0086}$
& \textbf{2} \\

\textbf{Wavelet Activation (epoch=1000)} 
& $\mathbf{1.05 \times 10^{-5}}$ 
& $\mathbf{0.0010}$ 
& $\mathbf{0.0055}$ 
& \textbf{4} \\
\bottomrule
\end{tabular}
\end{table}

\section*{Appendix B. Ablation study on the core components of W-PIRNNs}
\label{app:wpirnns_ablation}

The proposed W-PIRNNs framework incorporates two key design elements: (i) residual-based training to mitigate optimization difficulties associated with deep neural networks, and (ii) a custom wavelet-based activation function aimed at improving the representation of oscillatory flow dynamics. In this appendix, we quantify the individual and combined contributions of these components through a controlled ablation study.

The benchmark problem is the two-dimensional incompressible flow past a circular cylinder at Reynolds number $\mathrm{Re}=100$, with emphasis on the wake region where vortex shedding dominates the flow physics. Velocity data for both streamwise ($u$) and cross-stream ($v$) components are considered over the temporal interval $t \in [0.00, 19.90]\,\mathrm{s}$. A total of $10^6$ spatio-temporal samples are available, from which only $0.5\%$ are randomly selected and used as supervised training data.

To isolate the effect of each architectural component, four model variants are evaluated:
\begin{itemize}
    \item \textbf{SIREN PINNs}: vanilla PINNs equipped with sinusoidal activation functions.
    \item \textbf{Wavelet PINNs}: vanilla PINNs using the proposed custom wavelet activation without residual connections.
    \item \textbf{ResNet PINNs}: residual-based PINNs employing standard nonlinear activations (SIREN).
    \item \textbf{W-PIRNNs (proposed)}: the combination of residual connections and the custom wavelet activation.
\end{itemize}

All models share an identical network architecture, optimizer settings, learning rate schedule, and loss weighting strategy, consistent with those employed in the main flow reconstruction experiments across different sparsity levels. Each model is trained for a maximum of $1000$ epochs to enable a fair, one-to-one comparison.

The quantitative results are summarized in Table~\ref{tab:wpirnns_ablation}. The evaluation metrics include the final total loss, the relative $L_2$ errors of the predicted velocity components $u$ and $v$, and the total wall-clock training time measured in hours.

The results indicate that vanilla PINNs with sinusoidal activation (SIREN PINNs) perform poorly for this configuration, exhibiting comparatively higher losses and unbalanced velocity reconstruction errors. Introducing either residual connections or the wavelet-based activation leads to noticeable improvements, yielding better loss balance and reduced relative $L_2$ errors. The proposed W-PIRNNs achieve the best overall performance, consistently outperforming all baseline variants in accuracy.

Although wavelet-based models incur higher computational cost when trained to the full $1000$ epochs, it is important to note that both Wavelet PINNs and the proposed W-PIRNNs achieve comparable accuracy at substantially earlier stages of training. As discussed in Appendix~A, this early convergence implies that, in practical applications, the proposed framework can achieve high-fidelity reconstructions with reduced training time within fewer epochs. For consistency, the results reported here correspond to simulations run for the same number of epochs across all models.

\begin{table}[htbp]
\centering
\caption{
Ablation study on the core architectural components of W-PIRNNs for the $\mathrm{Re}=100$ cylinder wake reconstruction using $0.5\%$ supervised data. The table reports the final total loss, the relative $L_2$ errors of the velocity components $u$ and $v$, and the total wall-clock runtime in hours. All models are trained with identical hyperparameters for a maximum of $1000$ epochs.
}
\label{tab:wpirnns_ablation}
\renewcommand{\arraystretch}{1.2}
\begin{tabular}{lcccc}
\toprule
\textbf{Model} 
& \textbf{Loss} 
& \textbf{${L_2}(u)$} 
& \textbf{${L_2}(v)$} 
& \textbf{Runtime (hrs)} \\
\midrule
Wavelet PINNs 
& $3.48 \times 10^{-5}$ 
& $0.0023$ 
& $0.0076$ 
& 3.5 \\

SIREN PINNs 
& $1.35 \times 10^{-4}$ 
& $0.0086$ 
& $0.0108$ 
& 3 \\

ResNet-PINNs (SIREN + Skip Connections) 
& $4.35 \times 10^{-5}$ 
& $0.0039$ 
& $0.0089$ 
& 3 \\

\textbf{W-PIRNNs (Wavelet + ResNet)} 
& $\mathbf{1.05 \times 10^{-5}}$ 
& $\mathbf{0.0010}$ 
& $\mathbf{0.0055}$ 
& \textbf{4} \\
\bottomrule
\end{tabular}
\end{table}

    %============================= RESULTS AND DISCUSSION===============================
    % \input{wpirnns}
    % \input{results}
    % \input{casestudy}
    % \input{conclusion}
    % \newpage

% 	%============================= CONCLUSION===============================
	
		\clearpage	%============================= REFERENCE===============================
				\bibliographystyle{model1-num-names}
				% \bibliographystyle{cas-model2-names}
				
				\bibliography{Reference}

			\end{document}